\documentclass[10pt,twocolumn,showpacs,preprintnumbers,amsmath,amssymb,prb,nofootinbib]{revtex4-2}
\usepackage{graphicx}  %
\usepackage{bm}        %
\usepackage[utf8]{inputenc}
\usepackage{xcolor}
\usepackage{amsmath}
\usepackage{float}
\usepackage{url}
\usepackage{hyperref}

\usepackage{layouts}

\newcommand{\mle}{\text{mle}}

\newcommand{\hatx}{\hat{\bm{x}}}
\newcommand{\hatalpha}{\hat{\bm{\alpha}}}

\newcommand{\hatA}{\hat{\bm{A}}}
\newcommand{\hatB}{\hat{\bm{B}}}

\newcommand{\hatX}{\hat{\bm{X}}}
\newcommand{\hatZ}{\hat{\bm{Z}}}
\newcommand{\hSigma}{\hat{\Sigma}}
\newcommand{\hLambda}{\hat{\Lambda}}
\newcommand{\bLambda}{\bar{\Lambda}}
\newcommand{\nHalf}{\left \lfloor n/2 \right \rfloor}
\newcommand{\minusnHalf}{\left \lfloor -n/2 \right \rfloor}

\begin{document}

\newcommand{\HGF}[4]{{#1}\left(\genfrac.|{0pt}{}{#2}{#3} {#4} \right)}
\newcommand{\Foneone}[3]{\HGF{F}{#1}{#2}{#3} }
\newcommand{\Ftwoone}[3]{\HGF{F}{#1}{#2}{#3} }
\newcommand{\Fpq}[3]{\HGF{F}{#1}{#2}{#3} }

\title{Bayesian estimation of correlation functions}
\author{\'Angel Guti\'errez-Rubio\textsuperscript{1}}
\author{Juan S. Rojas-Arias\textsuperscript{1}}
\author{Jun Yoneda\textsuperscript{2}}
\author{Seigo Tarucha\textsuperscript{1,3}}
\author{Daniel Loss\textsuperscript{1,3,4}}
\author{Peter Stano\textsuperscript{2,5}}
\affiliation{\textsuperscript{1}RIKEN Center for Quantum Computing, 2-1 Hirosawa, Wako, Saitama, 351-0198 Japan}
\affiliation{\textsuperscript{2}Tokyo Tech Academy for Super Smart Society, Tokyo Institute of Technology, Meguro-ku, Tokyo 152-8552, Japan}
\affiliation{\textsuperscript{3}RIKEN Center for Emergent Matter Science, 2-1 Hirosawa, Wako, Saitama, 351-0198 Japan}
\affiliation{\textsuperscript{4}Department of Physics, University of Basel, Klingelbergstrasse 82, CH-4056 Basel, Switzerland}
\affiliation{\textsuperscript{5}RCQI, Institute of Physics, Slovak Academy of Sciences, Dubravska cesta 9, 845 11 Bratislava, Slovakia}

\date{\today}

\begin{abstract}
    We apply Bayesian statistics to the estimation of correlation functions.
    We give the probability distributions of auto- and cross-correlations as functions of the data.
    Our procedure uses the measured data optimally and informs about the certainty level of the estimation.
    Our results apply to general stationary processes and their essence is a non-parametric estimation of spectra.
    It allows one to better understand  the  statistical noise fluctuations, assess the correlations between two variables, and postulate parametric models of spectra that can be further tested.
    We also propose a method to numerically generate correlated noise with a given spectrum.
\end{abstract}
\maketitle

\section{Introduction}
\label{sec:introduction}

In this article, we propose a rigorous method to estimate correlation functions.
By ``rigorous'' we mean a non-parametric estimation based on Bayesian statistics, being free of ad-hoc procedures, ans\"atze, or assumptions going beyond the observed data. 
We use this notion to distinguish our approach from the orthodox methods, as we explain below.
To motivate our work, we briefly introduce the concept of correlation and discuss why its estimation can benefit from a Bayesian approach.

\emph{Correlation} is intimately related to causality and prediction.
For this reason, in physics one often deals with detecting, quantifying, explaining, and testing correlations.
A common way to analyze them is through correlation functions.
Roughly, a correlation function $\langle A(t)B(t') \rangle$ is the expected value of the product of two variables (observables) $A$ and $B$ measured at times $t$ and $t'$.

Correlation functions are ubiquitous in---and often the cornerstone of---different theories and formalisms.
They play a major role in the study of noise, particularly regarding \emph{spectra}.\cite{press2007numerical,priestley1981spectral,bretthorst1988bayesian}
Actually, a spectrum is the Fourier transform of a correlation function.
Spectral analysis has been the main tool for the study of noise in very different contexts for more than a century.\cite{schottky1918uber,johnson1925schottky,
gardner1978mathematical,milotti20021f,halford1968general,voss1978linearity,timmer1995on}
In the past, the application of spectral analysis has ranged from the pioneering works in vacuum tubes\cite{schottky1918uber} and electric circuits\cite{johnson1925schottky} to particle scattering,\cite{handel19751f} earthquakes and tides, music and geographic features,\cite{gardner1978mathematical,milotti20021f} to name a few examples.
Among the remarkable discoveries made along that path, one can highlight the almost universal appearance of $1/f$ or \emph{flicker} noise, whose origin is still under debate.\cite{milotti20021f,halford1968general,voss1978linearity,timmer1995on}
Today, noise is also in the spotlight in cutting-edge fields like quantum computing:
It requires using error-correction,\cite{epstein2014investigating} it affects coherence times, and drives spatial correlations between qubits.\cite{obrien2019quantum,boter2020spatial}
With emerging fields exploiting Noisy Intermediate-Scale Quantum (NISQ) devices,\cite{preskill2018quantum,bauer2020quantum} there is a need for a better understanding of the properties and origin of noise.

The \emph{estimation} of correlation functions emerges then as a relevant task.
However, the current standard techniques to estimate correlation functions can be subject to improvement.
By ``standard'', ``traditional'', or ``orthodox'' we mean non-Bayesian.
To cast more light on our motivations, we point out some issues in traditional parameter estimation.
Later, we introduce Bayesian statistics as the way to sort them out and achieve a better (and actually optimal, as we argue below) estimation of correlation functions, which is the main goal of our paper.

Standard statistics estimates a variable by yielding a ``best guess'' of its value, called an \emph{estimator}, and a \emph{confidence interval} that reflects its uncertainty.
Two familiar examples of estimators are the mean and the median.
In the context of spectral analysis, we can mention the \emph{periodogram}\cite{press2007numerical} (defined and discussed in the main text) as the most common estimator of the spectrum.
A fundamental issue in standard estimation is the arbitrariness in the choice of an estimator.
For example, there is no universally applicable ordering, preferring the mean over the median or vice versa.
Several arguments can support either according to different criteria, but still, in the traditional approach, one estimator has to be chosen.
Even the often adopted \emph{unbiased} estimators\cite{press2007numerical} have drawbacks under certain circumstances.\cite{jaynes2003probability}
Ultimately, there is no universal ``best estimator''.

A similar arbitrariness also affects the choice of parameters to quantify errors, test hypotheses, and, most importantly for the subject of this article, assess correlations.
For example, we will discuss in the main text the limitations of Pearson's $r$ coefficient\cite{lee-rodgers1988thirteen} to quantify the correlation between two variables.
There is always the doubt that a different parameter could serve that purpose better.
The search for ``better'' $r$ coefficients leads to a series of more complicated definitions\cite{press2007numerical}
like \emph{nonparametric} or \emph{rank correlation},\cite{press2007numerical}
that end up raising the same kind of doubts:
``It is precisely the uncertainty in interpreting the significance of the linear correlation coefficient $r$ that leads us to the important concepts of \emph{nonparametric} or \emph{rank correlation}.'' \cite{press2007numerical}
Moreover, the statistical fluctuations of these coefficients---necessary to assess the confidence interval---is another matter that quickly gets similarly nontransparent:
New (and potentially arbitrary) parameters are needed to analyze these fluctuations, making the problem scale in complexity.
The whole process becomes less transparent and more dependent on assumptions not always met in practice, such as the asymptotic conditions required by the central limit theorem.
For example, Ref.~\onlinecite{purcell1972variance} studies the statistical properties of $1/f$ noise analyzing the ``variance of the variance'' of the periodogram, namely ``the error of the error.''
Quantities like these are difficult to interpret and add elements of arbitrariness.
The variance of the variance of the variance would be the next variable to look at in a presumably never ending progression.
\footnote{On the other hand, if it is not a (two-point) correlation function but noise as a phenomenon which is in the focus, one should note that the former does not fully describe the latter. Various procedures have been put forward to test if the noise---or the noisy system---is equilibrium, Markovian, Gaussian, linear, time-reversal symmetric, or stationary \cite{voss1978linearity, nelkin1981deviation,milotti20021f}.}

This arbitrariness manifests a fundamental issue in traditional estimation:
One cannot guarantee the optimal usage of the data, or optimal processing of information, when performing any inference.
By inference, we mean, for instance, estimating a variable, assessing correlations, or testing a hypothesis.
And with ``optimal usage of the data,'' we mean that no relevant information contained in the data about the inference is lost in the inference process.

Bayesian probability theory,\cite{cox1946probability,jaynes2003probability} or Bayesian statistics, puts an end to these issues.
This theory is solidly built as an extension of logic, departing from the minimal Cox's axioms---which include Bayes' rule, hence the name.\cite{cox1946probability,jaynes2003probability}
Throughout its construction, there is no arbitrary assumption or ad hoc choice of parameters, of estimators, or of statistical tests.
Instead, Bayesian statistics focuses on the calculation of the \emph{probability distributions} that encode all our \emph{knowledge}---including measured data---, and nothing but our knowledge, about a variable.\footnote{
Remarkably, Bayesian statistics rejects the idea of observation as a sampling of the ``true'' probability distribution of a variable.
Actually, the ``true'' probability distribution is an ill-defined concept, in contrast to a probability distribution that just reflects our knowledge. These distinctions might seem irrelevant at first sight, but they become decisive on numerous occasions.
The method presented in this work is one of them, as we will explicitly point out.}
In this way, Bayesian statistics guarantees the optimal usage of the data when making inferences.
The interested reader is referred to Ref.~\onlinecite{jaynes2003probability} for a complete overview of fundamental historical problems and paradoxes of the orthodox theory that only a Bayesian approach can resolve.\footnote{
Another outstanding achievement of Bayesian statistics is the formulation of the whole statistical mechanics from information theory, requiring a minimal set of axioms \cite{jaynes1957information}.
}

Consequently, Bayesian statistics can enhance parameter estimation.
Surprisingly, beyond a few exceptions in some specific problems,\cite{bretthorst1988bayesian,granade2016practical} the Bayesian estimation of correlation functions has not been fully developed.
In this article, we take up this task.
Our work can find the following applications.
First and foremost, in the context of spectral analysis, our results allow one to estimate the spectrum \emph{including the quantification of the corresponding statistical estimation errors} (encoded in probability distributions or in the error bars calculated from them).
We emphasize that these statistical uncertainty measures (such as error bars) are produced also in the \emph{non-parametric} estimation, which is the topic of this article. 
Though we do not pursue it here, one could in turn use the results of a non-parametric estimation for a further parametric estimation, that is for estimation of parameters of a specific model of the spectrum.\cite{bretthorst1988bayesian}
Second, our results also allow one to discuss with generality several statistical properties of noise and the periodogram, like the signal-to-noise ratio (whose value we prove to be universal in the non-parametric estimation).
Third, we develop a formalism to judge---with its corresponding uncertainty---the correlation between two generic variables, without involving arbitrary parameters.
Finally, we provide a method to numerically generate correlated noise with an arbitrary spectrum.

The paper is organized as follows.
In Sec.~\ref{sec:onedim} we introduce the Bayesian formalism through a simple example and sets the basis for the estimation of correlation functions.
In Sec.~\ref{sec:autocorr} we study auto-correlations, namely the correlation functions of a time-dependent variable with itself.
This includes a subsection discussing the statistical properties of the periodogram and the generation of uncorrelated noise.
Sec.~\ref{sec:crosscorr} is parallel to Sec.~\ref{sec:autocorr} but focuses on the correlation between two variables, the cross-correlation.
It also proposes a method to generate correlated noise with an arbitrary spectrum.
In Sec.~\ref{sec:example} we illustrate our results numerically and in Sec.~\ref{sec:continuous} we give some guidelines on estimating continuous spectra.
Sec.~\ref{sec:conclusion} contains the conclusions.\\
To improve the flow of the main text, topics which are technical or can be discussed separately are delegated to appendices: 
In App.~\ref{app:gaussian} we prove that the multivariate Gaussian distribution maximizes entropy, in App.~\ref{app:priors} we derive the prior distributions from invariance principles, in App.~\ref{app:per_dist} we derive Eq.~\eqref{eq:per_dist}, App.~\ref{app:usefulIdentity} contains a useful identity for the zero-frequency case, and in App.~\ref{app:unnormalizedCrossCorrelation} we give the estimating distribution for unnormalized correlation strength. The remaining appendices are especially notable for practical usage of our formulas: in App.~\ref{app:pointAveraging} we explain how to merge points in a spectral plot, superseding artificial ``windowing'' or batching the data,  in App.~\ref{app:errorsOnInput} we explain how to rigorously correct the estimating distributions for errors on input, and in App.~\ref{app:centralLimits} we give the central-limit versions of the estimating distributions. 

\section{Variance in one dimension}
\label{sec:onedim}

In this section, we use a simple example to introduce Bayesian statistics and to set the basis of parameter estimation.
However, the results presented here go beyond mere illustration purposes and directly apply to spectral analysis.
The next sections will make the connection clear.

Before going into details, we recall a few definitions and conventions from probability theory and logic.
Everything inside the probability symbol $P$ is to be interpreted as a logical statement.
$P(a)$ represents the probability that the statement $a$ is true.
$P(a\cup b)$ is the probability that either $a$ or $b$ (or both) are true.
$P(a\cap b)$ is the probability that both $a$ and $b$ are true. However, usually the logical ``and'' operator is substituted by a comma or even omitted: $P(ab) = P(a, b) \equiv P(a\cap b)$.
The \emph{conditional probability} $P(a | b) \equiv P(ab)/P(b)$ denotes the probability that $a$ is true provided that $b$ is true.
When considering the probability that a variable $x$ has a value $x'$, one should write $P(x=x')$ (note that $x = x'$ is a logical statement), but the notation is usually simplified by omitting the variable name: $P(x') \equiv P(x=x')$.
At last, the mean of the expression $f(x)$ with the variable $x$ having a probability distribution $P(x)$ is
\begin{align}
    \langle f(x) \rangle \equiv
    \int_{-\infty}^{\infty}
    dx \,
    P(x) f(x) .
    \label{eq:mean_value}
\end{align}
The mean and variance of $x$, the two central quantities in this article, are defined as 
\begin{subequations}
\begin{align}
\nu &\equiv \langle x \rangle,\\
\Lambda &\equiv \langle (x-\nu)^2 \rangle,
\end{align}
\end{subequations}
respectively.

Having set the notation, consider a scalar variable $x$.
Assume we want to estimate its mean $\nu$ and variance $\Lambda$ given a data sample
\begin{align*}
    \hatx = \{x^{(0)}, \dots, x^{(M-1)}\}
\end{align*}
that contains $M$ values $x^{(m)}$.
One cannot directly apply the definitions $\nu = \langle x \rangle$ and $\Lambda = \langle (x-\nu)^2 \rangle$ because $P(x)$ in Eq.~\eqref{eq:mean_value} is unknown. One usually resorts to standard formulas like
\begin{align}
    \bar{\nu} = \frac{1}{M} \sum_{m=0}^{M-1}
    x^{(m)}, \quad
    \bar{\Lambda} = \frac{1}{M} \sum_{m=0}^{M-1}
    \big(x^{(m)}-\bar{\nu}\big)^2 .
    \label{eq:meanvar}
\end{align}
However, this estimation of $\nu$ and $\Lambda$ is not very informative.
To begin with, Eqs.~\eqref{eq:meanvar} do not report on the confidence of the result and how this confidence depends on the size $M$ of the sample.
In other words, $\bar{\nu}$ and $\bLambda$ are just specific estimator choices or ``best guesses'' of $\nu$ and $\Lambda$.
We aim at more than that.
Our goal is to get the probability distribution of $\nu$ and $\Lambda$ given the observed data, namely $P(\nu | \hatx)$ and $P(\Lambda | \hatx)$.
By \emph{estimating} the parameters $\nu$ and $\Lambda$, we mean computing these distributions.
We will refer to them as the \emph{estimating distributions} to make a clear distinction to the \emph{sampling distributions} defined below.
The estimating distributions encode all the information in the data, and nothing but the information in the data, that is relevant for the knowledge of $\nu$ and $\Lambda$.
This includes, for example, the size of the data sample $M$.

Our goal is to calculate the distributions $P(\nu | \hatx)$ and $P(\Lambda | \hatx)$, but we start by looking at the \emph{sampling distribution} $P(\hatx | \nu\Lambda)$.
Its meaning is the following: If we knew \emph{only} the mean $\nu$ and variance $\Lambda$ of a variable $x$, what would be the probability to get the data sample $\hatx$?

Shannon's entropy theorem\cite{shannon2001mathematical,jaynes1957information} is the key to construct sampling distributions.
It is about the entropy $S$ of a distribution $P(x)$, defined as
\begin{align*}
    S = -\int_{-\infty}^{\infty}
    dx \, P(x) \log P(x) .
\end{align*}
The theorem proves that $S$ measures the \emph{information} about the variable $x$ encoded in $P(x)$.
Therefore, assuming nothing about $x$ other than its mean $\nu$ and variance $\Lambda$, the distribution $P(x | \nu\Lambda)$ representing the knowledge about $x$ must have maximal entropy allowed by the constrains
\begin{align*}
    \nu = \int_{-\infty}^{\infty}
    dx \, P(x | \nu\Lambda) x,
    \quad
    \Lambda =  \int_{-\infty}^{\infty}
    dx \, P(x | \nu\Lambda) (x-\nu)^2 .
\end{align*}
In App.~\ref{app:gaussian} we prove that it is a Gaussian,
\begin{align}
    P(x | \nu\Lambda)
    = \frac{1}{\sqrt{2\pi \Lambda}}
    \exp\left(
    -\frac{(x-\nu)^2}{2\Lambda}
    \right) .
    \label{eq:Gauss}
\end{align}
It is important to emphasize the following. This result does not mean that the ``true'' distribution of $x$ is Eq.~\eqref{eq:Gauss}. 
It means that our \emph{knowledge} of $x$ is described by Eq.~\eqref{eq:Gauss}.
This interpretation is one of the core concepts of Bayesian statistics.
In contrast, standard statistics would consider Eq.~\eqref{eq:Gauss} an approximation or ansatz, even when it is not.
The rest of the estimation process---and its validity for a generic variable $x$---relies on Eq.~\eqref{eq:Gauss}, and hence on this subtle point.

Equation \eqref{eq:Gauss} gives the probability to observe the single value $x$ based on $\nu$ and $\Lambda$ as the only knowledge about $x$.
The probability to observe the whole data sample $\hatx$ can be obtained from the definition of conditional probabilities, that is, $P(ab) = P(a)P(b | a)$.
When $b$ is logically independent on $a$, namely when $P(b | a) = P(b)$, we have $P(ab) = P(a)P(b)$.
Therefore, if the values in the data sample $\hatx$ are logically independent,
\begin{align}
    P\big(\hatx \big| \nu\Lambda\big)
    = \prod_{m=0}^{M-1}
    P\big(x^{(m)} \big| \nu\Lambda\big) .
    \label{eq:sampling_x}
\end{align}
With this and Eq.~\eqref{eq:Gauss}, we have the sampling distribution $P(\hatx | \nu\Lambda)$.
However, this is not what we want to calculate.
For the estimation of $\nu$ and $\Lambda$, we rather need $P(\nu | \hatx)$ and $P(\Lambda | \hatx)$.
It is possible to get these estimating distributions from the sampling distribution $P(\hatx | \nu\Lambda)$ by using elementary probability theory.
We need to apply just two rules.
The first is
\begin{align}
    \begin{split}
        & \sum_i P(b_i) = 1 \ \text{and} \
        P(b_i \cap b_j) = 0 \ \forall i \neq j
        \\
        & \Rightarrow
        P(a) = \sum_i P(a \cap b_i) \quad \forall a ,
    \end{split}
    \label{eq:rule}
\end{align}
which can be easily derived from
(i) the axiom
\begin{align*}
    P(a \cap b) = 0
    \Rightarrow
    P\left(
        a \cup b
    \right)
    = P(a) + P(b) ;
\end{align*}
(ii) the equivalence between the logical statement $a$ and $a\cap(\cup_i b_i)$, because $\cup_i b_i$ is always true by virtue of $\sum_i P(b_i) = 1$ and $P(b_i \cap b_j) \propto \delta_{ij}$;
and (iii) the identity $a\cap(\cup_i b_i) = \cup_i (a\cap b_i)$.
The rule of Eq.~\eqref{eq:rule} allows us to write
\begin{align*}
    P(\nu | \hatx)
    & = \int_0^\infty
    d\Lambda \, P(\nu\Lambda | \hatx) ,
    \\
    P(\Lambda | \hatx)
    & = \int_{-\infty}^\infty
    d\nu \, P(\nu\Lambda | \hatx) .
\end{align*}
The second rule we need is the Bayes' rule,
\begin{align*}
    P(a | b)
    = \frac{P(b | a)P(a)}{P(b)} .
\end{align*}
When applied to the integrands, one gets
\begin{subequations}
\begin{align}
    P(\nu | \hatx)
    & = \int_0^\infty d\Lambda \,
    \frac{P(\hatx | \nu\Lambda) P(\nu\Lambda)}{P(\hatx)} ,\\
    P(\Lambda | \hatx)
    & = \int_{-\infty}^\infty d\nu \,
    \frac{P(\hatx | \nu\Lambda) P(\nu\Lambda)}{P(\hatx)} .
    \label{eq:int_nu}
\end{align}
    \label{eq:int_Lnu}
\end{subequations}
These equations express the estimating distributions as functions of the sampling distribution $P(\hatx | \nu\Lambda)$ and constitute
the essence of parameter estimation from the perspective of Bayesian statistics together with the Shannon's entropy theorem.

The steps that led to Eq.~\eqref{eq:int_Lnu} will be repeatedly used throughout the article.
We can look at $P(\hatx)$ in the denominators as a mere normalization constant.
$P(\nu\Lambda)$ is the \emph{prior} probability for the mean and variance, and it represents our knowledge about $\nu$ and $\Lambda$ previous to the acquisition or analysis of any data.
Assigning priors has been subject to extensive discussions, see Ref.~\onlinecite{jaynes2003probability}. We follow the approach advocated therein, and assign priors based on the requirements of symmetry (invariance under certain transformations).
Accordingly, in App.~\ref{app:priors} we prove that $P(\nu\Lambda) \propto 1/\Lambda$ keeps $P(\nu\Lambda)$ invariant under ``frame'' transformations of the variable $x$ being the independence of $P(\nu\Lambda)$ on the units and the reference value chosen to measure $x$ relative to.\footnote{To illustrate, assume that the variable $x$ is a voltage. The prior invariance that we impose means the requirement that the statistical inference on $x$ will not be changed upon assigning its mean to be  0 Volts instead of, say, 10 Volts, or measuring the fluctuations of $x$ away from its mean in milliVolts instead of Volts.}

Performing the integrations, we get Eq.~\eqref{eq:int_Lnu} in a closed form
\begin{subequations}
\begin{align}
    & P(\nu | \hatx)
    = \frac{
        \Gamma\left(
        \scriptstyle \frac{M}{2}
        \textstyle
        \right)}
        {
        \Gamma\left( 
        \scriptstyle \frac{M-1}{2}
        \textstyle
        \right)
        \sqrt{\pi \bLambda}\,
    }
    \left( 
    \frac{\bLambda}{
        \bLambda+(\nu-\bar{\nu})^2
    }\right)^{\frac{M}{2}} ,
        \label{eq:onevar1}
    \\
    & P(\Lambda | \hatx)
    = \frac{1}
    {
        \Gamma\left( 
        \scriptstyle \frac{M-1}{2}
        \textstyle
        \right)
    }
    \left(
        \frac{M}{2}
        \frac{\bLambda}{\Lambda}
    \right)^{\frac{M-1}{2}}
    \frac{1}{\Lambda}
    \exp\left(-\frac{M \bLambda}{2\Lambda}\right) .
    \label{eq:onevar2}
\end{align}
    \label{eq:onevar}
\end{subequations}
Remarkably, these distributions depend only on two variables that are functions of the data, or \emph{sufficient statistics}, namely $\bar{\nu}$ and $\bLambda$.
Not surprisingly, they are given by the well-known formulas of Eqs.~\eqref{eq:meanvar}.
However, the estimating distributions allow one to evaluate any desired estimator.
For example, the \emph{maximum-likelihood estimators} are the values of $\nu$ and $\Lambda$ that maximize $P(\nu | \hatx)$ and $P(\Lambda | \hatx)$, respectively. Equation \eqref{eq:onevar} gives
\begin{align}
    \mle(\nu) = \bar{\nu},
    \quad
    \mle(\Lambda) = \frac{M}{M+1}\bar{\Lambda} .
    \notag
\end{align}
Interestingly, $\mle(\Lambda)$ significantly differs from $\bLambda$ for low values of $M$. In a similar way, we could use the estimating distributions to calculate, for example, the confidence intervals.\footnote{As yet another illustration, keeping the lowest two terms in the Taylor expansion of $\ln P(\nu | \hatx)$ around its maximum, one gets the central limit expression $P(\nu | \hatx) \propto \exp[ - M (\nu-\bar{\nu})^2/2\bar{\Lambda}]$, showing that $\bar{\Lambda}/M$ is the variance of the posterior in the large-$M$ limit.}

Before illustrating the distributions, we examine Eq.~\eqref{eq:onevar} in certain special limits. The case $M=1$ corresponds to measuring a single value $x^{(0)}$, giving $\bar{\nu} = x^{(0)}$ and $\bLambda = 0$ by Eq.~\eqref{eq:meanvar}. Evaluating Eq.~\eqref{eq:onevar1} with $\bLambda=0$ and $M=1$ is undefined, but let us take the limits $\bLambda \to 0$ and then $M\to 1$:
\begin{equation*}
P(\nu | x^{(0)})
    = \lim_{M\to1}
     \frac{1}
        {
        \Gamma\left( 
        \scriptstyle \frac{M-1}{2}
        \textstyle
        \right)\,
    }
    \frac{1}{|\nu-\bar{\nu}|^M}.
\end{equation*}
That is, the estimating distribution for the mean that we deduce from a single data point is centered at that point and falls off as $1/|x|$ symmetrically to both sides. Keeping $M$ at a finite, albeit possibly small, distance above 1 is a way of regularization of this non-normalizable distribution.\footnote{Sometimes one can use directly the non-normalizable distribution, here $1/|\nu-\bar{\nu}|$, for example in algorithms using sampling where only likelihood ratios enter \cite{doucet2001sequential}.} The same happens with Eq.~\eqref{eq:onevar2}: For $M=1$, the formula should be interpreted as a limit towards a non-normalizable distribution $1/\Lambda$, with parameters $M \to 1$ and $\bLambda \to 0$ possibly providing effective cut-offs for the divergencies at both $\Lambda \to 0$ and $\Lambda \to \infty$ tails of the distribution. Since a single point delivers no information about the variance, our estimating distribution equals to the prior distribution. 

Remarkably, one can meaningfully interpret Eq.~\eqref{eq:onevar} even for $M=0$, even though the expressions in Eq.~\eqref{eq:meanvar} are undefined. Taking them as regularization parameters instead, we get that for $M \to 0$ also Eq.~\eqref{eq:onevar1} reduces to the corresponding prior, $P(\nu| \mathrm{no\,data}) = P(\nu)$. Recovering the prior distributions when the data deliver no information is an appealing property of Eq.~\eqref{eq:onevar}.

As a final example, consider that two points were measured, $M=2$, and they came out exactly the same, so that $\bLambda=0$. Equation \eqref{eq:onevar1} now becomes a regularized delta function centered at $\bar{\nu}$ with the width $\bLambda \to 0$. In reality, the values $x^{(m)}$ are measured with some finite precision. We therefore obtain another natural result that if $\bLambda=0$, one can be sure that $\nu=\bar{\nu}$, but only up to the precision with which $\bLambda=0$ is valid, which is never an infinite precision in reality.

As stressed in Ref.~\cite{jaynes2003probability}, this behavior is the hallmark of Bayesian statistics. The results such as Eq.~\eqref{eq:onevar}, which are based on the Bayes rule and other rules backed by the probability-theory axioms used in Eq.~\eqref{eq:int_Lnu}, remain valid even under ``extreme'' cases. We will not repeatedly alert the reader on this fact for the formulas to come. Nevertheless, in Sec.~\ref{sec:continuous} we discuss some paradoxes which occur for $M=1$ that are of a different type. 

To conclude the discussion of their general aspects, we note that having the estimating distributions $P(\nu | \hatx)$ and $P(\Lambda | \hatx)$, we know everything that can be known about $\nu$ and $\Lambda$ from the observed data $\hatx$ and---we emphasize again---nothing more than the data and the prior invariance principles: no model on $x$ or ansatz on the ``true'' $P(x)$ has been invoked.
These estimating distributions give more information than just the estimators $\bar{\nu}$ and $\bLambda$:
Spread or narrow, they express our confidence in the knowledge of $\nu$ and $\Lambda$. We will illustrate it with a figure shortly, but let us first consider a particular case relevant for spectral analysis.

Assume that the mean of $x$ is known, $\nu = \nu_0$. Proceeding as before, we get the estimating distribution of the variance in the form
\begin{equation}
\begin{split}
    P(\Lambda | \nu=\nu_0, &\hatx)
    = \frac{
        P(\hatx | \nu=\nu_0, \Lambda)P(\Lambda)
    }{P(\nu=\nu_0, \hatx)} 
    \\
    &=
    \frac{1}
    {
        \Gamma\left( 
        \scriptstyle \frac{M}{2}
        \textstyle
        \right)
    }
    \left(
        \frac{M}{2}
        \frac{\bLambda}{\Lambda}
    \right)^{\frac{M}{2}}
    \frac{1}{\Lambda}
    \exp\left(-\frac{M \bLambda}{2\Lambda}\right) .
    \label{eq:zeroMean}
\end{split}
\end{equation}
Compared to Eq.~\eqref{eq:onevar2}, there is a difference in the power exponent. 
Also, $\bLambda$ should be here calculated according to the second expression in Eq.~\eqref{eq:meanvar} replacing $\bar{\nu} \to \nu_0$.  Though not needed here, for later convenience we note that this short calculation can be performed also in the following way,
\begin{equation}
    P(\Lambda | \nu=\nu_0, \hatx)
    =
    \int \mathrm{d}\nu 
     \frac{
        P(\hatx | \nu \Lambda)P(\Lambda) 
    }{P(\nu=\nu_0, \hatx)}P(\nu),
\end{equation}
and using the prior $P(\nu)=\delta(\nu-\nu_0)$.

\begin{figure}
    \centering
    \includegraphics{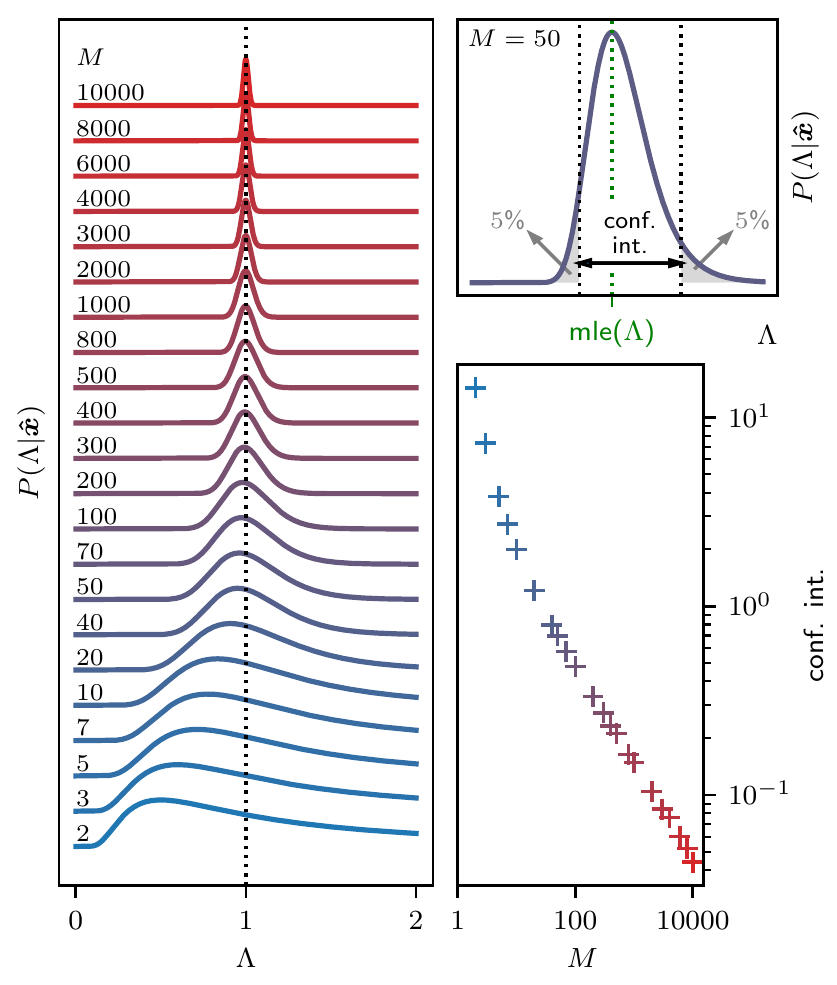}
    \caption{Left: Estimating distributions---Eq.~\eqref{eq:zeroMean}---for $\bLambda = 1$ and different sizes $M$ of the data sample.
    The curves have been shifted and scaled in the vertical axis for the sake of visibility.
    Right: 90\%-confidence intervals for the estimating distributions.
    The top figure shows how the maximum-likelihood estimator and confidence intervals are defined.
    The bottom figure plots the relation between confidence and sample size.
    The color code is consistent with the plot on the left.}
    \label{fig:var}
\end{figure}

As more important for spectral estimation, we plot Eq.~\eqref{eq:zeroMean} in Fig.~\ref{fig:var} for different values of $M$ and $\bLambda = 1$. The information content of the full estimating distribution can be appreciated from the figure.  
Despite all the plotted distributions corresponding to the same variance estimator ($\bLambda = 1$), they clearly convey a different knowledge on $\Lambda$. One way to illustrate it is to look at the dependence of the confidence interval size on the sample size $M$, plotted in the right panel of Fig.~\ref{fig:var}. The change of the information content with the size of the data sample $M$ is obvious.

As a final remark, we emphasize that in this section, with the main result being Eqs.~\eqref{eq:onevar} and \eqref{eq:zeroMean}, we have indeed estimated a correlation function:
The variance $\Lambda = \langle (x-\langle x \rangle)^2 \rangle$ is the correlation function of $x$ with itself. 
What we do next is to extend this procedure to correlation functions of (possibly two different) variables that evolve in time.
The essence of the parameter-estimation formalism remains identical: First, we find the sampling distribution that maximizes the entropy.
Then, we apply Bayes' rule to calculate the estimating distribution.
The main difference is the passage to Fourier space in order to simplify both distributions.

\section{Auto-correlations} \label{sec:autocorr}

This section extends the estimation of Sec.~\ref{sec:onedim} to a real variable $A$ that evolves in time.
This scenario includes the study of noise.
The results that we present allow one to estimate the auto-correlation matrix of $A$ (defined below) with Bayesian statistics.
We establish connections with the common estimators used in the literature, such as the periodogram.
Along the lines of Sec.~\ref{sec:onedim}, our results contain more information.

Consider $M$ independent batches of $n$ measurements of $A$ in time:
\begin{align}
    & \hat{\bm{A}}
    \equiv
    \big(\bm{A}^{(0)}, \dots, \bm{A}^{(M-1)}\big) , 
    \notag \\
    & \bm{A}^{(m)}
    \equiv \big(A^{(m)}_0, \dots, A^{(m)}_{n-1}\big)^T .
    \notag
\end{align}
The super- and subscripts are the batch and time index, respectively.
Different batches correspond to uncorrelated copies or runs of the experiment.
Let the time interval $\Delta$ between $A^{(m)}_j$ and $A^{(m)}_{j+1}$ be the same for all $m$ and $j$.
Let 
\begin{subequations}
\begin{equation}
\mu = \langle A_j \rangle
\end{equation}
be the mean of the variable $A$ (the same for all $j$) and $\hSigma$ the correlation matrix, with components 
\begin{equation} \hSigma_{jk} = \langle A_j A_k \rangle-\mu^2. 
\label{eq:covariance}
\end{equation}
\end{subequations}
Our goal now is to estimate these parameters.

We adopt two approximations.
First, we impose periodic boundary conditions: $A_{j+n} = A_j$.
These conditions are implicit in the discrete Fourier transforms used throughout this paper.
Second, we consider \emph{stationary noise} so that
$\hSigma_{jk}$ only depends on $j-k$.
The scalar function $\Sigma_j$ defined by  $\hSigma_{ij} = \Sigma_{i-j}$ is called the \emph{auto-correlation function}, or simply correlation function.
Stationary noise implies that $\Sigma_j = \Sigma_{-j}$.
These two commonly adopted approximations are necessary for the correlation matrix to be diagonal in Fourier space.

In parallel with Sec.~\ref{sec:onedim}, we start by giving the sampling distribution of the variable $A$.
Next, we estimate the mean $\mu$ and the correlation matrix $\hSigma$, and conclude by discussing some spectral properties of noise.

\subsection{Sampling distribution}
\label{sec:autocorr_sampling}

Since by definition the $M$ data batches are independent, the sampling distribution given only $\mu$ and $\hSigma$ reads
\begin{align}
    P(\hat{\bm{A}} | \mu\hSigma)
    = \prod_{m=0}^{M-1}
    P\big(\bm{A}^{(m)} \big| \mu\hSigma\big),
    \label{eq:acor_sampling_hatA}
\end{align}
with, as we prove in App.~\ref{app:gaussian}, the multivariate Gaussian
\begin{align}
    \begin{split}
        P\big(& \bm{A}^{(m)} \big| \mu\hSigma\big)
        = \frac{1}{\sqrt{(2\pi)^n|\hSigma|}}
        \\
        & \times \exp\left(
            -\frac{1}{2}\big(\bm{A}^{(m)}-\bm{\mu}\big)^T
            \hSigma^{-1}\big(\bm{A}^{(m)}-\bm{\mu}\big)
        \right)
    \end{split}
    \label{eq:acor_sampling}
\end{align}
and $\bm{\mu} = \mu (1,\dots,1)^T$.
These expressions are analogous to Eqs.~\eqref{eq:Gauss} and \eqref{eq:sampling_x}.
We insist on the fact that this is not an approximation of the distribution of $A$.
Rather, it describes our knowledge based on the mean $\mu$ and correlation matrix $\hSigma$.

It is convenient to perform a coordinate change to simplify the sampling distribution.
The translational invariance encoded in $\hSigma_{ij} = \Sigma_{j-i}$ hints on switching to Fourier space.
The Fourier transform of each batch $(m)$ reads
\begin{align}
    & \alpha_j^{(m)} \equiv \frac{1}{\sqrt{n}}
    \sum_{k=0}^{n-1} A_k^{(m)}
    \exp\left(
        -i \frac{2\pi jk}{n}
    \right) .
\end{align}
In agreement with our previous notation, we define
\begin{align}
    & \hatalpha
    \equiv \big(
        \bm{\alpha}^{(0)}, \dots,
        \bm{\alpha}^{(M-1)}
    \big) , 
    \notag \\
    & \bm{\alpha}^{(m)}
    \equiv \big(
        \alpha^{(m)}_0, \dots,
        \alpha^{(m)}_{n-1}
    \big)^T .
    \notag
\end{align}
Denoting by $\hat{F}$ the Fourier-transform matrix, with components
\begin{align}
    \hat{F}_{jk} \equiv \frac{1}{\sqrt{n}}
    \exp\left(
        -i \frac{2\pi jk}{n}
    \right)
    \label{eq:FT}
\end{align}
and thus unitary, we can write $\bm{\alpha}^{(m)} = \hat{F} \bm{A}^{(m)}$.
Similarly, we define the vector $\bm{\nu} \equiv \hat{F} \bm{\mu}$. All components of this vector except the first are zero, $\nu_k = \nu \delta_{k,0}$, where, for later convenience, we define $\nu \equiv \mu\sqrt{n}$. 
As mentioned, with periodic boundary conditions the correlation matrix is diagonal in Fourier space:
\begin{align}
    & \hat{\Lambda} \equiv \hat{F} \hat{\Sigma} \hat{F}^\dagger ,
    \quad
    \hat{\Lambda}_{jk}
    = \delta_{jk} \Lambda_k,
    \label{eq:hLambdak_acor}
    \\
    & \Lambda_k \equiv
    \sum_{j=0}^{n-1} \Sigma_j e^{-i 2\pi jk/n} .
    \notag
\end{align}
Notice the difference in notation between the correlation matrix $\hLambda$, with components $\hLambda_{jk}$, and the scalar function (of $k$) $\Lambda_k$.
$\Lambda_k$ gets the same name as $\Sigma_j$, namely auto-correlation function or simply correlation function.
Since it is the Fourier transform of $\Sigma_j$, it is also called spectrum.\footnote{Another usual name for $\Lambda_k$ is the power spectrum. If the frequency index is continuous rather than discrete (See Sec.~\ref{sec:continuous} for more on models with continuous $k$), the usual name is power spectral density (PSD). In this case, one might add a qualifier, calling $\Lambda_k$ the auto-PSD, to distinguish it from the cross-PSD discussed in the next section.}

Another notation caveat is pertinent.
The bijections between $\bm{A}^{(m)}$ and $\bm{\alpha}^{(m)}$, and between $\hSigma$ and $\hLambda$, allow one to write the sampling distribution in the following equivalent forms:
\begin{align*}
    P(\hatA | \mu\hSigma) =
    P(\hatA | \mu\hLambda) =
    P(\hatalpha | \mu\hSigma) =
    P(\hatalpha | \mu\hLambda) .
\end{align*}
Basically, data (in real or Fourier space) is on the left of the bar, whereas the mean and correlation matrices (in real or Fourier space) are on the right.
These changes are trivial in the Bayesian formalism as introduced in Sec.~\ref{sec:onedim}.
We will express the sampling distribution in any of these forms according to our convenience.
Also, distributions like $P\big(\alpha^{(m)}_k \big| \mu\Lambda_k\big)$ may be referred to as ``sampling distributions'' with no risk of confusion.
Analogously, the estimating distributions will be written in different equivalent forms, easily identifiable because the mean and correlation matrix will lie on the left of the bar, and the data on the right. In a similar spirit, we can replace the mean $\mu$ by the vector $\bm{\nu}$ or its only non-zero component $\nu$.

The sampling distribution of Eq.~\eqref{eq:acor_sampling_hatA} becomes much simpler in the $\alpha$ coordinates:
\begin{align}
    P(\hatalpha | \mu\hLambda)
    = \prod_{k=0}^{\nHalf}
        \prod_{m=0}^{M-1}
        P\big(\alpha_k^{(m)} \big| \nu_k\Lambda_k\big) ,
    \label{eq:sam_fac_M}
\end{align}
with $\lfloor x \rfloor$ the largest integer smaller or equal to $x$,  
\begin{align}
    \begin{split}
        P\big(\alpha_k^{(m)} \big| \nu_k \Lambda_k\big)
        & = \left(\frac{d_k}{\pi\Lambda_k}\right)^{d_k}
        \\
        & \times
        \exp\left(
            -d_k\frac{\big|\alpha_k^{(m)}-\nu_k\big|^2}
              {\Lambda_k}
        \right),
    \end{split}
    \label{eq:nts1}
\end{align}
and\footnote{As a mnemonic, $d_k$ equals one half of the number of ``(real) degrees of freedom" of the (in general complex) variable $\alpha_k$.}
\begin{align*}
    d_k \equiv
    \begin{cases}
        1/2 & \text{if $k = 0$}, \\
        1/2 & \text{if $n$ is even and $k = n/2$}, \\
        1   & \text{otherwise.}
    \end{cases}
\end{align*}
The $k$-th component of the vector $\bm{\nu}$ is the mean of the variable $\alpha_k$; out of these, the only one with a non-zero mean is $\alpha_{0}$. 
This remark will be relevant in the next subsection for the choice of priors.
It also allows us to write $P\big(\alpha_k^{(m)} \big| \nu_k \Lambda_k\big) = P\big(\alpha_k^{(m)} \big| \Lambda_k\big)$ for $k \neq 0$.
A last detail worth discussing about Eq.~\eqref{eq:sam_fac_M} is that $k$ ranges from $0$ to $\nHalf$ (and not $n$).
This, as well as the origin of the factor $d_k$, is due to (i) $\alpha_{k}^{(m)} = \big(\alpha_{-k}^{(m)}\big)^*$, because the variable $A$ is real and (ii) $\Lambda_{-k} = \Lambda_{k}$, with $\Lambda_k \in \mathbb{R}$, because $\Sigma_j \in \mathbb{R}$ is even in $j$. (In these relations, the subscript indexes are understood mod $n$.)
In other words, the independent coordinates in $\bm{\alpha}^{(m)}$ can be taken as $\alpha^{(m)}_k$ for $k = 0, \dots, \nHalf$, and, similarly, $\hLambda$ is fully parametrized by the independent values $\Lambda_k$ with $k = 0, \dots, \nHalf$.
The sampling distribution already encodes this information.

\subsection{Estimation of auto-correlations}
\label{sec:est_autocorr}

The estimation of the correlation matrix $\hSigma$ requires the calculation of $P(\hSigma | \hat{\bm{A}})$.
Equivalently, we can estimate its transform $\hLambda$ with $P(\hLambda | \hatalpha)$.
This is easier because $P(\hatalpha | \mu\hLambda)$ manifestly factors in $k$, see Eq.~\eqref{eq:sam_fac_M}.
Assuming a factored prior (as argued in App.~\ref{app:priors})
\begin{align}
    P(\mu \hLambda) = \prod_{k=0}^{\nHalf} P(\nu_k \Lambda_k)
    \notag
\end{align}
and using Bayes' rule, it is easy to see that $P(\hLambda | \hatalpha)$ also factors:
\begin{align}
    \begin{split}
        &P(\mu \hLambda | \hatalpha)
        =
        \frac{P(\hatalpha |\mu \hLambda) P(\mu \hLambda)}
        {P(\hatalpha)}
        \propto
        P(\hatalpha | \mu \hLambda) P(\mu \hLambda)
        \\
        &
        =
        \prod_{k = 0}^{\nHalf}
        P(\bm{\alpha}_k |\nu_k  \Lambda_k) P(\nu_k \Lambda_k)
        \propto
        \prod_{k = 0}^{\nHalf}
        P(\nu_k \Lambda_k | \bm{\alpha}_k) ,
    \end{split}
    \label{eq:est_factor}
\end{align}
where we denoted $\bm{\alpha}_k \equiv \big\{\alpha_k^{(0)}, \dots, \alpha_k^{(M-1)}\big\}$.
Equation \eqref{eq:est_factor} greatly simplifies the estimation of $\hLambda$: It decouples the estimation of the matrix $\hLambda$, dependent on all the (Fourier transformed) data $\hatalpha$, into the estimation of the diagonal elements $\Lambda_k$ for $k = 0, \dots, \nHalf$, each of which only depends on $\bm{\alpha}_k$.
In this way, the problem reduces to the one solved in Sec.~\ref{sec:onedim}. This connection is not unexpected: Since the change to Fourier space diagonalizes $\hSigma$, the multivariate Gaussian of Eq.~\eqref{eq:acor_sampling} is mapped to a product of one-variable Gaussians.
Thus, spectral analysis reduces to the estimation of the variance in each of the $\nHalf+1$ independent one-dimensional problems labeled by $k = 0, \dots, \nHalf$.

One could now in principle read off the result from Eqs.~\eqref{eq:onevar} and \eqref{eq:zeroMean}, noting that Eq.~\eqref{eq:Gauss} gives Eq.~\eqref{eq:nts1} upon replacing $1/2 \to d_k$. However, it might be easier to repeat the calculation: $\Lambda_k$ can be estimated with Bayes' rule,
\begin{align}
    \begin{split}
        P(\Lambda_k &| \bm{\alpha}_k)
         = \int \mathrm{d} \nu_k \frac{P(\bm{\alpha}_k | \nu_k \Lambda_k) P(\nu_k \Lambda_k)}
               {P(\bm{\alpha}_k)}
        \\
        &
        \propto
        P(\Lambda_k)
        \prod_{m=0}^{M-1}
        \int \mathrm{d} \nu_k P\big(\alpha_k^{(m)} \big| \nu_k \Lambda_k\big) P(\nu_k),
    \end{split}
    \label{eq:bbb}
\end{align}
the sampling distribution $P\big(\alpha_k^{(m)} \big| \nu_k \Lambda_k\big)$ of Eq.~\eqref{eq:nts1} and the prior (cf.\ App.~\ref{app:priors})
\begin{align*}
    P(\nu_k \Lambda_k) &\propto \delta(\nu_k) \times 1/\Lambda_k &\mathrm{if\,} k \neq 0,\\
     P(\nu_k \Lambda_k) &\propto \mathrm{const} \times 1/\Lambda_k &\mathrm{if}\, k = 0 .
\end{align*}
The resulting estimating distribution of $\Lambda_k$ reads
\begin{align}
    \begin{split}
        P(\Lambda_k | \bm{\alpha}_k)
        = & \frac{1}{\Gamma\left(
        \scriptstyle M d_k-\frac{\delta_{k,0}}{2} \textstyle
        \right)}
        \left(
            Md_k \frac{\bLambda_k}{\Lambda_k}
        \right)^{
        \scriptstyle M d_k-\frac{\delta_{k,0}}{2} \textstyle
        }
        \\
        & \times \frac{1}{\Lambda_k}
        \exp
        \left(
            -Md_k \frac{\bLambda_k}{\Lambda_k}
        \right) ,
    \end{split}
    \label{eq:autocorr}
\end{align}
with the sufficient statistics
\begin{subequations}
\begin{align}
    & \bLambda_k \equiv \frac{1}{M}
    \sum_{m=0}^{M-1}
    \left|
        \alpha^{(m)}_k
        -\delta_{k,0} \bar{\alpha}_0
    \right|^2 ,
    \label{eq:periodogram} \\
    &
    \bar{\alpha}_0
    \equiv \frac{1}{M} \sum_{m=0}^{M-1}
    \alpha_0^{(m)}.
    \label{eq:langle_alpha}
\end{align}
\label{eq:suff}
\end{subequations}
This is the main result of this section.
In the literature, the quantities $\bLambda_k$ (sometimes up to a factor) are called the \emph{periodogram}.
Whereas $\Lambda_k$ is commonly estimated with $\bLambda_k$, our distribution $P(\Lambda_k | \bm{\alpha}_k)$ contains more information.
Figure \ref{fig:var} and its discussion in Sec.~\ref{sec:onedim} apply to $\Lambda_k$
and make this evident.
One important application of our results can take place in the design of experiments:
As the right bottom plot of Fig.~\ref{fig:var} reflects, one can figure out the amount of data needed in order to reach a certain confidence level in $\Lambda_k$ (that is, in the spectrum).
This precise assessment of confidence would be difficult without the estimating distributions.
One would be limited, at most, to the asymptotic analysis of the periodogram fluctuations to give a rough idea of the error.

The maximum-likelihood estimator for $\Lambda_k$ is
\begin{align*}
    \mle(\Lambda_k)
    = \frac{M d_k}{M d_k + 1 -\delta_{k,0}/2} \bLambda_k ,
\end{align*}
and we repeat that it tends to the periodogram value $\bLambda_k$ only for large $M$ and that the estimating distribution $P(\Lambda_k | \bm{\alpha}_k)$ allows the calculation of the corresponding confidence interval as explained in Fig.~\ref{fig:var}.

The estimation of the mean is completely identical to that of Sec.~\ref{sec:onedim} with the result given in Eq.~\eqref{eq:onevar1}. We copy it here converting to the notation used in this section,
\begin{align}
    P(\nu | \hatalpha)
    = \frac{
        \Gamma\left( 
        \scriptstyle \frac{M}{2}
        \textstyle
        \right)}
        {
        \Gamma\left( 
        \scriptstyle \frac{M-1}{2}
        \textstyle
        \right)
        \sqrt{\pi \bLambda_0}\,
    }
    \left(
        \frac{\bLambda_0}{\bLambda_0+(\nu-\bar{\alpha}_0)^2}
    \right)^{\frac{M}{2}} .
    \label{eq:autocorr_nu}
\end{align}
For small values of $M$, this Lorentzian-like function significantly differs from the Gaussian that one might conjecture under (an abuse of) the central limit theorem.
The maximum-likelihood estimator of $\mu$ is
\begin{align*}
    \mle(\mu) = \frac{\mle(\nu)}{\sqrt{n}}
    = \frac{\bar{\alpha}_0}{\sqrt{n}}
    = \frac{1}{M n}
    \sum_{m=0}^{M-1}
    \sum_{j=0}^{n-1}
    A_j^{(m)} .
\end{align*}
As expected, $\mle(\mu)$ is the mean of the variable $A$ taken over all batches and all times.

\subsection{Statistical properties of noise}
\label{sec:generate_autocorr}

Not only is the sampling distribution $P\big(\alpha_k^{(m)} \big| \mu\Lambda_k\big)$ necessary to estimate $\Lambda_k$ as in Sec.~\ref{sec:est_autocorr}, but it is a key to understand statistical properties of noise.
In this section, we discuss two applications.
On the one hand, we show how to generate noise with a given mean $\mu$ and spectrum $\Lambda_k$.
On the other, we find the exact expression for the signal-to-noise ratio (SNR) of the periodogram.
Our derivation will clarify some aspects of the SNR that can be found in the literature\cite{caloyannides1974microcycle,timmer1995on,priestley1981spectral,press2007numerical} and make extensions.
In particular, we elaborate on its value of order $1$ irrespective of the data-sample size, arguing that it is universal for any kind of noise.
We offer a simple general proof that avoids relying on asymptotic limits, since these are not necessarily met in practice.

We focus first on the generation of noise with a given mean $\mu$ and spectrum $\Lambda_k$.
Since $\alpha_k^{(m)} \in \mathbb{C}$, it is convenient to get the sampling distributions of its real and imaginary parts, $\mathfrak{R}\alpha_k^{(m)}$ and $\mathfrak{I}\alpha_k^{(m)}$, separately.
From Eq.~\eqref{eq:nts1},
\begin{align*}
    & P\big(\mathfrak{R}\alpha_k^{(m)}
    \big| \nu_k \Lambda_k\big)
    = \sqrt{\frac{d_k}{\pi\Lambda_k}}
    \exp\left(
        -d_k
        \frac{
            \big(\mathfrak{R}\alpha_k^{(m)}-\nu_k\big)^2
        }{\Lambda_k}
    \right) ,
    \\
    & P\big(\mathfrak{I}\alpha_k^{(m)}
    \big| \nu_k \Lambda_k\big)
    = \sqrt{\frac{d_k}{\pi\Lambda_k}}
    \exp\left(
        -d_k
        \frac{
            \big(\mathfrak{I}\alpha_k^{(m)}\big)^2
        }{\Lambda_k}
    \right) .
\end{align*}
The distribution for $\mathfrak{I}\alpha_k^{(m)}$ only applies to $k \neq 0, \nHalf$; $\alpha_k^{(m)} \in \mathbb{R}$ otherwise.
These expressions are all we need: To generate the batch $(m)$, generate the normally distributed and independent variables $\mathfrak{R}\alpha_k^{(m)}$ and $\mathfrak{I}\alpha_k^{(m)}$ for $k = 0, \dots, \nHalf$.
Then, Fourier transform $\bm{\alpha}^{(m)}$ to get $\bm{A}^{(m)}$.
Repeat the process to generate $M$ batches.

This is in agreement with Ref.~\onlinecite{timmer1995on}, which restricts itself to $1/f^\beta$ noise 
and departs from the statistical properties of white noise ($\Lambda_k \propto 1$).\footnote{Another method to generate 1/f noise from white noise is discussed in Ref.~\cite{press1978flicker}.}
Our approach does not impose any condition on noise other than being stationary.
In fact, the statistical properties of noise, given by the sampling distribution $P(\hatalpha | \mu\hLambda)$ and common to every function $\Lambda_k$, should not be confused with the particular shape of the function $\Lambda_k$, which describes a specific kind of noise.
With our sampling distribution, one can generate noise with arbitrary spectra.
This discussion sets the basis for Sec.~\ref{sec:generate_crosscorr}, which will extend to the generation of correlated noise.

We proceed to discuss some statistical properties of the periodogram.
The literature reports that the periodogram $\bLambda_k$ for one batch, given by Eq.~\eqref{eq:periodogram} with $M=1$, has a $\text{SNR}\sim 1$ irrespective of the size $n$ of the data sample.
In other words, its value fluctuates a lot for different measurements of the same phenomenon (i.e.\ when repeating an experiment or measuring a new batch), displaying a relative error around $1$.
We offer a simple and general proof that this holds for any kind of noise.
Once again, the key point is to deal with distributions instead of estimators.
According to App.~\ref{app:per_dist}, the distribution for the periodogram value $\bLambda_k$ depends only on $\Lambda_k$ and reads
\begin{align}
    \begin{split}
        P\big(\bLambda_k | \Lambda_k)
        =
        & \frac{1}{\Gamma(Md_k)}
        \left(
            Md_k
            \frac{\bLambda_k}{\Lambda_k}
        \right)^{Md_k}
        \\
        &
        \times
        \frac{1}{\bLambda_k}
        \exp\left(
        -M d_k \frac{\bLambda_k}{\Lambda_k}
        \right) .
    \end{split}
    \label{eq:per_dist}
\end{align}
This distribution\footnote{For example, Ref.~\cite{strasilla1974narrow-band} investigated this distribution experimentally and compared the findings with a model where the distribution is Gaussian (which it is not).} has the moments\footnote{As a side remark, we note that from Eqs.~\eqref{eq:per_dist} and \eqref{eq:autocorr} we can get the prior distribution for the periodogram value $P(\bLambda_k) \propto P(\bLambda_k | \Lambda_k) / P(\Lambda_k | \bLambda_k) \propto 1/\bLambda_k$ for $k\neq0$ and $\propto 1/\bLambda^{1/2}$ for $k=0$. Thus, the prior for a finite frequency element of the periodogram is $1/\bLambda_k$, as expected for a scale variable. However, the symmetry between the priors for $\Lambda_k$ and $\bLambda_k$ is not valid for $k=0$.}
\begin{align*}
    \langle \bLambda_k^n \rangle
    = \left(\frac{\Lambda_k}{M d_k}\right)^n \frac{\Gamma(n+M d_k)}{\Gamma(M d_k)},
\end{align*}
the integral converging for $n+M d_k >0$. We observe that while the maximum of the distribution is not at $\bLambda_k=\Lambda_k$, the mean of the observed values of $\bLambda_k$ is equal to $\Lambda_k$. For the signal to noise ratio, we get
\begin{align}
    \text{SNR}
    = \frac{\Lambda_k}{[\text{var}(\bLambda_k)]^{1/2}}
    = \sqrt{Md_k} .
    \label{eq:snr}
\end{align}
For $k \neq 0$ and taking $M=1$, we finally recover the result $\text{SNR} = 1$.
Our derivation confirms that the SNR does not depend on the batch size $n$, but only on the number of data batches $M$.
The  paradox that the SNR does not increase with $n$ is discussed in Sec.~\ref{sec:paradox1} in the context of continuous spectra.
We also point out that Eq.~\eqref{eq:snr} agrees with the dependence $\sqrt{M}$ when $M \to \infty$, expected from the law of large numbers \cite{priestley1981spectral}.

\section{Cross correlations}
\label{sec:crosscorr}

This section extends the estimation of Sec.~\ref{sec:autocorr} to two real variables $A$ and $B$ that evolve in time.
Such estimation includes the study of their noise which might be correlated.
The results that we present allow the estimation of the correlation strength and phase (defined below) with Bayesian statistics.
We discuss the connection with common estimators used in the literature, like Pearson's $r$ coefficient.
Similarly to the previous cases, our results contain more than just suggesting a specific estimator.

The notation of the variable $A$ is identical to that of Sec.~\ref{sec:autocorr} and trivially extends to the variable $B$.
Let us use $X = (A, B)$ as a composite variable.
The batch $(m)$ contains the data
\begin{align}
    & \bm{X}^{(m)}
    \equiv \big(A_0^{(m)}, \dots, A_{n-1}^{(m)},
    B_0^{(m)}, \dots, B_{n-1}^{(m)}\big)^T ,
    \notag
\end{align}
and the whole data set $\hatX$ contains $M$ batches,
\begin{align}
    \hatX
    = \big(
        \bm{X}^{(0)}, \dots, \bm{X}^{(M-1)}
    \big).
    \notag
\end{align}
Let the means of $A$ and $B$ be $\mu_A$ and $\mu_B$, respectively, and write the correlation matrix as
\begin{equation}
    \hSigma \equiv
    \begin{bmatrix}
        \hSigma^A & \hSigma^{AB} \\
        \hSigma^{BA} & \hSigma^B
    \end{bmatrix} .
    \notag
\end{equation}
Here,
$\hSigma^A_{jk} \equiv \langle A_j A_k \rangle-\mu_A^2$ and $\hSigma^{AB}_{jk} \equiv \langle A_j B_k \rangle-\mu_A\mu_B$, with analogous expressions for $\hSigma^B_{jk}$ and $\hSigma^{BA}_{jk}$. 
As in Sec.~\ref{sec:autocorr}, our only approximations are (i) periodic boundary conditions: $A_{j+n} = A_j$ and the same for $B$; and (ii) stationary noise: $\hSigma_{jk}^C = \Sigma^C_{j-k}$, for $C = A, B, AB, BA$.
As before, it follows that the scalar functions $\Sigma_j^A$ and $\Sigma_j^B$ are even in $j$. Here, 
$\Sigma_j^A$, $\Sigma_j^B$ are auto-correlation functions and $\Sigma_j^{AB}$, $\Sigma_j^{BA}$ the so-called \emph{cross-correlation functions}.
Often, all of them are called correlation functions without the risk of confusion.
We note that from now on, the matrix $\hSigma$ refers to the variable $X$, and not to $A$ or $B$ separately.

As in Sec.~\ref{sec:autocorr}, we first give the sampling distribution of the variable $X$, then estimate the correlation matrix $\hSigma$ and the means $\mu_A$, $\mu_B$, and finish with the procedure to generate correlated noise.

\subsection{Sampling distribution}
\label{sec:crosscorr_sampling}

Since the $M$ data batches are independent, the sampling probability given only by $\mu_A$, $\mu_B$, and $\hSigma$ factors in the batch index,
\begin{align}
    P(\hat{\bm{X}} | \bm{\mu}\hSigma)
    = \prod_{m=0}^{M-1}
    P\big(\bm{X}^{(m)}
    \big| \bm{\mu}\hSigma\big).
    \notag
\end{align}
As we prove in App.~\ref{app:gaussian}, $P\big( \bm{X}^{(m)} \big| \bm{\mu}\hSigma\big)$ is a multivariate Gaussian,
\begin{align}
    \begin{split}
        P\big(& \bm{X}^{(m)} \big| \bm{\mu}\hSigma\big)
        = \frac{1}{\sqrt{(2\pi)^{2n}|\hSigma|}}
        \\
        & \times \exp\left(
            -\frac{1}{2}\big(\bm{X}^{(m)}-\bm{\mu}\big)^T
            \hSigma^{-1}\big(\bm{X}^{(m)}-\bm{\mu}\big)
        \right) .
    \end{split}
    \label{eq:ccor_sampling}
\end{align}
The vector $\bm{\mu}$ has the components
\begin{align}
    \mu_j =
    \begin{cases}
        \mu_A & \text{for } 0 \leq j < n, \\
        \mu_B & \text{for } n \leq j < 2n.
    \end{cases}
    \notag
\end{align}

As in Sec.~\ref{sec:autocorr_sampling}, it is convenient to Fourier transform the variables for each batch $(m)$.
We denote by $Z$ the conjugate variable to $X$:
\begin{align}
    & \bm{Z}^{(m)} = \tilde{F} \bm{X}^{(m)}, \quad
    \tilde{F} =
    \begin{bmatrix}
        \hat{F} & 0 \\
        0 & \hat{F}
    \end{bmatrix} .
    \notag
\end{align}
$\hat{F}$ is the Fourier-transform matrix given by Eq.~\eqref{eq:FT}, so $\tilde{F}$ is unitary.
For each batch, the coordinates $\bm{Z}^{(m)}$ gather the Fourier transform of $\bm{A}^{(m)}$ and $\bm{B}^{(m)}$,
\begin{align}
    \bm{Z}^{(m)} =
    \big(
        \alpha_0^{(m)}, \cdots,
        \alpha_{n-1}^{(m)},
        \beta_0^{(m)}, \cdots,
        \beta_{n-1}^{(m)}
    \big)^T .
    \notag
\end{align}
The notation for $\alpha^{(m)}_k$ is also shared with Sec.~\ref{sec:autocorr} and trivially extends to $\beta^{(m)}_k$ for $B$.

In the transformed coordinates $Z$, the correlation matrix takes the form
\begin{align}
    \hLambda \equiv
    \tilde{F} \hSigma \tilde{F}^\dagger =
    \begin{bmatrix}
        \hLambda^A & \hLambda^{AB} \\
        \hLambda^{BA} & \hLambda^B
    \end{bmatrix} ,
    \notag
\end{align}
with
\begin{align}
    & \hLambda^A_{jk} = \delta_{jk} \Lambda^A_k,
    & \Lambda^A_k 
    \equiv
    \sum_j \Sigma^{A}_j e^{-i 2\pi jk/n} ,
    \notag
    \\
    & \hLambda^{AB}_{jk} = 
    \delta_{jk} \Lambda^{AB}_k ,
    & \Lambda^{AB}_k \equiv
    \sum_j \Sigma^{AB}_j e^{-i 2\pi jk/n}
    \notag
\end{align}
and analogous expressions for $B$ and $BA$.
Therefore, $\hLambda$ is block diagonal in $k$, with blocks
\begin{align}
    & \hat{\Lambda}_k = \begin{bmatrix}
        \Lambda_k^A & \Lambda_k^{AB} \\
        \Lambda_k^{BA} & \Lambda_k^{B}
    \end{bmatrix} .
    \label{eq:hLambdak}
\end{align}
To prevent confusion, the reader should take note of the difference in notation for the $2n\times 2n$ matrix $\hLambda$, the $2\times 2$ matrix $\hLambda_k$ and the scalar functions (of $k$) $\Lambda_k^C$ for $C = A, B, AB, BA$.
The functions $\Lambda_k^C$ are correlation functions in Fourier space and are thus termed spectra.
$\Lambda_k^{AB}$ and $\Lambda_k^{BA}$ are called \emph{cross spectra} or simply spectra.

Below, we need also the Fourier transform of the vector $\bm{\mu}$, which is a vector with $2n$ components, only two of which are nonzero. In line with the notation for $\Lambda_k$, we collect the two nonzero components into $\bm{\nu}_0 = \sqrt{n}(\mu_A, \mu_B)^T$ and put $\bm{\nu}_k = (0,0)^T$ for $k\neq 0$. 

For a convenient parametrization of $\hLambda_k$, it is necessary to consider the following constraints on its matrix elements:
(i) $\Lambda^C_{k} = \big(\Lambda^C_{-k}\big)^*$ for $C = A, B, AB, BA$ since the variables $A$ and $B$ are real;
(ii) $\Lambda_k^A, \Lambda_k^B \in \mathbb{R}$ because, moreover, $\Sigma_j^A$ and $\Sigma_j^B$ are even in $j$ due to the assumption of stationary auto-correlations;
and (iii) $\Lambda_k^{AB} = (\Lambda_k^{BA})^*$, which follows from $\hSigma_{jk}^{AB} = \hSigma_{kj}^{BA}$ together with the assumption of stationary cross-correlations, namely from $\Sigma_j^{AB} = \Sigma_{-j}^{BA}$.
The relations (i)-(iii) imply that the only independent variables in $\hLambda$ are the values $\{\Lambda^A_k, \Lambda^B_k, \Lambda^{AB}_k\}$ for $k = 0, \dots, \nHalf$.
It also follows that the elements of $\hLambda_k$ are complex in general, but $\Lambda_k^A, \Lambda_k^B, \Lambda_0^{AB},$ and $\Lambda_{n/2}^{AB}$ for even $n$ are real.

We can now parametrize $\hLambda_k$.
Writing $\Lambda_k^{AB} = |\Lambda_k^{AB}| \exp(i\phi_k)$ giving $\Lambda_k^{BA} = |\Lambda_k^{AB}| \exp(-i\phi_k)$, each block $\hLambda_k$ of the correlation matrix is given by the real parameters $\{\Lambda_k^A, \Lambda_k^B, |\Lambda_k^{AB}|, \phi_k\}$:
the auto-correlation functions or spectra $\Lambda_k^A$, $\Lambda_k^B$, the cross-correlation modulus $|\Lambda_k^{AB}|$, and the cross-correlation \emph{phase} $\phi_k$ of the (cross) spectrum.
Alternatively, we can consider the real parameters $\{\Lambda_k^A, \Lambda_k^B, s_k, \phi_k\}$, with
\begin{align}
    s_k \equiv
    \frac{\big|\Lambda_k^{AB}\big|}
    {\sqrt{\Lambda_k^A \Lambda_k^B}} .
    \label{eq:sk}
\end{align}
The dimensionless parameter $s_k$ quantifies the degree of correlation between $A$ and $B$, but normalized to the auto-correlations of $A$ and $B$.
We will refer to $s_k$ as the \emph{correlation strength}.
It hints at a connection with \emph{Pearson's r} coefficient or \emph{linear correlation coefficient}, which will be defined in Eq.~\eqref{eq:barsk}.
Far from being an arbitrary choice to quantify cross-correlations, we stress that the definition of $s_k$ has emerged in a natural parametrization of the matrix $\hLambda_k$.

The  caveat regarding notations made in Sec.~\ref{sec:autocorr} also applies here: The sampling distribution will be denoted by the equivalent forms
\begin{align*}
    P(\hatX | \bm{\mu}\hSigma) =
    P(\hatX | \bm{\nu}\hLambda) =
    P(\hatZ | \bm{\mu}\hSigma) =
    P(\hatZ | \bm{\nu}\hLambda) ,
\end{align*}
according to  convenience. 
In these forms, the data lies on the left of the bar, and the means and correlation function on the right.
In this sense, it should not be confusing to refer to distributions such as $P(Z^{(m)}_k | \bm{\nu}_k\hLambda_k)$ as sampling distributions.
Analogous considerations, with sides swapped, apply to the estimating distribution.

After the passage to the Fourier space, the sampling probability of Eq.~\eqref{eq:ccor_sampling} takes the form
\begin{subequations}
\begin{align}
    P(\hatZ | \bm{\nu} \hLambda)
    =
    \prod_{k=0}^{\nHalf}
        \prod_{m=0}^{M-1}
        P\big(\bm{Z}_k^{(m)} \big|
        \bm{\nu}_k \hat{\Lambda}_k\big) .
    \label{eq:ccor_sampling_F}
\end{align}

Here,
\begin{equation}
\begin{split}
    P\big(
        & \bm{Z}_k^{(m)} \big|
        \bm{\nu}_k \hLambda_k
    \big)
    = \frac{1}{
        \big[
            (\pi/d_k)^2 \Lambda_k^A \Lambda_k^B
            (1-s_k^2)
        \big]^{d_k}}
    \label{eq:sampling_zk} \\
    & \times \exp\left[
        -d_k
        {\Big(
            \bm{Z}_k^{(m)}}-\bm{\nu}_k
        \Big)^\dagger
        \hLambda_k^{-1}
        \Big(\bm{Z}_k^{(m)}-\bm{\nu}_k\Big)
    \right] ,
\end{split}
\end{equation}
with the obvious notation $\bm{Z}_k^{(m)} \equiv \big(\alpha_k^{(m)}, \beta_k^{(m)}\big)^T$.
The inverse of the block $\hLambda_k$ reads
\begin{align}
    \hLambda_k^{-1} = \frac{1}{1-s_k^2}
    \begin{bmatrix}
        \frac{1}{\Lambda_k^A}
        & -\frac{
            s_k \exp(i\phi_k)
        }{\sqrt{\Lambda_k^A\Lambda_k^B}}
        \\
        -\frac{
            s_k \exp(-i\phi_k)
        }{\sqrt{\Lambda_k^A\Lambda_k^B}}
        & \frac{1}{\Lambda_k^B}
    \end{bmatrix} .
    \label{eq:LambdakI}
\end{align}
\label{eq:pZuL}
\end{subequations}
Equation \eqref{eq:pZuL} is the main result of this section.
As in Eq.~\eqref{eq:nts1}, note that in Eq.~\eqref{eq:sampling_zk} the means $\mu_A$, $\mu_B$ (inside $\bm{\nu}$) only appear for $k = 0$.
This fact is important for choosing the priors, see App.~\ref{app:priors}.
Finally, we note that $k$ ranges from $0$ to $\nHalf$, and not $n$, in Eq.~\eqref{eq:ccor_sampling_F}.
The reason is that 
$\gamma_{k}^{(m)} = \big(\gamma_{-k}^{(m)}\big)^*$ for $\gamma = \alpha, \beta$.
That is, the only independent coordinates in $\bm{Z}^{(m)}$ are $\bm{Z}^{(m)}_k$ for $k = 0, \dots, \nHalf$.
We also recall that in the parametrization of $\hLambda$, only the values of $k = 0, \dots, \nHalf$ are needed.

\subsection{Estimation of cross-correlations}
\label{sec:crosscorr_estimation}

In this section, we estimate the correlation matrix $\hSigma$, or equivalently its transform $\hLambda$, which in turn contains the spectra and cross spectra.
As previously [see Eq.~\eqref{eq:est_factor}], the factoring in $k$ of the sampling distribution---Eq.~\eqref{eq:ccor_sampling_F}---implies the factoring of the estimating distribution:
\begin{align}
    P(\bm{\nu}\hLambda | \hatZ)
    =
    \prod_{k = 0}^{\nHalf}
    P( \bm{\nu}_k \hLambda_k | \bm{Z}_k) .
    \label{eq:ccor_factor}
\end{align}
This means that the estimation of the correlation matrix $\hLambda$ decouples into $\nHalf+1$ independent estimations, one for each $k = 0, \dots, \nHalf$, of the set of four parameters $\{\Lambda_k^A, \Lambda_k^B, s_k, \phi_k\}$.

We start by estimating the spectrum $\Lambda_k^A$.
The estimation of $\Lambda_k^B$ is identical upon the exchange of $A$ and $B$.
Our goal is to calculate the estimating distribution $P(\Lambda_k^A | \hatX)$---equivalent to $P(\Lambda_k^A | \bm{Z}_k)$ by virtue of the $k$ factoring, see Eq.~\eqref{eq:ccor_factor}.
We do not expect the result to be exactly the same as Eq.~\eqref{eq:autocorr}, because now, due to possible correlations, the data about $B$ might give extra information about $A$ and vice versa.
The estimating distribution is:
\begin{align}
    P&(\Lambda_k^A | \bm{Z}_k)
    = 
    \int_{-\infty}^\infty \mathrm{d} \mu_A \int_{-\infty}^\infty \mathrm{d} \mu_B
    \int_0^{\infty} d\Lambda_k^B
    \int_0^1 ds_k
    \notag
    \\
    & \qquad \qquad \qquad \times
    \int_0^{2\pi} d\phi_k \,
    P(\Lambda_k^A \Lambda_k^B s_k\phi_k \bm{\nu}_k | \bm{Z}_k)
    \notag
    \\
    &
    \propto
    \int_0^{\infty} d\Lambda_k^B
    \int_0^1 ds_k
    \int_0^{2\pi} d\phi_k \,
    \overline{P(\bm{Z}_k | \bm{\nu}_k \hLambda_k)} P(\hLambda_k) .
    \label{eq:LA_cross_prev}
\end{align}
In these two steps, we applied Eq.~\eqref{eq:rule} and the Bayes' rule.
Note also the obvious replacement $\Lambda_k^A \Lambda_k^B s_k \phi_k \to \hLambda_k$ inside the probability sign $P$, since both contain the same information.
We have also introduced the notation 
\begin{equation}
\overline{f(\bm{\nu})} = \int_{-\infty}^\infty \mathrm{d} \mu_A \int_{-\infty}^\infty \mathrm{d} \mu_B f(\bm{\nu}) P(\mu_A) P(\mu_B),
\end{equation}
which makes it easier to provide general formulas covering both $k=0$ and $k \neq 0$ cases.
The sampling distribution $P(\bm{Z}_k | \bm{\nu}_k \hLambda_k)$ entering Eq.~\eqref{eq:LA_cross_prev} is given by 
Eq.~\eqref{eq:pZuL}.
As for the priors, according to App.~\ref{app:priors},
\begin{align}
    & P(\mu_A, \mu_B, \hLambda)
    =
    \prod_{k=0}^{\nHalf}
    P\big(\bm{\nu}_k, \hLambda_k\big) ,
    \label{eq:factor_ccor}
\end{align}
with
\begin{align}
    & P(\bm{\nu}_k, \hLambda_k)
    =
    P\big(\nu^A_k\Lambda^A_k\big)
    P\big(\nu^B_k\Lambda^B_k\big)
    P(s_k) P(\phi_k)
    \label{eq:factor_mat}
\end{align}
for all $k$ and with the notation $\bm{\nu}_k \equiv (\nu^A_k,\nu^B_k)$.
In turn,
\begin{align}
    P(\nu^A_k \Lambda^A_k) = P(\nu^A_k) P(\Lambda^A_k) =
    \begin{cases}
        \mathrm{const}\times 1/\Lambda^A_k
        & \text{for } k=0,
        \\
        \delta(\nu^A_k)\times 1/\Lambda^A_k
        & \text{otherwise,}
    \end{cases}
        \label{eq:lambdaPriors}
\end{align}
with analogous expressions for the variable $B$, and
\begin{align*}
    &
    P(s_k) = 1 \ \forall k, 
    \\
    &
    P(\phi_k) =
    \begin{cases}
        [\delta(\phi_0)+\delta(\phi_0-\pi)]/2
        & \text{for } k=0, n/2 ,
        \\
        1/(2\pi)
        & \text{otherwise.}
    \end{cases}
\end{align*}
Inserting all this into Eq.~\eqref{eq:LA_cross_prev} and performing the integrations in $\Lambda^B_k$ and $\phi_k$, we get:
\begin{align}
    &P(\Lambda^A_k | \bm{Z}_k)
    \propto  \frac{1}{(\Lambda_k^A)^{(M-\delta_{k,0}) d_k+1}}
    \notag
    \\
    &\times \int_0^1 ds_k \,
    \label{eq:LA_cross}
    \exp\left(
        \frac{-Md_k}{1-s_k^2}
        \frac{\bLambda^A_k}{\Lambda^A_k}
    \right)
    \\
    & \qquad \times
    \Foneone{(M-\delta_{k,0}) d_k}{d_k}
    {Md_k \frac{\bar{s}_k^2 s_k^2}{1-s_k^2} \frac{\bLambda^A_k}{\Lambda^A_k}}.
    \notag
\end{align}
Here, $F$ is the hypergeometric function.\footnote{Here and in further we use the hypergeometric function notation putting the upper parameters up, the lower parameters down, and the function argument behind a vertical bar. In case there are multiple upper or lower parameters, they are separated by commas. We do not use the $p$ and $q$ indexes such as $_p F_q$ as the number of the upper and lower parameters can be easily counted.}
We could not find a primitive for the integral in $s_k$ and leave it for numerical evaluation.
The sufficient statistics in Eq.~\eqref{eq:LA_cross} are the periodograms $\bLambda^A_k$ and $\bLambda^B_k$, given by applying Eq.~\eqref{eq:periodogram} to $A$ and $B$, respectively. Further,
\begin{align}
    \bar{s}_k =
    \frac{\big|\bLambda_k^{AB}\big|}
    {\sqrt{\bLambda_k^A \bLambda_k^B}} ,
    \label{eq:barsk}
\end{align}
where
\begin{align}
    & \bLambda^{AB}_k \equiv
    \frac{1}{M} \sum_{m=0}^{M-1}
    \big(
        \alpha^{(m)}_k 
        - \delta_{k,0}\bar{\alpha}_0
    \big)
    \big(
        \beta^{(m)}_k
        - \delta_{k,0}\bar{\beta}_0
    \big)^* .
    \label{eq:bLambdaABk}
\end{align}
$\bar{\alpha}_0$ is defined in Eq.~\eqref{eq:langle_alpha} and $\bar{\beta}_0$ analogously.
The sufficient statistic $\bar{s}_k$ is the Pearson's $r$ coefficient, a parameter commonly used in the literature to analyze cross-correlations. We have arrived at it as the natural parameter of the estimating distribution.

\begin{figure}
    \centering
    \includegraphics{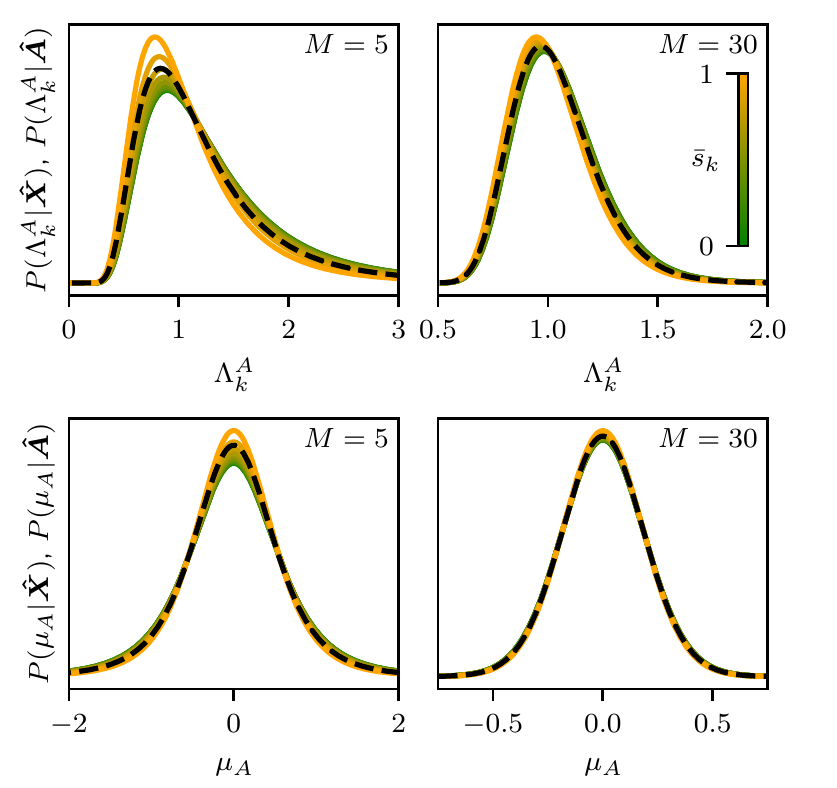}
    \caption{
        Comparison of estimating distributions of the variable $A$ when considering the partial data $\bm{\hat{A}}$ (dashed lines) versus the whole data $\bm{\hat{X}}$ (colored solid lines).
        The different colors indicate the values of $\bar{s}_k$, ranging from $0$ to $0.9$ in steps of $0.1$.
        The batch size $M$ appears on the top right corners, $\bLambda_k^A = 1$ and $\bar{\alpha}_0 = 0$.
        Top: Estimating distributions of $\Lambda_k^A$, namely Eq.~\eqref{eq:autocorr} versus Eq.~\eqref{eq:LA_cross}.
        Bottom: Estimating distributions of $\mu_A$, namely Eq.~\eqref{eq:autocorr_nu} versus Eq.~\eqref{eq:P_LambdaA_D}.
        The deviations from the dashed lines are significant for low values of $M$ and $\bar{s}_k$ close to 0 or 1.
}
    \label{fig:A_vs_AB}
\end{figure}

As follows from Eq.~\eqref{eq:LA_cross}, the data about $B$ enter the estimation of the auto-correlations $\Lambda_k^A$, through $\bar{s}_k$ and the integration in $s_k$. On the other hand, the noise of $B$ is integrated out in the estimation, see Eq.~\eqref{eq:LA_cross_prev}.
In contrast, applying Eq.~\eqref{eq:autocorr} would implicitly assign all the noise observed in the data on $A$ directly to $\Lambda^A_k$.
It is reassuring to note that Eq.~\eqref{eq:LA_cross} reduces to Eq.~\eqref{eq:autocorr} in the absence of correlations, that is in the limit of $\bar{s}_k \to 0$ and $M \to \infty$.
Figure \ref{fig:A_vs_AB} visualizes the difference between the two formulas. One can see that Eq.~\eqref{eq:LA_cross} corrects Eq.~\eqref{eq:autocorr} appreciably when $M$ is small, while the difference diminishes for larger $M$.
Nevertheless, the difference proves the need to consider \emph{all} the available data ($\bm{\hat{X}}$) to perform an estimation, even when it may seem that only a part of it ($\bm{\hat{A}}$) is relevant.
After these remarks, our study of auto-correlations is complete.

So far, we have estimated the diagonal elements of $\hLambda_k$, namely the auto-correlations of the variables $A$ and $B$.
Now, we focus on estimating the remaining parameters, $s_k$ and $\phi_k$, which account for cross-correlations.
To do so, we calculate the estimating distribution $P(s_k \phi_k | \hatX)$---equivalently $P(s_k \phi_k | \bm{Z}_k)$---once again by virtue of the $k$ factoring in Eq.~\eqref{eq:ccor_factor}.
Performing the same manipulations used to get Eq.~\eqref{eq:LA_cross_prev}, we can write
\begin{equation}
\begin{split}
    P(s_k\phi_k | \bm{Z}_k)
    &= 
    \int_0^{\infty} d\Lambda_k^A
    \int_0^{\infty} d\Lambda_k^B
 \overline{P(\Lambda_k^A \Lambda_k^B s_k\phi_k \bm{\nu}_k| \bm{Z}_k)}
    \\
    &
    \propto
    \int_0^{\infty} d\Lambda_k^A
    \int_0^{\infty} d\Lambda_k^B
    \overline{P(\bm{Z}_k | \bm{\nu}_k\hLambda_k)} P(\hLambda_k) .
    \label{eq:skphik_Zk}
\end{split}
\end{equation}
The sampling distribution %
is given by $P(\bm{Z}_k | \bm{\nu}_k \hLambda_k) = \prod_m P\big(\bm{Z}_k^{(m)} | \bm{\nu}_k  \hLambda_k\big)$ and Eq.~\eqref{eq:sampling_zk} and the prior in Eq.~\eqref{eq:lambdaPriors}.
Performing the integrations 
one gets
\begin{align}
    \begin{split}
     P(s_k \phi_k | \bm{Z}_k)
    \propto &\ P(\phi_k) (1-s_k^2)^{m}(1-q_{s_k\phi_k})^{\frac{1}{2}-2m}
    \\
    & \times 
    \Ftwoone{\scriptstyle\frac{1}{2},\frac{1}{2}}{\scriptstyle 2m+\frac{1}{2}}
    {\frac{1+q_{s_k\phi_k}}{2}}
    \end{split}
    \label{eq:skphik}
\end{align}
Here, 
\begin{align}
    & q_{s_k\phi_k} \equiv
    s_k\bar{s}_k \cos\big(\phi_k - \bar{\phi}_k\big),
    \label{eq:qskphik}
\end{align}
where the angle $\bar{\phi}_k$ is the argument of the complex number $\bLambda^{AB}_k$.
We have also introduced 
\begin{equation}
m=(M-\delta_{k,0}) d_k,
\label{eq:m}
\end{equation}
to ease the formulas in this section.
In sum, the only sufficient statistics relevant to inferences about cross-correlations are 
$\bar{s}_k$ and $\bar{\phi}_k$.\footnote{That the scale variables $\bLambda^A_k$ and $\bLambda^B_k$ are not needed is another advantage of the parameterization through the dimensionless variable $s_k$. See App.~\ref{app:unnormalizedCrossCorrelation} for a different choice.}

The estimating distribution for the correlation strength, $P(s_k | \bm{Z}_k)$, can be obtained by integrating out $\phi_k$ in Eq.~\eqref{eq:skphik}, what yields
\begin{align}
    \begin{split}
        P(s_k | \bm{Z}_k)
        \propto 
        &
        (1-s_k^2)^{m}
        \times
        \Ftwoone{m,m}{d_k}{s_k^2\bar{s}_k^2}.
    \end{split}
    \label{eq:sk_Zk}
\end{align}
We note in passing that for $d_k=1$, so that $m$ is a positive integer, this hypergeometric function can be rewritten in terms of the Legendre polynomial $L_{m-1}$ of order $m-1$:
\begin{align*}
    P(s_k | \bm{Z}_k)
    \propto 
    &
    \left(
        \frac{1-s_k^2}{1-\bar{s}_k^2 s_k^2}
    \right)^m
    L_{m-1}
    \left(
        \frac{1+\bar{s}_k^2 s_k^2}{1-\bar{s}_k^2 s_k^2}
    \right) .
\end{align*}
The advantage of this form is that for small $m$ one gets the result in the form of a simple polynomial in the two rational expressions appearing in the previous equation.

Integrating out $s_k$ instead gives the estimating distribution for the phase,
\begin{align}
  \label{eq:phik}
    P&(\phi_k | \bm{Z}_k)
    \propto
    P(\phi_k)
    \bigg[
    \Ftwoone{\scriptstyle m, m}{\scriptstyle m+\frac{3}{2}}{q_{\phi_k}^2}
    \\
    & +\frac{2 q_{\phi_k}}{\sqrt{\pi}}
    \frac{
        \Gamma
        \left(
            \scriptstyle m+\frac{3}{2} 
            \textstyle
        \right)
    }{\Gamma(m+2)}
    \left(
        \frac{
         \Gamma\left(
            \scriptstyle m+\frac{1}{2}
            \textstyle
        \right)
        }
        {\Gamma(m)}
    \right)^2
      \Fpq{\scriptstyle m+\frac{1}{2},m+\frac{1}{2},1}{\scriptstyle m+2,\frac{3}{2}}{q_{\phi_k}^2}
    \bigg] ,
    \notag
\end{align}
with %
\begin{align*}
    & q_{\phi_k} \equiv
    \bar{s}_k \cos\big(\phi_k - \bar{\phi}_k\big) .
\end{align*}

\begin{figure}
    \centering
    \includegraphics{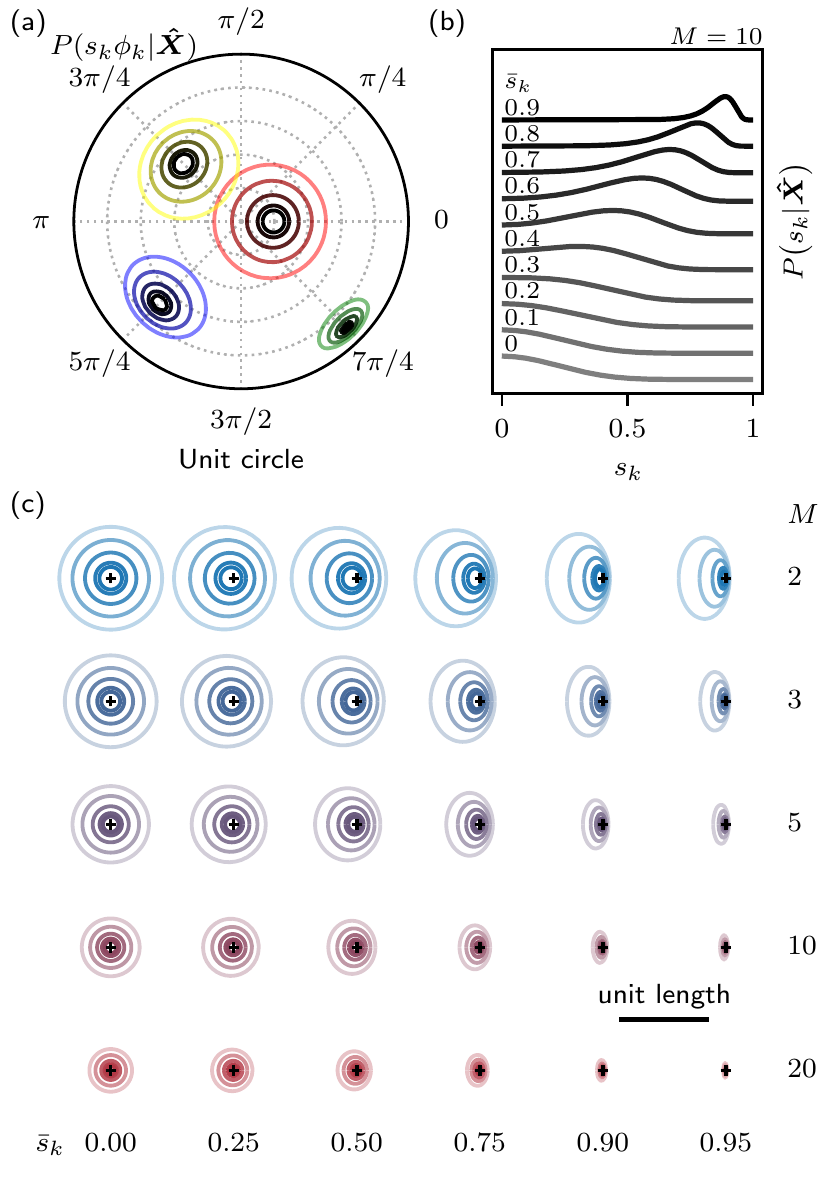}
    \caption{
    Distributions of cross-correlations as a function of $\bar{s}_k$, $\bar{\phi}_k$, and $M$.
    (a) Contours of $P(s_k \phi_k | \hatX)$---Eq.~\eqref{eq:skphik}---enclosing the 75, 50, 25, 10, and 5\%-probability regions.
    The plot is in the complex plane, with $s_k$ as the radial coordinate and $\phi_k$ as the polar angle.
    The unit length is given by the radius of the circle.
    For red, yellow, blue, and green, $\bar{s}_k = \{0.2, 0.5, 0.7, 0.9\}$ and $\bar{\phi}_k = \{0, 3\pi/4, 5\pi/4, 7\pi/4\}$, respectively.
    In all cases, $M = 10$.
    (b) Dependence of $P(s_k | \hatX)$---Eq.~\eqref{eq:sk_Zk}---on $\bar{s}_k$ for $M = 10$.
    The curves have been shifted and scaled in the vertical axis.
    (c) Contours like in (a) as a function of $\bar{s}_k$ and $M$.
    The complex-plane axes have been omitted (but note the given length scale) and the plots arranged in a matrix for the sake of visibility and ease of comparison.
    We take $\bar{\phi}_k = 0$, but recall that the value of $\bar{\phi}_k$ is irrelevant for the shape of $P(s_k \phi_k | \hatX)$ up to polar rotations.
    The crosses mark the maximum-likelihood estimators.
}
    \label{fig:ccor_theory_bsk}
\end{figure}

\begin{figure}
    \centering
    \includegraphics{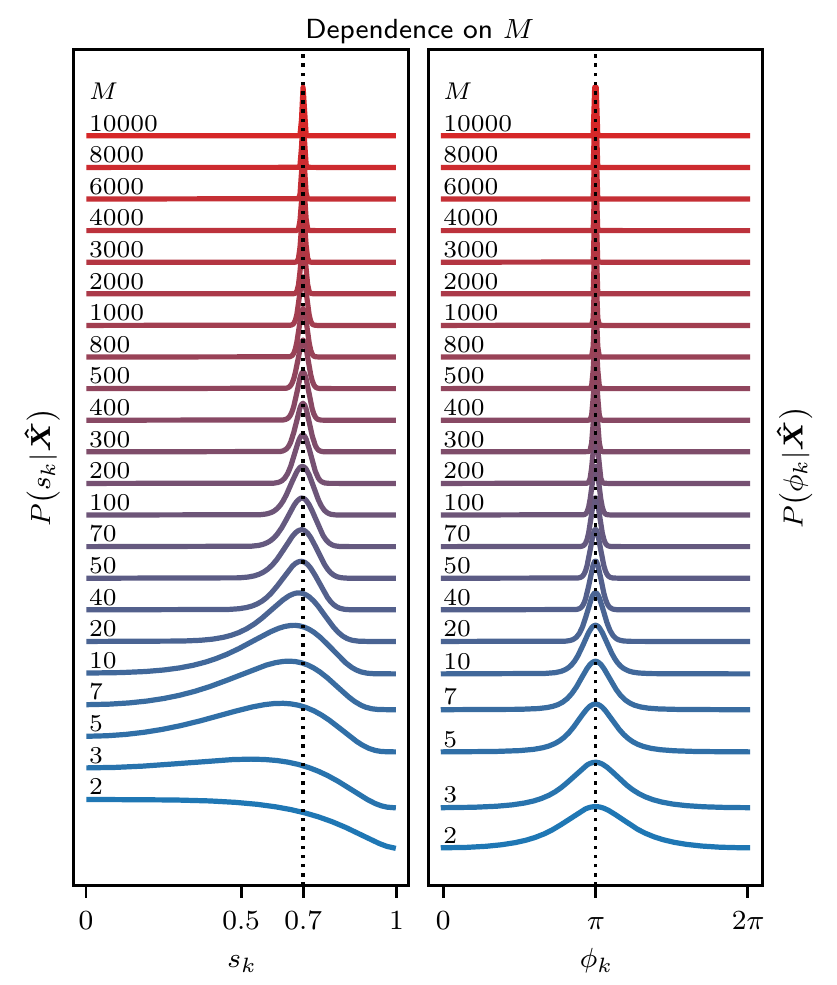}
    \caption{
    Distributions of cross-correlations as a function of $M$ for $\bar{s}_k = 0.7$ and $\bar{\phi}_k = \pi$.
    The curves are shifted and scaled in the vertical axis for visibility.
}
    \label{fig:ccor_theory_M}
\end{figure}

With Eqs.~\eqref{eq:LA_cross}, \eqref{eq:skphik}, \eqref{eq:sk_Zk} and \eqref{eq:phik}, the estimation of the correlation matrix $\hLambda$ is complete.\footnote{One might be interested in different parameterizations, most often using the unnormalized correlation strength $\Lambda_k^{AB}$ instead of $s_k$. The single additional estimating distribution needed for this parameterization is given in App.~\ref{app:unnormalizedCrossCorrelation}.}
Since its diagonal elements (auto-correlations) were already discussed, now we focus on the off-diagonal ones (cross-correlations).
We proceed to describe the estimating distributions of $s_k$ and $\phi_k$ in terms of $M$ and of the sufficient statistics $\bar{s}_k$, $\bar{\phi}_k$.
As expected from the dependence on $q_{s_k\phi_k}$---defined in Eq.~\eqref{eq:qskphik}---, $P(s_k\phi_k | \bm{Z}_k)$ is radially symmetric:
Its shape does not depend on $\bar{\phi}_k$ but only on $\bar{s}_k$ and $M$.
The angle $\bar{\phi}_k$ is just a polar shift, see Fig.~\ref{fig:ccor_theory_bsk}a.
The dependence on $\bar{s}_k$ is analyzed with $P(s_k | \hatX)$ in Fig.~\ref{fig:ccor_theory_bsk}b and with the contour lines of $P(s_k\phi_k | \bm{Z}_k)$ in Fig.~\ref{fig:ccor_theory_bsk}c.
We point out that the estimating distribution is ``conservative'' in the assessment of the correlation strength.
We mean that, (i) as Fig.~\ref{fig:ccor_theory_bsk}b shows, for low or moderate values of $\bar{s}_k$---say,  in the range $0 < \bar{s}_k \lesssim 0.5$---$P(s_k | \bm{Z}_k)$ is quite spread along the $s_k$ axis (the radial axis);
and (ii) $P(s_k \phi_k | \bm{Z}_k)$ is skewed towards low values of $s_k$, see Fig.~\ref{fig:ccor_theory_bsk}c.
As for the dependence on $M$ for a fixed $\bar{s}_k$, Fig.~\ref{fig:ccor_theory_bsk}c shows that greater values reduce the spread of $P(s_k \phi_k | \bm{Z}_k)$.
Figure \ref{fig:ccor_theory_M} displays the transition upon changing $M$ for the estimating distributions $P(s_k | \bm{Z})$ and $P(\phi_k | \bm{Z})$.

The case $M = 1$ does not appear in Fig.~\ref{fig:ccor_theory_M} but it is worth of discussion.
For any $k$, Eq.~\eqref{eq:sk_Zk} reduces to a uniform distribution, $P(s_k | \bm{Z}_k) \propto 1$. It follows from identities\cite{abramowitz2013handbook} ${}_2F_1(1,1,1,x) \propto (1-x)^{-1}$, ${}_2F_1(1/2,1/2,1/2,x) \propto (1-x)^{-1/2}$, and ${}_2F_1(0,0,1/2,x) =1$, for $m=1$, $m=1/2$, and $m=0$, respectively.
Consequently, a single batch of data does not allow one to infer anything about the degree of correlation between the variables $A$ and $B$.
This can be explained as follows:
The variables $\alpha$ and $\beta$ (the transforms of $A$ and $B$) are decoupled in $k$, and for each $k$, we only have the values $\{\alpha_k^{(0)}, \beta_k^{(0)}\}$.
With only two values, it is impossible to tell whether $\alpha$ and $\beta$ are correlated, anticorrelated or not correlated at all.
Hence the posterior for $s_k$ is equal to its prior.
Section \ref{sec:paradox2} contains more on this point.

In parallel with the previous sections, we discuss next the importance and possible applications of our results.
To begin with, one could study cross-correlations with coefficients like Pearson's $r$, which is equal to $\bar{s}_k$---cf.\ Eq.~\eqref{eq:barsk}---, and with the phase $\bar{\phi}_k$ of $\bLambda^{AB}_k$.
But we stress once again that it is more informative to make Bayesian inferences, that is, describing $s_k$ and $\phi_k$ with $P(s_k \phi_k | \hatZ)$, or their marginals $P(s_k | \hatZ)$ and $P(\phi_k | \hatZ)$.
Figures \ref{fig:ccor_theory_bsk}c and \ref{fig:ccor_theory_M} illustrate this point:
Even for the same values of the sufficient statistics $\bar{s}_k$ and $\bar{\phi}_k$, the estimating distributions convey a different degree of confidence depending on the amount of available data.
Not only the confidence level, but also other features of the estimating distribution (like the skewness that we have pointed out) are possibly relevant pieces of information that our estimation retains.
As we discussed with Fig.~\ref{fig:var}, the explicit dependence of the estimating distributions on the number of batches $M$ can help to design experiments aiming at a certain confidence level.
In turn, this enables to detect low correlations or correlations contaminated by strong noise:
When a variable (like $s_k$) is close to $0$, one can only resolve it from its fluctuations by knowing how its error scales with the amount of data.

Moreover, there is an important conceptual difference between choosing a coefficient like Pearson's $r$ (or, for example, \emph{Kendall's $\tau$})\cite{press2007numerical} and our approach.
Sometimes, these coefficients are not defined from the parametrization of the correlation matrix $\hLambda_k$, but instead from other arguments.
For example, Pearson's $r$ is introduced in the context of linear regressions, hence the name ``linear correlation coefficient.''\cite{press2007numerical}
Consequently, one might wonder whether using other parameters would be better to analyze cross-correlations.
However, this concern does not apply to our estimation:
The sufficient statistic $\bar{s}_k$ appeared as a native parametrization of the correlation matrix $\hLambda_k$ in Sec.~\ref{sec:crosscorr_sampling}.
Actually, with our set of parameters $\{\Lambda^A_k, \Lambda^B_k, s_k, \phi_k\}$ for $k = 0, \dots, \nHalf$, we are sure not to loose any information about the correlation matrix $\hLambda$.

The estimation of cross-correlations is now complete.
We conclude this section by discussing the estimation of $\mu_A$.
It holds identically for $\mu_B$.
Similarly to what we pointed out about $\Lambda^A_k$, the result for the estimating distribution $P(\mu_A | \hatX)$ is not expected to be exactly the same as $P(\mu_A | \hatA)$---Eq.~\eqref{eq:autocorr_nu}.
The estimating distribution $P(\mu_A | \hatX)$---equivalent to $P(\mu_A | \bm{Z}_0)$ due to the $k$ factoring in Eq.~\eqref{eq:ccor_factor}---reads
\begin{align}
    P&(\mu_A | \bm{Z}_0)
    = 
    \int_0^{\infty} d\Lambda_0^A
    \int_0^{\infty} d\Lambda_0^B
    \sum_{\phi_0\in \{0,\pi\}}
    \notag
    \\
    & \times
    \int_0^1 ds_0
    \int_{-\infty}^{\infty} d\mu_B \,
    P(\mu_A \mu_B \Lambda_0^A \Lambda_0^B
    s_0\phi_0 | \bm{Z}_0)
    \notag
    \\
    &
    \quad \quad
    \propto
    \int_0^{\infty} d\Lambda_0^A
    \int_0^{\infty} d\Lambda_0^B
    \sum_{\phi_0\in \{0,\pi\}}
    \notag
    \\
    &
    \times
    \int_0^1 ds_0
    \int_{-\infty}^{\infty} d\mu_B \,
    P(\bm{Z}_0 | \bm{\nu}\hLambda_0)
    P(\bm{\nu}\hLambda_0) .
    \label{eq:muA_cross_prev}
\end{align}
Once more, we have applied Eq.~\eqref{eq:rule} and Bayes' rule to get this expression.
As for the presence of the means $\bm{\nu}$ in this formula, recall our previous remark that $\bm{\nu}$ only appears in the sampling distribution for $k=0$.
The sampling distribution $P(\bm{Z}_0 | \bm{\nu}\hLambda_0)$ is given by Eq.~\eqref{eq:ccor_sampling_F}, and the prior $P(\bm{\nu}\Lambda_0)$ by Eq.~\eqref{eq:factor_mat}.
After these substitutions, we can integrate analytically in all variables but $s_0$, which remains for numerical evaluation:
\begin{equation}
\begin{split}
    P(& \mu_A | \bm{Z}_0)
    \propto
    \int_0^1
    ds_0
    \left( \sqrt{\frac{1}{1-s_0^2}+\frac{(\bar{\alpha}_0-\nu_A)^2}{\bar{\Lambda}^A_0}} \right)^{-1} \\
    & \times
        \sum_{\kappa=\pm1} \left( \frac{ \kappa s_0 \bar{s_0}}{\sqrt{1-s_0^2}} +\sqrt{\frac{1}{1-s_0^2} +\frac{(\bar{\alpha}_0-\nu_A)^2}{\bar{\Lambda}^A_0}} \right)^{1-M}
 \end{split}
    \label{eq:P_LambdaA_D}
\end{equation}
There is a clear resemblance to the estimating distribution of Sec.~\ref{sec:autocorr}:
As a check, Eq.~\eqref{eq:P_LambdaA_D} reduces to Eq.~\eqref{eq:autocorr_nu} in the absence of cross-correlations, namely using the prior $P(s_0) =\delta(s_0-0)$.
The comparison between both results appears in Fig.~\ref{fig:A_vs_AB}.
An identical discussion to that comparing Eqs.~\eqref{eq:autocorr} and \eqref{eq:LA_cross} applies also here.

\subsection{Generation of noise}
\label{sec:generate_crosscorr}

While the method to generate noise given in Sec.~\ref{sec:generate_autocorr} was previously reported in the literature for a specific kind of noise \cite{timmer1995on}, we are not aware of proposals of how
to generate \emph{correlated noise}, namely variables with given means $\mu_A$, $\mu_B$ and spectra $\Lambda_k^A$, $\Lambda_k^B$, and $\Lambda_k^{AB}$ (or equivalently, $\Lambda_k^A$, $\Lambda_k^B$, $s_k$, and $\phi_k$).
In this section, we propose a general method for this purpose.

We proceed analogously to Sec.~\ref{sec:generate_autocorr}, using now the sampling distribution given by Eq.~\eqref{eq:sampling_zk}.
However, it is necessary to transform the variables $\alpha^{(m)}_k$ and $\beta^{(m)}_k$---inside $\bm{Z}_k^{(m)}$---into two independently distributed scalars.
To do so, we perform the transformation
$
    \bm{W}_k^{(m)}
    = \hat{T}_k
    \big(
        \bm{Z}_k^{(m)}-\bm{\nu}_k
    \big)
    \notag
$
that diagonalizes the matrix inside the sampling distribution: 
\begin{align}
    \hat{T}_k\hLambda_k^{-1}\hat{T}_k^{-1}
    =
    \begin{bmatrix}
        H_k & 0 \\
        0   & J_k
    \end{bmatrix} .
    \label{eq:D}
\end{align}
$\hat{T}_k$ exists and $H_k, J_k \in \mathbb{R}$ because $\hLambda_k^{-1}$ is Hermitian, see Eq.~\eqref{eq:LambdakI}.
In the coordinates $\bm{W}_k^{(m)} = \big(h_k^{(m)}, j_k^{(m)}\big)$, the sampling distribution reads
\begin{align}
    P\big(
        & h_k^{(m)} \big| \bm{\nu}_k\hLambda_k
    \big)
    = \left(\frac{d_k}{\pi H_k}\right)^{d_k}
    \exp\left(
        -d_k
        \frac{\big|h_k^{(m)}\big|^2}{H_k}
    \right) .
    \notag
\end{align}
The formula applies for $j_k^{(m)}$ upon the substitution $H_k \to J_k$.
Splitting to the real and imaginary parts,
\begin{align}
    & P\big(\mathfrak{R}h_k^{(m)}
    \big| \bm{\nu}_k H_k\big)
    = \sqrt{\frac{d_k}{\pi H_k}}
    \exp\left(
        -d_k
        \frac{
            \big(\mathfrak{R}h_k^{(m)}\big)^2
        }{H_k}
    \right) ,
    \label{eq:Rehk}
\end{align}
the same for $\mathfrak{I}h_k^{(m)}$, and also for $\mathfrak{R}j_k^{(m)}$ and $\mathfrak{I}j_k^{(m)}$ after the change $H_k \to J_k$.
The distribution for the imaginary parts only applies to $k \neq 0, n/2$, because $h^{(m)}_k$ and $j^{(m)}_k$ are real otherwise.

These distributions are all we need to generate two variables $A$ and $B$ with given means $\mu_A$, $\mu_B$ and spectra parametrized by $\Lambda_k^A$, $\Lambda_k^B$, $s_k$, and $\phi_k$.
The procedure is as follows.
To generate a batch $(m)$,
(i) for each $k$, diagonalize the matrix $\hLambda_k$, finding the coordinate-change matrix $\hat{T}_k$ and the eigenvalues $H_k$ and $J_k$;
(ii) generate the independent and normally distributed variables $\mathfrak{R}h_k^{(m)}$ and $\mathfrak{I}h_k^{(m)}$ with their sampling distributions---Eq.~\eqref{eq:Rehk} and similar---, and do the same for $j_k^{(m)}$;
finally, (iii) perform the transformation $\bm{W}_k^{(m)} \to \bm{Z}_k^{(m)}-\bm{\nu}_k$ with $\hat{T}_k^{-1}$, and $\bm{Z}_k^{(m)} \to \hatA^{(m)}, \hatB^{(m)}$ with $\tilde{F}^{-1}$ (the inverse Fourier transform from the frequency space back to the time space).

\section{Numerical example}
\label{sec:example}

\begin{figure}
    \centering
    \includegraphics{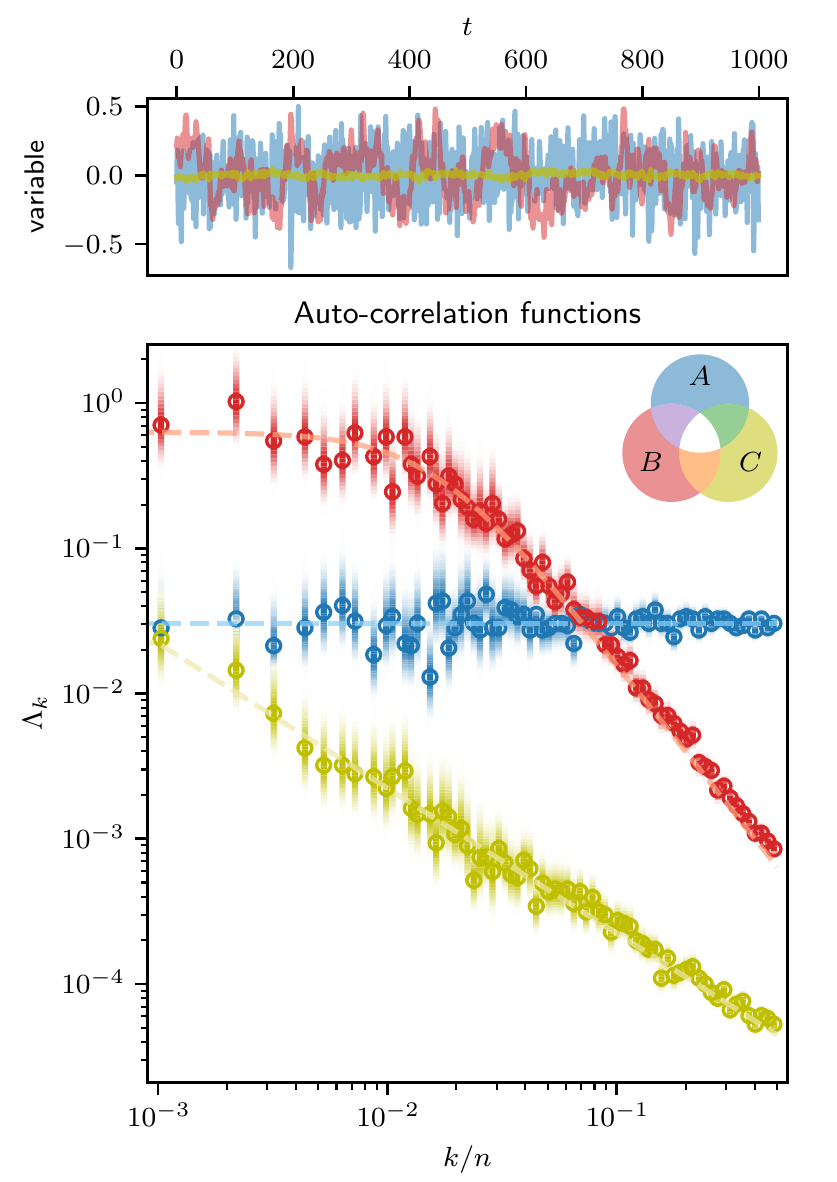}
    \caption{
    Top: Time evolution of the generated variables $A$, $B$ and $C$ for batch $m = 1$.
    Bottom: Estimation of the spectra $\Lambda^A_k$ with Eq.~\eqref{eq:autocorr}; and of $\Lambda^B_k$, $\Lambda^C_k$ with Eq.~\eqref{eq:LA_cross}.
    The maximum-likelihood estimators appear circled, and the estimating distributions for $\Lambda_k^A$, $\Lambda_k^B$ and $\Lambda_k^C$ are overlaid as a linear color gradient.
    The dashed lines plot the spectral functions used to generate the data, see Eqs.~\eqref{eq:noises}.
     Note that adopting the log scale for the horizontal axis makes the distributions for high $k$ overlap. To declutter the plot and keep its information content, we locally average the points using the procedure described in App.~\ref{app:pointAveraging}.
    }
    \label{fig:acorr_clustered_AB}
\end{figure}

In this section, we test and illustrate our analytical Bayesian estimation formulas applying them on numerically generated uncorrelated and correlated noise. Specifically, we generate $M = 10$ batches of noisy signals for three variables $A$, $B$, and $C$.
Each batch contains $n = 1000$ data points.
We work in arbitrary units for all magnitudes, including time.
We choose spectra,
\begin{align}
    \Lambda^A_k = \text{const.} ,
    \quad
    \Lambda^B_k \propto \frac{1}{\lambda^2 + k^2} ,
    \quad
    \Lambda^C_k \propto \frac{1}{k} ,
    \label{eq:noises}
\end{align}
corresponding to white noise; to noise with an exponentially decaying correlation function, $\langle x(s) x(s+t) \rangle \propto e^{-t/\lambda}$, taking $\lambda = 10$; and to $1/f$ noise, respectively.
For $1/f$ noise, we take a frequency cutoff smaller than the lowest nonzero $k$ frequency.
$A$ is generated independently from $B$ and $C$ with the method presented in Sec.~\ref{sec:generate_autocorr}.
$B$ and $C$ are anti-correlated with a constant strength $s_k = 0.7$ and phase $\phi_k = \pi$.
They are generated according to Sec.~\ref{sec:generate_crosscorr}.
Finally, $A$ and $B$ have the same total spectral power (that is, the same area under the spectral-power curve), while $C$ is hundred times weaker.

\begin{figure*}[tbp]
    \centering
    \includegraphics{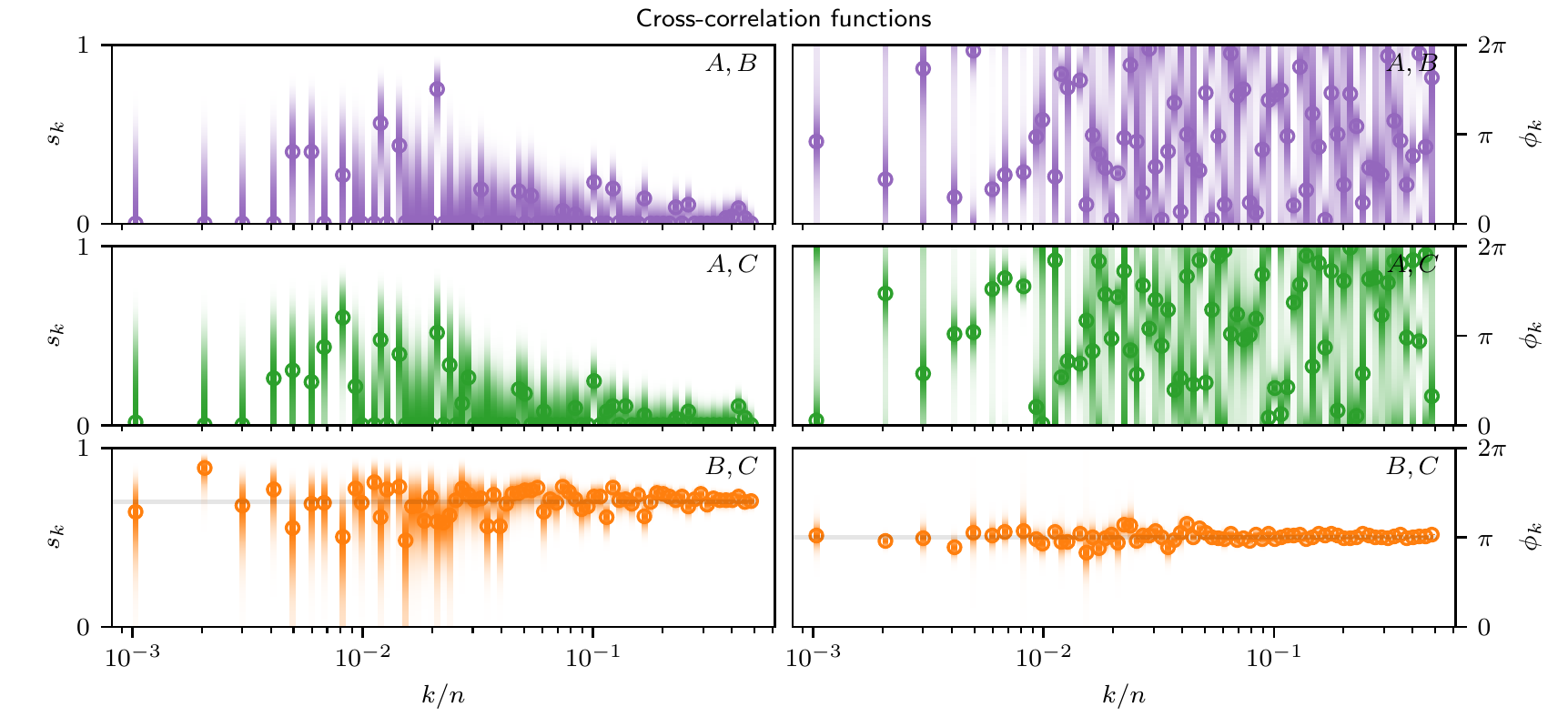}
    \caption{
    Left: Estimation of the correlation strength for all possible pairs in the set $\{A, B, C\}$, with $P(s_k | \hatX)$ as a linear color gradient.
    The legend appears in the upper right corner and the color code is consistent with Fig.~\ref{fig:acorr_clustered_AB}.
    The maximum-likelihood estimators appear circled.
    The results reflect that only the variables $B$ and $C$ are correlated.
    The value $s_k = 0.7$, used to generate the data, is marked with a faint gray line.
    Right: The same for the correlation phase with the estimating distribution $P(\phi_k | \hatX)$.
    The gray line marks $\phi_k = \pi$. The points were averaged according to the procedure described in App.~\ref{app:pointAveraging} for better visibility.
}
    \label{fig:sf}
\end{figure*}

We chose to simulate this data according to the following purposes.
First, we aim at analyzing auto-correlations for some typical noise types.
Second, we intend to check that the cross-correlation between the variables $B$ and $C$ is correctly quantified in strength and phase, and also assess the lack of correlation between $A$ and $B$, and $A$ and $C$.
In other words, we set a test for the positive and negative assessment of correlation.
At last, we want to check that our estimation of discrete spectra is sensible for data generated from continuous dispersions, at least qualitatively.
The next section discusses this point further.

Figure \ref{fig:acorr_clustered_AB} plots the time evolution of the first batch (top panel) and the resulting estimated auto-correlations (bottom panel).
The estimation of $\Lambda_k$ uses Eq.~\eqref{eq:LA_cross} of Sec.~\ref{sec:crosscorr_estimation} for $B$ and $C$, and Eq.~\eqref{eq:autocorr} of Sec.~\ref{sec:est_autocorr} for $A$.
These estimating distributions are plotted in a color scale.
The results fit well the continuous curves used to generate the data.

Figure \ref{fig:sf} analyzes cross-correlations.
The estimating distributions for $s_k$ and $\phi_k$ appear plotted in a color scale for every pair in the set $\{A, B, C\}$.
The degree of correlation is correctly estimated in all cases:
While $P(s_k | \hatX)$ accumulate around $0.7$ for $\{B, C\}$, they peak at 0 for the uncorrelated pairs $\{A, B\}$ and $\{A, C\}$ at majority of the frequencies.\footnote{But not at every $k$: as a result of statistical fluctuations, the noise of uncorrelated variables shows false correlations (a peak away from $s_k=0$) at some frequencies. This behavior is to be expected. At these values, the estimating distribution of the phase is localized (not spread to whole $[0,2\pi)$ interval), which is consistent with a finite value of MLE$(s_k)$. One could come up with a statistical distribution for their number analogous to Eq.~\eqref{eq:per_dist}. However, one can judge whether the given two variables are or are not correlated even without such tests, see Sec.~\ref{sec:paradox2}.}
The results for $P(\phi_k | \hatX)$ are consistent with it those for the correlation strength:
For the uncorrelated cases of $\{A, B\}$ and $\{A, C\}$, the estimating distributions peak at random values and mostly are spread over $[0, 2\pi)$.
For the correlated case $\{B, C\}$, the estimating distribution is narrow and peaks at $\pi$.

Our results succeed in estimating the parameters that were used to generate the data.
Some comments regarding the convergence from discrete to continuous spectra, oriented towards the application to actual experimental data, appear in the next section.

\section{Discussion}

We now discuss several aspects of our results concerning their application. In addition, a reader with experience in spectral estimation might find some of our results counter-intuitive. We clarify those seeming contradictions, assigning them mostly to the difference between parametric and non-parametric estimation. The technical results of our article have been already given in previous sections and they can be used without following the discussion here.

\subsection{Non-parametric versus parametric estimation}

Let us return to a single variable of time $A(t)$ observed at discretized times $t_i=i \Delta$, resulting in a discrete set $\{ A_i \equiv A(t_i) \}_{i=1}^n$. This discrete set allows one to infer properties of the auto-correlation $\Sigma(\tau)$ evaluated at discrete values of the time delay $\tau_d =d \Delta$:  $\{ \Sigma_d \equiv \Sigma(\tau_d) \}_{d=\minusnHalf+1}^{\nHalf}$. Due to the discrete sampling of $A$ one has no access to $\Sigma(\tau)$ for other values of $\tau$. Since the Fourier transform is a bijection, the discrete set of the Fourier coefficients $\{\Lambda_k \}$ is equivalent to $\{ \Sigma_d \}$: It is a minimal-size set of parameters required to describe what can be known about $\Sigma(t)$ from the discretized values of $A$. However, one can introduce a different set of parameters, as a \emph{model} to explain or predict $\{ \Sigma_d \}$.
Any such model can be then tested with respect to the data, in the sense that the parameters of the model can be estimated. 

If the function $\Sigma(\tau)$ is defined for any real $\tau$, what requires $A(t)$ to be defined for any real $t$, it is natural to consider $\Lambda(\kappa)$, the continuum of the Fourier components of $\Sigma(\tau)$, as such an extended model. Introducing a continuum of redundant variables, there would be no hope of estimating them from $\{ \Sigma_d \}$. The redundancy is compensated by considering simple functions for $\Lambda(\kappa)$,  describable by only a few parameters. A typical example would be $\Lambda(\kappa) = c /\kappa^\alpha$ for $\kappa \in \langle \kappa_\mathrm{low}, \kappa_\mathrm{high} \rangle$ and zero otherwise. Here, the overall strength $c$, the power $\alpha$ and the cutoffs $\kappa_\mathrm{low}$ and $\kappa_\mathrm{high}$ would be the four (instead of a continuum of) parameters to be estimated. In general, one considers a family of continuous curves $\mathcal{C}_{\vec{p}}(\kappa)$ defined by $N$ parameters gathered in vector $\vec{p}$. Since the goal is to estimate the parameters $\vec{p}$, one calls the procedure \emph{parametric} estimation. We do not introduce such functional models. We estimate the natural parameterization of $\{ \Sigma_d \}$, namely the discrete set $\{\Lambda_k \}$. To distinguish it from the other approach, we call it \emph{non-parametric} estimation.

Now we come to the central point of this section. The motivation to introduce the continuum version of $\Lambda_k$ is not so much the existence of $\Sigma(\tau)$ as a ``more-true'' entity than its discretized version $\{ \Sigma_d \}$.\footnote{Indeed, a continuum model can be adopted even if $A_i$ are undefined for intermediate ``times'' (or if $i$ does not represent time, but an index that is inherently integer, such as the rank). There is no difference from the point of view of the estimation. The difference would be in the interpretation: it might be unclear what a continuum function $\Lambda(\kappa)$, even if fitting a simple shape well, represents. In this case, one might be better off postulating a candidate shape directly in the discrete space, with the benefit that no $\mathcal{F}$ discussed in the next subsection arises.}
Rather, the motivation is to compress the information contained in the results of non-parametric estimation. \footnote{While Ref.~\cite{good1950probability} discusses a slightly different ``frequency distribution'', the following citation from page 60 there is exactly to the point here: ``But it is often possible to find a simple form that fits the frequency distribution approximately. If this can be done it has the advantage of describing the results of the statistics briefly. In some cases it is suggestive of the causes that lie behind the results. But the main reason, in general, for looking for a simple mathematical 'law' of this type, is that if it is found it is believed to have predictive value. That is to say the simple law, if it is a very good approximation to the distribution function $F$ of the original sample, is likely to describe the distribution function of another sample (or of the whole population) \textit{even better than $F$ would}. This is partly because it is likely that there are a few predominating causes lying behind the statistics, even though these cases are unknown''.}
The latter typically results in hundreds to millions of data points in a plot such as Fig.~\ref{fig:acorr_clustered_AB}. To interpret such a plot requires that one or a few of the typical shapes are recognized in it.\footnote{The typical shapes are: a constant representing a white noise, that is a Markovian environment; a Lorentzian for an environment equivalent to a two-level system with fixed transition rates \cite{machlup1954noise}, a $1/f$ for a collection of two-level systems with a specific distribution of transition rates (Ref.~\cite{milotti20021f} attributes it to Johnson 1925 and Schottky 1926); $1/f^2$ for a random walk \cite{wang1945on}. Further, a substantial volume of results exists on ARMA models \cite{kay1981spectrum}, with spectrum a general rational function of $k$.} Whether knowingly or not, anyone looking at a figure such as Fig.~\ref{fig:acorr_clustered_AB} performs intuitively at least a rough parametric estimation and draws conclusion from there. That is, 
 the global properties of the spectral data are judged, meaning relations between different frequencies. \emph{Our formulas do not do this; they describe each frequency $k$ in isolation.} Once this point is appreciated, the ``paradoxes'' that we discuss below quickly resolve.

\subsection{Continuous spectra}
\label{sec:continuous}

Let us now shortly comment on the parametric estimation, which is fitting the parameters of a continuous curve $\Lambda(\kappa) \equiv \mathcal{C}_{\vec{p}}(\kappa)$. Unfortunately, the issue with it is that, unlike the relation of $\{ A_i \}$ to $A(t)$ and $\{ \Sigma_d \}$ to $\Sigma(\tau)$ which are trivial, the relation of $\{ \Lambda_k \}$ to $\Lambda(\kappa)$ is not: $\Lambda_k$ is not equal to $\Lambda(\kappa_k)$ evaluated at the corresponding real frequency $\kappa_k=k/n\Delta$. Namely, the function $\Lambda$ has to be additionally transformed:
\begin{equation}
\Lambda(\kappa) \to \mathcal{F}[\Lambda] (\kappa) \to \{\mathcal{F}[\Lambda] (\kappa_k)\}_{k=1}^n \equiv \{ \Lambda_k\}_{k=1}^n,
\label{eq:FC}
\end{equation} 
The transformation $\mathcal{F}$ reflects two well-known complications: the continuous frequencies which are beyond the Nyquist frequency $|\kappa| > 1/2\Delta$ are aliased (folded over) into this range and the values of $\Lambda(\kappa^\prime)$ for which $n \Delta \kappa^\prime$ is not integer are smeared throughout the discrete set $\{ \Lambda_k\}$ with a weight falling off only as $1/(\kappa-\kappa^\prime)^2$. We do not give formulas for $\mathcal{F}$, they can be found in Ref.~\cite{press2007numerical}. Equation \eqref{eq:FC} then states that $\Lambda(\kappa)$ has to be first aliased and smeared (the first arrow) and then sampled (the second arrow), to arrive at a discrete set $\{ \Lambda_k \}$, the equivalent discrete spectrum.

Nevertheless, once the candidate model function $\mathcal{C}_{\vec{p}}(\kappa)$ has been decided for and the above transformation is performed, one can estimate the parameters $\vec{p}$ from the likelihood 
\begin{align}
    L \equiv
    P\big(\mathcal{F}[\mathcal{C}_{\vec{p}}]
    \big| \hatX\big)
    = \prod_{k=0}^{\nHalf}
    P\big(
    \Lambda_k = \mathcal{F}[\mathcal{C}_{\vec{p}}]\left(\kappa_k\right)
    \big| \hatX
    \big) .
    \label{eq:likelihood}
\end{align}
Here, $P\big(\Lambda_k = \mathcal{F}[\mathcal{C}_{\vec{p}}](\kappa_k) \big| \bm{\hat{X}} \big)$ is given by Eq.~\eqref{eq:autocorr} or Eq.~\eqref{eq:LA_cross} evaluated at $\Lambda_k = \mathcal{F}[\mathcal{C}_{\vec{p}}](\kappa_k)$.
In practice, the parametric fitting is often done either simply ignoring the $\mathcal{F}$ in Eq.~\eqref{eq:likelihood}, or moving it on the data, by changing the equation $\Lambda = \mathcal{F}[\mathcal{C}]$ into $\mathcal{F}_\mathrm{inv} [\Lambda] = \mathcal{C}$, using an abstract notation without indexes. In the first approach the quality of the approximation $\mathcal{F} \approx 1$ relies on the properties of the function $\mathcal{C}_{\vec{p}}(\kappa_k)$ being estimated, and is thus unknown. 
In the second approach, the effective inverse $\mathcal{F}_\mathrm{inv}$ is implemented by windowing,\cite{harris1978on} filtering, and procedures of similar kind.\footnote{As an illustration, Ref.~\cite{caloyannides1974microcycle} which measured 1/f noise down to one of the lowest frequencies ever (for a semiconductor), found the need for ``pre-whitening'', ``windowing'', ``de-aliasing'', and ``post-greening'' of the measured data.} Again, since $\mathcal{F}$ certainly does not have an exact inverse (being a map from a real axis to a finite set), the quality of the effective inverse is difficult to predict and will depend on the function $\mathcal{C}$ itself. Therefore, we suggest using Eq.~\eqref{eq:likelihood} instead: there, the data are, in principle, processed independently from the parametric model, by a non-parameteric estimation. The likelihood is given by evaluating the resulting estimating distributions at points given by the model function transformed into an effective discrete spectrum.\footnote{The evaluation of $\mathcal{F}[\mathcal{C}](\kappa_k)$ does not need high precision: a precision comparable to the error bar of $\Lambda_k$ suffices.}
The advantage is that this likelihood is exact, and thus allows one to assign rigorous error bars to the estimated parameters $\vec{p}$.   

We note that one can do parametric estimation directly, skipping the non-parametric one. Reference \onlinecite{bretthorst1988bayesian} is one example, where the model likelihood is calculated in the time domain (that is, directly from the data $A_i$). Nevertheless, the key to be able to do so is to have the candidate model, equivalent to $\mathcal{C}_{\vec{p}}$ in the notation here.\footnote{In the nomenclature of Ref.~\onlinecite{bretthorst1988bayesian}, it requires ``prior knowledge~...~to supplement the data''.} Without that, to propose a reasonable candidate $\mathcal{C}_{\vec{p}}$ one has to start with a plot such as Fig.~\ref{fig:acorr_clustered_AB} anyway. Our approach can, therefore, be viewed as splitting the parametric estimation to two steps, what allows one to delay conjecturing about $\mathcal{C}_{\vec{p}}$ till we can take a more informed decision.

Finally, analogous considerations hold for the cross-correlation, where formulas such as Eq.~\eqref{eq:likelihood} will apply for $\Lambda_k^A$, $\Lambda_k^B$, and $\Lambda_k^{AB}$ together. We point out here that the estimating distribution in Eq.~\eqref{eq:ccor_factor} is actually non-separable in the four variables. There are thus dependencies in the plausible values of the parameters. Going to marginals, which we have done in calculating, for example, the estimating distributions $P(s_k)$ and $P(\phi_k)$, means that those dependencies were lost. One could consider fitting parametric models to the more-variable estimating functions such as $P(s_k \phi_k$) and similar, instead of the marginals. Nevertheless, we leave investigations of parametric estimation as outlined in this subsection for future works. 

\subsection{Paradox 1: Precision does not increase upon acquiring more data?}
\label{sec:paradox1}

Let us again focus on a single variable with discrete spectrum $\Lambda_k$.
Consider that we increase the batch size $n$ keeping the number of batches $M$ and the time interval $\Delta$ fixed.
While the number of $k$ points, being $\nHalf+1$, increases, the number of data points $\big\{\alpha_k^{(0)}, \dots, \alpha_k^{(M-1)}\big\}$ for each $k$ does not, remaining at $M$.
Given that our estimating distributions (and the error that they convey) depend only on $M$---see e.g.\ $P(\Lambda_k^A | \bm{\alpha}_k)$ in Eq.~\eqref{eq:autocorr}---, it looks paradoxical that the acquisition of more data by increasing $n$ does not have any effect on the precision of the estimation (given by $M$).

This  paradox is resolved by considering our previous discussion of Eq.~\eqref{eq:likelihood}:
While it is true that after increasing $n$, the precision of $\Lambda_k^A$ does not increase \emph{for given $k$}, the fact that there are more values of $k$ results in a more precise parametric estimation. 
For example, in Fig.~\ref{fig:acorr_clustered_AB}, doubling $n$ but keeping $M$ fixed would correspond to roughly doubling the number of red distributions---and circles---, but keeping their shape---and thus the length of the corresponding error bar---fixed. 
We could fit a continuous curve better than with a lower $n$.
The same point has arisen in Sec.~\ref{sec:generate_autocorr}:
It might have looked paradoxical that the signal to noise ratio given in Eq.~\eqref{eq:snr} does not increase with the number of measured data, proportional to $n$. Again, the increase of the \emph{number} of $k$ points delivers a higher precision of a parametric fit. 

\subsection{Paradox 2: A single trace can not reveal anything about cross-correlations?}
\label{sec:paradox2}

The paradox from the previous subsection has a variant appearing for cross-correlations. Namely, we found that the estimating distribution of the correlation strength $s_k$, Eq.~\eqref{eq:sk_Zk}, is constant,  $P(s_k | \bm{Z}_k)=1$ irrespective of the data, if $M=1$. This seems to imply that irrespective of what was measured, from a single batch one can not draw any conclusions 
about a possible correlation of two variables $A$ and $B$. One can easily find an example where the intuition 
points against such a statement: For example, imagine that a million of pairs ${B_i,C_i}$ were measured in a single batch, and while wildly fluctuating individually, they happen to be opposite in all instances $B_i=-C_i$. One can obviously judge the perfect anti-correlation of $B$ and $C$ from such data. 

\begin{figure*}
    \centering
    \includegraphics{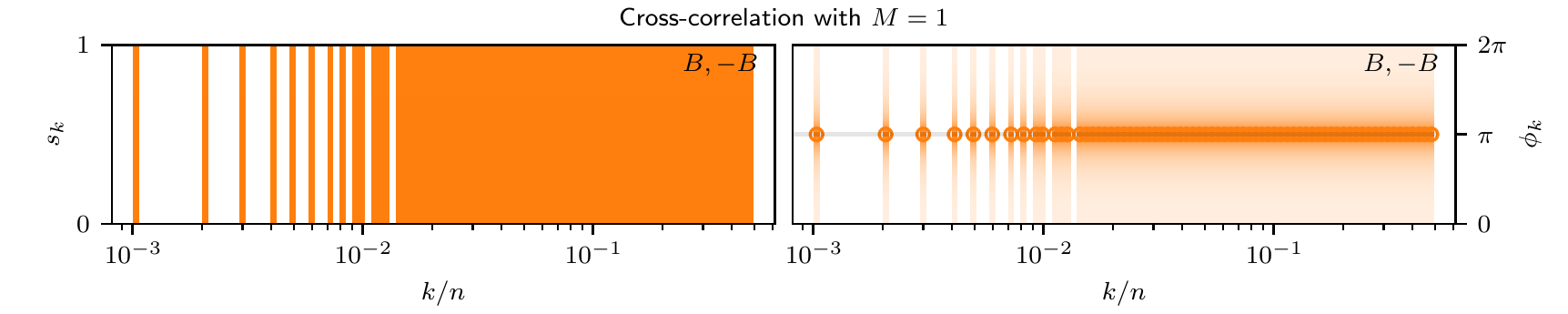}
        \includegraphics{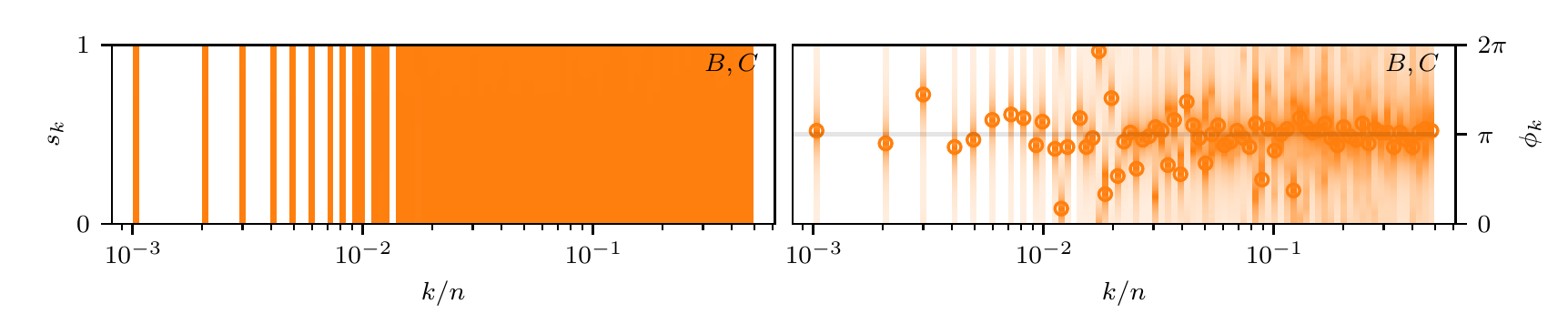}
    \caption{
    The analog to Fig.~\ref{fig:sf} for fully or partially anti-correlated signals for a single batch, $M=1$. The left panels show the estimating distributions for the correlation strength $s_k$, the right ones for the correlation phase $\phi_k$. We generate $B$ and $C$ with spectra as shown in Fig.~\ref{fig:acorr_clustered_AB}, with $s_k=0.7$ and $\phi_k=\pi$ for all $k$. In the first row we calculate correlation of $B$ with its inverse $-B$, while in the second row the correlation of $B$ and $C$. To declutter the plot the horizontal axis was divided into 100 bins equally spaced in the logarithmic scale and all distributions falling within the same bin were arithmetically averaged. To deliver the point of the figure, we intentionally do not use here a more efficient averaging described in App.~\ref{app:pointAveraging}.
}
    \label{fig:sf_M1}
\end{figure*}

The explanation is, again, in the global properties of the estimating distributions, that is, their relations for different frequencies $k$. While it is true that looking at a single $k$, one can conclude nothing about $s_k^{BC}$, the fact that all the estimated phases $\phi_k$ are the same in the given example makes it obvious that $B$ and $C$ are correlated. The parametric estimation would confirm it, for example examining a model such as $B(t)=a A(t) + c C(t)$, finding $c \approx -1, a \approx 0$. To illustrate, Fig.~\ref{fig:sf_M1} shows the resulting estimating distributions in the first row. A similar situation arises if the correlation is imperfect, see the second row of the figure. Again, while nothing can be concluded from the estimating distributions for $s_k$ as they are all constant, the clustering of phases $\phi_k$ shows that they are not randomly distributed and thus the two signals are not uncorrelated.

\subsection{Paradox 3: Are formulas for $M>1$ ever relevant?}

In practice one has seldom (if ever) access to statistically independent batches. In majority of cases, the whole set of $n \times M$ data points is measured on a single system, after which the data are separated into $M$ batches. There might or might not be time delays between measurements taking data assigned to different batches $m$. Nevertheless, data measured in this way correspond to a single batch, possibly with some portions of the data omitted. As omitting valid data never improves any estimation, the best estimation one can hope for in this scenario is to retain all the data without omissions and use it as a single batch, not splitting it into artificial batches. This is the best (most informative) procedure for a subsequent parametric estimation: Despite our formulas taking trivial forms (for $M=1$) at each $k$, for example $s_k=1$ irrespective of the data, there is no issue in using the $M=1$ non-parametric estimation results for the parametric estimation; we have explained it in the previous two subsections.

Nevertheless, calculating the likelihood in the parametric estimation, Eq.~\eqref{eq:likelihood}, is not the only use of the non-parametric estimation. As we pointed out in Sec.~\ref{sec:continuous}, the latter is required to assist in formulating the parametric model in the first place. This task might be difficult based on seeing only the $M=1$ plots. The artificial splitting of a single measurement into $M$ batches can then be seen as a visualization tool.\footnote{The ``windowing'', et cetera, can also be viewed as useful visualization tools.} Once the parametric model is formulated, one should use the true number of batches, without splitting them artificially to smaller ones, even for $M=1$.

A natural question is the error stemming from splitting a single batch into several smaller ones which are then treated as statistically independent (even though they are not). While we have investigated this question, we conclude that quantifying the error arising in this way in the non-parametric estimation requires additional assumptions on the correlation function. We have not been able to find assumption(s) which would be general and lead to useful formulas. We leave this question open. In any case, we expect that the ``averaging'' described in App.~\ref{app:pointAveraging} makes this issue irrelevant, since it replaces the artificial splitting of batches by a more efficient procedure.

\section{Conclusion}
\label{sec:conclusion}

In this article, we have applied Bayesian statistics to estimate correlation functions.
We have given the estimating probability distributions to calculate (i) the variance for scalar variables; (ii) the auto-correlation functions for time-dependent variables, which includes the study of noise spectra; and (iii) the cross-correlations for possibly coupled pairs of variables, which covers the study of correlated noise.
Our approach guarantees an optimal usage of the information contained in the data, makes no assumptions on the distribution of the variables, reduces to a minimum the assumptions on the priors, and avoids the arbitrary choice of estimators.
Every result is expressed in terms of a distribution, from which confidence intervals can be calculated afterwards.

The relevance of our results is manifest when applied to spectral analysis.
They allow one to estimate the spectrum in terms of the estimating distributions.
This inference is more informative than using standard estimators such as the periodogram.
It allows one to assess the certainty of the spectrum estimation (for example, by error bars) and fit continuous spectra in terms of maximum likelihood.
Our analysis also covers the study of cross-correlations from a Bayesian perspective:
We offer a systematic way to assess the correlation between two variables without choosing arbitrary parameters to quantify it and with distributions that assess the certainty level.
We presented some numerical tests that support our calculations.

Beyond these fundamental results, our work discusses other aspects of noise from a Bayesian perspective.
As an example concerning the statistical properties of the periodogram, we derived its exact signal-to-noise ratio and proved that it is universal.
We also proposed a method to numerically simulate correlated stationary noise with an arbitrary spectrum.

\acknowledgments

We acknowledge the support from CREST JST (JPMJCR15N2 and JPMJCR1675), Swiss National Science Foundation (SNSF), NCCR SPIN, JST PRESTO grant No. JPMJPR21BA, MEXT Quantum Leap Flagship Program (MEXT Q-LEAP) grant No. JPMXS0118069228, JST Moonshot R\&D Grant Number JPMJMS2065, JSPS KAKENHI grant Nos. 16H02204, 17K14078, 18H01819, 19K14640, and 20H00237, The Precise Measurement Technology Promotion Foundation, Suematsu Fund, and Advanced Technology Institute Research Grants.

\appendix

\section{Multivariate Gaussians}
\label{app:gaussian}

In this appendix, we prove that for the set of variables $\vec{x} = \{x_0, \dots, x_{n-1}\}$ and given the expected values $\mu_i = \langle x_i \rangle$ and $\hat{\Omega}_{ij} = \langle x_i x_j \rangle$, the multivariate Gaussian distribution maximizes the entropy.\footnote{While the entropy-maximization property of Gaussian distributions is known \cite{jaynes2003probability,contributors2021differential}, we include the proof for the article completeness.}
Note that giving $\hat{\Omega}_{ij}$ is equivalent to giving $\hat{\Sigma}_{ij}$, since $ \hSigma_{ij} = \hat{\Omega}_{ij}-\mu_i\mu_j$.

We start by the simple case $n = 1$, for which $\vec{x}$, $\vec{\mu}$, $\hSigma$ and $\hat{\Omega}$ are the scalars $x$, $\mu$, $\Sigma$ and $\Omega$.
For notational simplicity, only in this section we denote $P(x | \mu\Sigma)$ by $P(x)$.
The goal is to find the distribution $P(x)$ that maximizes the entropy
\begin{align}
    S = \int_{-\infty}^{\infty} dx\, P(x) \log P(x) ,
    \label{eq:S_1var}
\end{align}
subject to the conditions
\begin{align}
    \begin{split}
        & \int_{-\infty}^{\infty} dx\, P(x) = 1 ,
        \quad
        \mu = \int_{-\infty}^{\infty} dx\, P(x) x ,
        \\
        & \Omega
        = \int_{-\infty}^{\infty} dx\, P(x) x^2 .
        \end{split}
    \label{eq:conds_1var}
\end{align}
We apply the method of Lagrange to $P(x)$ with multipliers $\lambda-1$, $\lambda_1$ and $\lambda_{11}$:
\begin{align*}
    \int_{-\infty}^{\infty}
    \delta P(x)
    \left(
        \log P(x) + \lambda
        +  \lambda_1 x
        +  \lambda_{11} x^2
    \right) = 0 ,
\end{align*}
and thus
\begin{align}
    P(x)
    = \frac{1}{e^\lambda}
    \exp
    \left(
        - \lambda_1 x
        - \lambda_{11} x^2
    \right) .
    \label{eq:Px_mults}
\end{align}
To determine the multipliers $\lambda$, $\lambda_1$ and $\lambda_{11}$, we insert Eq.~\eqref{eq:Px_mults} in Eqs.~\eqref{eq:conds_1var}.
With this, we get
\begin{align}
    P(x) = \frac{1}{\sqrt{2\pi\Sigma}}
    \exp
    \left(
        -\frac{(x-\mu)^2}{2\Sigma}
    \right) ,
    \label{eq:Gauss_proof}
\end{align}
which ends the proof for $n=1$.

Now, consider $n>1$.
Once again, only in this section for notational simplicity, we denote $P(\vec{x} | \vec{\mu}\hat{\Sigma})$ by $P(\vec{x})$.
The goal is to find the distribution $P(\vec{x})$ that maximizes the entropy
\begin{align}
    S = \int_{\mathbb{R}^n} d^nx \,
    P(\vec{x}) \log P(\vec{x}) ,
    \label{eq:S}
\end{align}
subject to the conditions
\begin{align}
    \begin{split}
        & \int_{\mathbb{R}^n} d^nx \, P(\vec{x}) = 1 ,
        \quad
        \mu_i = \int_{\mathbb{R}^n} d^nx \, P(\vec{x}) x_i ,
        \\
        & \hat{\Omega}_{ij}
        = \int_{\mathbb{R}^n} d^nx \,
        P(\vec{x}) x_i x_j
        \quad
        \text{for } i, j = 0, \dots, n-1 .
    \end{split}
    \label{eq:conds_1}
\end{align}
It is convenient to perform a coordinate change $\vec{y} = U \vec{x}$, with $U$ unitary such that $\hat{\Xi} = U \hat{\Omega} U^\dagger$ is diagonal.
Because $\hat{\Omega}$ is symmetric and real, $U$ exists and can be chosen real.
Define $\vec{\nu} = U \vec{\mu}$ and note that $P(\vec{x}) = P(\vec{y})$, since the Jacobian of the change $\vec{x} \to \vec{y}$ is $|U| = 1$.
Rewriting Eqs.~\eqref{eq:S} and \eqref{eq:conds_1} in the $\vec{y}$ coordinates is then trivial: It only requires the changes $\mu_i \to \nu_i$ and $\hat{\Omega}_{ij} \to \hat{\Xi}_{ij}$.

We apply the method of Lagrange to $P(\vec{y})$ with multipliers $\lambda-1$, $\lambda_i$ and $\lambda_{ij}$:
\begin{align*}
    \int_{\mathbb{R}^n} d^ny \,
    &
    \delta P(\vec{y})
    \bigg(
        \log P(\vec{y}) + \lambda
        \\
        &
        + \sum_i \lambda_i y_i
        + \sum_{i,j} \lambda_{ij} y_i y_j
    \bigg) = 0 ,
\end{align*}
and thus
\begin{align*}
    P(\vec{y})
    = \frac{1}{e^\lambda}
    \exp
    \bigg(
        -\sum_i \lambda_i y_i
        -\sum_{i,j} \lambda_{ij} y_i y_j
    \bigg) .
\end{align*}
To determine the multipliers, we insert this expression into Eqs.~\eqref{eq:conds_1} (rewritten for the $y$ variable):
$\int d^ny \, P(\vec{y}) = 1$ allows us to write
\begin{align}
    P(\vec{y})
    = \frac{
        \exp
        \big(
            -\sum_i \lambda_i y_i
            -\sum_{i,j} \lambda_{ij} y_i y_j
        \big)
    }{
        \int d^ny'
        \exp
        \big(
            -\sum_i \lambda_i y_i'
            -\sum_{i,j} \lambda_{ij} y_i' y_j'
        \big)
    } .
    \label{eq:Py_1}
\end{align}
$\hat{\Xi}_{ij} = \int d^ny \, P(\vec{y}) y_i y_j$ for $i \neq j$ implies
\begin{align*}
    \frac{\partial}{\partial \lambda_{ij}}
    \int d^ny
    \exp\bigg(
        -\sum_i \lambda_i y_i
        -\sum_{ij} \lambda_{ij} y_i y_j
    \bigg) = 0 ,
\end{align*}
so the exponentials in Eq.~\eqref{eq:Py_1} are independent on $\lambda_{ij}$ for $i \neq j$.
Therefore,
\begin{align}
    P(\vec{y})
    = \frac{
        \exp
        \big[
            -\sum_i (\lambda_i y_i + \lambda_{ii} y_i^2)
        \big]
    }{
        \int d^ny'
        \exp
        \big[
            -\sum_i (\lambda_i y_i' + \lambda_{ii} {y_i'}^2)
        \big]
    } .
    \label{eq:Py_2}
\end{align}
From this equation,
\begin{align*}
    P(\vec{y}) = \prod_i P(y_i) ,
\end{align*}
with
\begin{align*}
    P(y_i)
    = \frac{
        \exp(-\lambda_i y_i - \lambda_{ii} y_i^2)
    }{
        \int d^ny'
        \exp
        \big(
            -\lambda_i y_i' - \lambda_{ii} {y_i'}^2
        \big)
    } ,
\end{align*}
which in turn yields
\begin{align*}
    S = \sum_i S_i, \quad
    S_i \equiv \int_{-\infty}^{\infty} dy_i \,
    P(y_i) \log P(y_i) .
\end{align*}
In other words, the maximization of $S$ can be decoupled in $n$ independent maximizations of $S_i$, each of them respective to $P(y_i)$ and subject to the conditions $\langle y_i \rangle = \nu_i$ and $\langle y_i^2 \rangle = \hat{\Xi}_{ii}$.
The solution for each $i$ is given by Eq.~\eqref{eq:Gauss_proof}:
\begin{align*}
    P(y_i) =
    \frac{1}{\sqrt{2\pi \Lambda_{y_i}}}
    \exp
        \left(
            -\frac{
            (y_i-\nu_i)^2
            }
            {2\Lambda_{y_i}}
        \right) ,
\end{align*}
with $\Lambda_{y_i} = \langle y_i^2 \rangle -\langle y_i \rangle^2$.
Transforming back to the original $\vec{x}$ coordinates, we get
\begin{align*}
    P(\vec{x} | \mu\hSigma)
    = \frac{1}{\sqrt{(2\pi)^n |\hat{\Sigma}|}}
    \exp\left(
        -\frac{1}{2}
            (\vec{x}-\vec{\mu})
            \hat{\Sigma}^{-1}
            (\vec{x}-\vec{\mu})
    \right) ,
\end{align*}
with $\hat{\Sigma}_{ij} = \langle x_i x_j \rangle-\mu_i\mu_j$.
This completes the proof.

\section{Priors}
\label{app:priors}

In this appendix, we discuss how to calculate non informative priors.
By a non informative prior, we mean a prior distribution that contains no information other than certain symmetry arguments.
Rather than being assumptions, these symmetry arguments, encoded in transformation groups, are necessary for a consistent description of the problem.
This point will become clear throughout the calculation.
We start by computing the priors of the mean and the variance of a single variable, needed for Sec.~\ref{sec:onedim}.
After that, we give the priors of the mean and the correlation matrix that appear in Secs.~\ref{sec:autocorr} and ~\ref{sec:crosscorr}.

\subsection{Single variable}

Let us address the case of a single scalar variable $x$.
We aim at estimating the prior probability distribution $P(\nu, \Lambda)$ of its mean $\nu$ and variance $\Lambda$.
The only condition or ``symmetry'' that we impose on $P(\nu, \Lambda)$ is that it should not depend on the arbitrary choice of the reference frame and units to measure $x$. In other words, $P(\nu, \Lambda)$ must be invariant under the ``frame'' transformations $x \to x'$ with 
\begin{equation}
x' = a(x-\langle x \rangle)+ \nu + b.
\label{eq:frameTransformation}
\end{equation}
Here, $b$ shifts the origin of the $x$ reference frame, whereas the scaling factor $a$ changes the units in which we measure the displacements of $x$ from its mean.
Under this transformation, the mean and variance change to
\begin{align*}
    \nu' = \nu + b,
    \quad
    \Lambda' = a^2 \Lambda .
\end{align*}
The invariance of the prior is expressed as
\begin{align*}
    P(\nu, \Lambda) d\nu d\Lambda
    = P(\nu', \Lambda') d\nu' d\Lambda' ,
\end{align*}
which implies
\begin{align}
    P(\nu, \Lambda)  = a^2 P(\nu + b, a^2\Lambda)
    \label{eq:muL_cond}
\end{align}
for all $a$, $b$.
Assuming that $P(\nu, \Lambda)$ is differentiable in the interior of its domain ($\nu \in \mathbb{R}$, $\Lambda \geq 0$), Eq.~\eqref{eq:muL_cond} is equivalent to the system of differential equations
\begin{align}
    \begin{cases}
    & \partial_{\nu} P(\nu, \Lambda) = 0,
    \\
    &
    2\Lambda
    \partial_{\Lambda} P(\nu, \Lambda)
    + 2P(\nu, \Lambda) = 0 .
    \end{cases}
    \label{eq:eqs_diff}
\end{align}
This equivalence can be proved by differentiating Eq.~\eqref{eq:muL_cond} with respect to $a$ and $b$ and then setting the values $a = 1$, $b = 0$.
The only solution reads
\begin{align}
    P(\nu, \Lambda)
    = \frac{\text{const}}{\Lambda}.
    \label{eq:prior_new}
\end{align}
The result can be alternatively written as
\begin{align}
    \begin{split}
        & P(\nu, \Lambda) \propto P(\nu) P(\Lambda),
        \\
        & P(\nu) \propto 1,
        \quad
        P(\Lambda) \propto \frac{1}{\Lambda} .
    \end{split}
    \label{eq:prior_factor_new}
\end{align}
We note that the scale-invariant prior that we have obtained for the variance is known as the \emph{Jeffreys prior} \cite{jaynes2003probability}. We still need to repeat the calculation for a case relevant for spectral analysis, namely when it is know that the mean $\nu$ of $x$ is $0$. We thus consider the transformation involving only scaling, $ x' = a x $. Proceeding identically to the previous case, we again get Eq.~\eqref{eq:prior_new}. The derivation of the priors for the one-dimensional case is complete.\footnote{We note in passing that an affine transformation, $x \to x' = ax +b$, would give a different prior for the case where the mean is nonzero, namely $P(\nu,\Lambda) \propto 1/\Lambda^{3/2}$.}

\subsection{Many variables, factorizable priors}

Let us consider the priors for the multivariate case, that is priors of $\mu$ and $\hLambda$---defined by Eq.~\eqref{eq:hLambdak_acor}---used in Sec.~\ref{sec:autocorr}.
We proceed to argue that
\begin{align}
    P(\mu, \hLambda)
    = P(\mu, \Lambda_0) \prod_{k=1}^{\nHalf} P(\Lambda_k),
    \label{eq:factor}
\end{align}
with $P(\Lambda_k)$ given by Eq.~\eqref{eq:prior_factor_new}.
Eq.~\eqref{eq:factor} seems reasonable due to the decoupling of the variables $A_j$ into the $\nHalf+1$ independent variables $\alpha_k$, with $k = 0, \dots, \nHalf$---see the discussion after Eq.~\eqref{eq:nts1}.
This decoupling is manifest in the sampling distribution $P(\hatalpha | \mu\hLambda)$---Eq.~\eqref{eq:sam_fac_M}---and suggests to consider priors that also factor in $k$.
However, let us take up a more systematic approach to get Eq.~\eqref{eq:factor}.
The invariance of the prior under the frame transformation Eq.~\eqref{eq:frameTransformation} of $A$ reads:
\begin{align}
    P(\mu', \hLambda')
    d\mu' d\Lambda_0' \dots d\Lambda_{\nHalf}'
    = 
    P(\mu, \hLambda)
    d\mu\, d\Lambda_0 \dots d\Lambda_{\nHalf} ,
    \notag
\end{align}
which implies
\begin{align}
    P(\mu, \hLambda)
    =
    a^{2\nHalf+2}
    P(\mu+b, a^2\hLambda) .
    \label{eq:invariance_many_var}
\end{align}
It is immediate to see that Eq.~\eqref{eq:factor} together with Eqs.~\eqref{eq:prior_new} fulfills this condition.
However, we now demonstrate that it is not the only solution. We proceed as before in this section:
We rewrite Eq.~\eqref{eq:invariance_many_var} as a system of differential equations and solve it.
The differential equations read:
\begin{align}
    \begin{cases}
        & \partial_{\mu} 
        P(\mu, \hLambda) = 0,
    \\
    &
    2\hLambda \cdot
    \partial_{\hLambda} P(\mu, \hLambda)
    + (2\nHalf+2) P(\mu, \hLambda) = 0 .
    \end{cases}
    \notag
\end{align}
Here, we used the obvious notation $\hLambda \cdot \partial_{\hLambda} \equiv \sum_k \Lambda_k \partial_{\Lambda_k}$.
The first equation is trivial.
It implies the independence of $P(\mu, \hLambda)$ on $\mu$.
We can thus focus on the second,
\begin{align}
    2\hLambda \cdot
    \partial_{\hLambda} P(\hLambda)
    + (2\nHalf+2) P(\hLambda) = 0 .
    \label{eq:eqs_diff_2}
\end{align}
A canonical way to solve linear differential equations in manifolds is the method of separation of variables \cite{haberman1983elementary}.
It starts by a factored ansatz like Eq.~\eqref{eq:factor}, leaving for the end the proof that it is the unique solution.
Here, the ansatz leads to the system of uncoupled equations
\begin{align}
    \Lambda_k \partial_{\Lambda_k} P(\Lambda_k)
    +\lambda_k P(\Lambda_k) = 0,
    \quad k = 0, \dots, \nHalf,
    \label{eq:eqs_diff_3}
\end{align}
with the constants $\lambda_k$ fulfilling
\begin{align}
    \sum_{k=0}^{\nHalf} \lambda_k = \nHalf + 1.
    \label{eq:lambdas}
\end{align}
We can solve these equations to conclude that any function of the form
\begin{align}
    P(\mu, \hLambda)
    \propto \prod_{k=0}^{\nHalf}
    \frac{1}{\Lambda_k^{\lambda_k}}
    \label{eq:P_lambdas}
\end{align}
is a solution of Eq.~\eqref{eq:invariance_many_var}.
That is, such a function is invariant under the frame transformations of $A$.
We also impose $\lambda_k > 0$ to have $\lim_{\Lambda_k\to\infty} P(\mu, \hLambda) = 0$.

In many classical problems that require solving a differential equation like the one in Eq.~\eqref{eq:eqs_diff}, one has a set of boundary conditions (along a so-called characteristic curve)\cite{haberman1983elementary} that (i) narrow down the solution to unique values of $\lambda_k$ and (ii) allow one to prove that the factored ansatz is actually the only solution.
However, we do not have such boundary conditions for the prior $P(\mu, \hLambda)$ and thus we find that the frame invariance does not lead to a unique factorizable prior. 

We now invoke an additional invariance argument: the prior for the noise at a fixed \emph{physical} frequency should not depend on how many data points the batch contains. The Fourier index $k$ corresponds to the frequency $k/n \Delta$, where $\Delta$ is the time step with which the data are measured. Crucially, changing $n$ changes the assignments of the indexes $k$ to the physical frequencies. If the reassignement is to leave the priors the same, it must be that the constants $\lambda_k$ for $k\neq 0$ actually do not depend on $k$. Requiring in addition that the prior for $\Lambda_{k=0}$, is described by Eq.~\eqref{eq:prior_new}, we arrive at $\lambda_k=1$ for all $k$. With this additional invariance requirement, we conclude that there is a unique factorizable prior,
\begin{align}
    P(\mu, \hLambda)
    \propto \frac{1}{\Lambda_0}\prod_{k=1}^{\nHalf}
    \frac{1}{\Lambda_k},
    \label{eq:P_lambdas_new}
\end{align}
where we separated $\Lambda_{k=0}$ which is a somewhat different variable than all other $\Lambda_{k\neq0}$. 

\subsection{Many variables, non-factorizable priors}

It is not too difficult to find a non-factorizable prior fulfilling Eq.~\eqref{eq:eqs_diff_2}. We first note the following function family,
\begin{equation}
P(\hLambda) = \left( (\Lambda_0^p + \Lambda_1^p + \cdots +\Lambda_{\nHalf}^p)^{1/p}\right)^{-\nHalf-1}.
\label{eq:nonfactorizablePrior}
\end{equation}
Here, $p$ can be any real number except zero, though probably only $p\geq1$, or at least $p>0$, can ever be relevant for scenarios realistic in physics. As the expression in the outer bracket is the p-norm of the vector $\hLambda$, this family already includes appealing choices, such as an expression which depends only on the sum of $\Lambda_k$ (for $p=1$), the Euclidean distance in the space of vectors $\hLambda$ (for $p=2$), or the maximum  $\mathrm{max}_k\,\Lambda_k$ (for $p \to \infty)$. Yet another possibility worth considering is 
\begin{equation}
P(\hLambda) = \frac{1}{\Lambda_0}\left( (\Lambda_1^p + \cdots +\Lambda_{\nHalf}^p)^{1/p}\right)^{-\nHalf},
\label{eq:nonfactorizablePrior2}
\end{equation}
a prior factorizable in $\Lambda_0$ and the rest.

A second look on Eq.~\eqref{eq:eqs_diff_2} reveals that it specifies the prior $P(\hLambda)$ as a homogeneous function of variables $\hLambda$ of degree $-\nHalf-1$, by the Euler theorem on homogeneous functions \cite{contributors2021homogeneous}. Equations \eqref{eq:P_lambdas}, \eqref{eq:P_lambdas_new}, \eqref{eq:nonfactorizablePrior}, and \eqref{eq:nonfactorizablePrior2} are instances of such homogeneous functions, and there are many more. Still, we speculate that the requirement of the symmetry of the prior, either factorizable or not, with respect to permutation of the indexes $k$ within the set $k\neq 0$, is a relevant invariance requirement. It leads to a unique factorizable prior and  reduces the number of possibilities to consider for a non-factorizable one, suggesting Eqs.~\eqref{eq:nonfactorizablePrior} or \eqref{eq:nonfactorizablePrior2} as (perhaps the only) viable candidates. 

To conclude, without imposing extra conditions beyond the frame invariance, the prior is not unique. 
We adopt Eq.~\eqref{eq:P_lambdas_new} and its analogs as the prior for the main text, following the common practice of the separation of variables technique. We leave the investigations of the consequences of non-factorizable priors as a future research avenue.

\subsection{Priors for cross-correlations}

At last, we make some remarks about the priors needed for Sec.~\ref{sec:crosscorr}.
Following the notation of the main text, beware that now $\hLambda$ refers to Eq.~\eqref{eq:hLambdak} and not Eq.~\eqref{eq:hLambdak_acor}.
An analogous discussion to that of Eq.~\eqref{eq:factor} leads to 
\begin{subequations}
\begin{align}
    & P(\mu_A, \mu_B, \hLambda)
    =
    \prod_{k=0}^{\nHalf}
    P\big(\mu_A, \mu_B, \hLambda_k\big) ,
    \label{eq:factor_ccor_app_a}
\end{align}
with
\begin{align}
    & P(\mu_A, \mu_B, \hLambda_k)
    =
    P\big(\Lambda^A_k\big)
    P\big(\Lambda^B_k\big)
    P(s_k) P(\phi_k),
        \label{eq:factor_ccor_app_b}
\end{align}
        \label{eq:factor_ccor_app}
\end{subequations}
for all $k$.
$P\big(\Lambda_k^A\big)$ and $P\big(\Lambda_k^B\big)$ are given by Eq.~\eqref{eq:prior_factor_new}.
Note that we have assumed the factoring of the prior in $k$ and also in the independent quantities $\{\Lambda_k^A, \Lambda_k^B, s_k, \phi_k\}$ that parametrize $\hLambda_k$ for each $k$.
Analogously to the previous case, the factorization does not follow from the invariance requirements that we impose. 

In Eq.~\eqref{eq:factor_ccor_app}, it remains to determine the priors $P(s_k)$ and $P(\phi_k)$.
The correlation strength $s_k$ and the correlation phase $\phi_k$ are the modulus and the argument of $\Lambda^{AB}_k/(\Lambda_k^A \Lambda_k^B)^{1/2}$, respectively.
Similarly to the quantity $\Lambda$ of the one-dimensional case, discussed in the beginning of this appendix, $\Lambda_k^A$, $\Lambda_k^B$ and $\Lambda_k^{AB}$ transform by a multiplicative factor under the frame transformation. 
Therefore, $s_k$ and $\phi_k$ remain invariant and there are no nontrivial conditions such as Eqs.~\eqref{eq:muL_cond} that apply to $P(s_k)$ and $P(\phi_k)$.
We can impose only entropy maximization on these priors.
The result, of a trivial proof, is a uniform distribution:
\begin{align*}
    &
    P(s_k) = 1 \ \forall k, 
    \\
    &
    P(\phi_k) =
    \begin{cases}
        [\delta(\phi_0)+\delta(\phi_0-\pi)]/2
        & \text{for } k=0, n/2,
        \\
        1/(2\pi)
        & \text{otherwise.}
    \end{cases}
\end{align*}
The phase prior for $k = 0, n/2$ follows from $\Lambda^{AB}_0$ and $\Lambda^{AB}_{n/2}$ (for even $n$) being real.

\section{Derivation of the periodogram distribution}
\label{app:per_dist}

In this appendix, we derive the distribution $P(\bLambda_k | \Lambda_k)$, necessary to understand the statistical properties of the periodogram $\bLambda_k$.
The distribution can be calculated as follows.
We start by applying some probability-theory rules on $P(\bLambda_k | \Lambda_k)$:
\begin{align*}
    & P\big(\bLambda_k | \Lambda_k) = 
    \int_{-\infty}^{\infty}
    d\mu
    \prod_{m=0}^{M-1}
    d\alpha_k^{(m)}
    P(
        \bLambda_k
        \bm{\alpha}_k
        \mu
        | \Lambda_k
    )
    \notag
    \\
    & %
    = \int_{-\infty}^{\infty}
    d\mu
    \prod_{m=0}^{M-1}
    d\alpha_k^{(m)}
    P(\mu | \Lambda_k)
    P\big(
        \bm{\alpha}_k
        | \mu\Lambda_k
    \big)
    P\big(
    \bLambda_k | \bm{\alpha}_k
    \mu\Lambda_k\big) .
    \notag
\end{align*}
The first equality uses Eq.~\eqref{eq:rule}.
In the second, we apply the conditional-probability definition $P(ab) = P(a)P(b|a)$ twice. 
The differentials in the variables $\alpha$ read
\begin{align*}
    & d\alpha_k^{(m)}
    =
    \begin{cases}
        d\mathfrak{R}\alpha_k^{(m)}
        & \text{for } k = 0, n/2,
        \\
        d\mathfrak{R}\alpha_k^{(m)}
        d\mathfrak{I}\alpha_k^{(m)}
        & \text{otherwise},
    \end{cases}
\end{align*}
because $\alpha_k^{(m)}$ is real for $k = 0, n/2$ and complex (in general) otherwise.
Consequently, the number of one-dimensional integrals over $\alpha$ is $2Md_k$.
Next, we use that $P(\mu|\Lambda_k)=P(\mu)$ and define the average
\begin{equation}
\overline{P\big(\bm{\alpha}_k | \mu\Lambda_k\big)} = \int d\mu \, P(\mu) P\big(\bm{\alpha}_k | \mu\Lambda_k\big).
\label{eq:C1}
\end{equation}
Also, using the definition of Eq.~\eqref{eq:periodogram}, we write $P(\bLambda_k | \bm{\alpha}_k \mu\Lambda_k)$ as a delta function. These two replacements give
\begin{equation}
\begin{split}
    P\big(\bLambda_k & | \Lambda_k)
    = \int \prod_{m=0}^{M-1}
    d\alpha_k^{(m)}
        \overline{P\big(\bm{\alpha}_k | \mu\Lambda_k\big)}\\
    & \times
    \delta\left(
    \bLambda_k - \frac{1}{M}\sum_{m=0}^{M-1}
    \big|
        \alpha_k^{(m)}-\delta_{k,0}\bar{\alpha}_0
    \big|^2\right) .
    \notag
\end{split}
\end{equation}
We now consider $k \neq 0$ and $k = 0$ separately. 
For $k\neq0$ the 
sampling distribution $P(\bm{\alpha}_k | \mu\Lambda_k)$ is independent on $\mu$. The integral in $\mu$ is then trivial, giving $\int d\mu P(\mu)=1$ by the normalization of the probability distribution $P(\mu)$ and the overline can be simply omitted.
We can now substitute the condition imposed by the delta function in $P(\bm{\alpha}_k | \mu\Lambda_k)$, in turn given by $\prod_m P\big(\bm{\alpha}^{(m)}_k \big| \mu\Lambda_k\big)$ and Eq.~\eqref{eq:nts1}.
This yields
\begin{align*}
    P& \big(\bLambda_k | \Lambda_k)
    =
    \exp\left(
    -M d_k \frac{\bLambda_k}{\Lambda_k}
    \right)
    \left( 
    \frac{d_k}{\pi \Lambda_k}
    \right)^{M d_k}
    \\
    & \times \int \prod_{m=0}^{M-1}
    d\alpha_k^{(m)}
    \delta\left(
    \bLambda_k - \frac{1}{M}\sum_{m=0}^{M-1}
    \big|
        \alpha_k^{(m)} -\delta_{k,0}\bar{\alpha}_0
    \big|^2\right) .
    \notag
\end{align*}
For $k=0$, a simple integration in $\mu$ is needed to evaluate Eq.~\eqref{eq:C1} and confirm that the previous equation holds also for $k=0$, and thus for all $k$.
Using $\delta(x) = d\Theta(x)/dx$ and taking the derivative out of the integrals, we recognize the formula for the volume of a $2Md_k$-dimensional sphere and write
\begin{align*}
    P\big(\bLambda_k & | \Lambda_k)
    \propto
    \exp\left(
    -M d_k \frac{\bLambda_k}{\Lambda_k}
    \right)
    \frac{\partial}{\partial \bLambda_k}
    V_{2Md_k}\left( \sqrt{ M \bLambda_k} \right) ,
    \notag
\end{align*}
where the $N$-dimensional volume $V_N(r)$ reads
\begin{align*}
    V_N(r) =
    \frac{\pi^{\frac{N}{2}}}
    {\Gamma\left( \frac{N}{2}+1 \right)}
    r^N .
\end{align*}
Substituting this expression, taking the derivative with respect to $\bLambda_k$ and normalizing the result, we get Eq.~\eqref{eq:per_dist}.

\section{A useful identity to integrate out the mean}
\label{app:usefulIdentity}

Concerning zero-frequency spectral components which have finite mean, the following formula is useful. The $k=0$ sampling probability from Eq.~\eqref{eq:sampling_zk} is
\begin{subequations}
\begin{equation}
\begin{split}
    &P\big(
         \bm{Z}_0 \big|
        \bm{\nu}_0 \hLambda_0
    \big)
    = \frac{1}{
        \big[
            (2\pi)^2 \Lambda_0^A \Lambda_0^B
            (1-s_0^2)
        \big]^{M/2}}
    \label{eq:sampling_z0} \\
    & \times \exp\left[
        -\frac{1}{2}\sum_m
        \left(\begin{tabular}{c} $\alpha^{(m)}_0-\nu_A$ \\ $\beta^{(m)}_0-\nu_B$ \end{tabular} \right)^\dagger
        \cdot \hLambda_0^{-1} \cdot
        \left(\begin{tabular}{c} $\alpha^{(m)}_0-\nu_A$ \\ $\beta^{(m)}_0-\nu_B$ \end{tabular} \right)
    \right] ,
\end{split}
\end{equation}
where we have used $d_k=1/2$, and, with $s_\phi=\exp(i\phi)$,
\begin{equation}
   \hLambda_0^{-1}=\left( \begin{tabular}{cc}
        $\frac{1}{\Lambda_0^A}$
        & $-\frac{
            s_0 s_\phi
           }{\sqrt{\Lambda_0^A\Lambda_0^B}}$
        \\
        $-\frac{
            s_0 s_\phi^*
        }{\sqrt{\Lambda_0^A\Lambda_0^B}}$
        & $\frac{1}{\Lambda_0^B}$
    \end{tabular}
    \right),
\end{equation}
\end{subequations}
We note that all quantities in the above equations are real. It means that $s_\phi$ is either $+1$ or $-1$. The variable $\bm{Z}_0$ denotes together the two sets $\{\alpha^{(m)}_0\}$ and $\{\beta^{(m)}_0\}$ that contain $M$ real numbers each. It is now an algebraic exercise to cast the expression into the following form
\begin{subequations}
\label{eq:pz0}
\begin{equation}
\begin{split}
    P\big(
        & \bm{Z}_0 \big|
        \bm{\nu}_0 \hLambda_0
    \big)
    = \frac{
    \exp\left[-\frac{M}{2 (1-s_0^2)} X \right]
    }{
        \big[
            (2\pi)^2 \Lambda_0^A \Lambda_0^B
            (1-s_0^2)
        \big]^{M/2}},
\end{split}
\end{equation}
where 
\begin{equation}
\begin{split}
X = & \,
\frac{\bLambda_0^A}{\Lambda_0^A} 
+ \frac{\bLambda_0^B }{\Lambda_0^B} 
-\frac{2 s_0 s_\phi \bLambda_0^{AB}}{\sqrt{\Lambda_0^A \Lambda_0^B}} 
+\frac{1-s_0^2}{\Lambda_0^A}(\nu_A-\bar{\alpha}_0)^2\\
 & + \frac{1}{\Lambda_0^B}\left(\nu_B- \bar{\beta}_0 -s_0 s_\phi \sqrt{\frac{\Lambda_0^B}{\Lambda_0^A}} (\nu_A-\bar{\alpha}_0) \right)^2.
\end{split} 
\end{equation}
In the derivation, it is helpful to notice that $\bLambda_0^A$ and $\bLambda_0^B$ are the variances of the sets $\{\alpha^{(m)}_0\}$ and $\{\beta^{(m)}_0\}$, and $\bLambda_0^{AB}$ is the covariance of the two sets.
Particularly, the (co)variance is not changed upon a uniform shift of all input numbers.
\end{subequations}
One can use Eq.~\eqref{eq:pz0} to calculate the sampling distributions averaged over the mean, $\overline{P(\cdots | \nu_0 \hLambda_0)}$. Indeed, setting $s=0$ and ignoring the $B$-related variables, one can easily integrate out the mean $\mu_A$ in Eq.~\eqref{eq:pz0}, deriving Eq.~\eqref{eq:autocorr} from Eq.~\eqref{eq:bbb} for $k=0$. Similarly, integrating out both $\nu_A$ and $\nu_B$ in Eq.~\eqref{eq:pz0} gives the $k=0$ sampling distribution entering Eq.~\eqref{eq:LA_cross_prev}.

\section{Posterior for unnormalized cross-correlation strength}
\label{app:unnormalizedCrossCorrelation}

Here we consider an alternative parametrization of the cross-correlation, using $\{\Lambda_k^A, \Lambda_k^B, |\Lambda_k^{AB}|, \phi_k\}$ instead of the set $\{\Lambda_k^A, \Lambda_k^B, s_k, \phi_k\}$ considered in the main text. These sets differ only in one variable, being $|\Lambda_k^{AB}|$ instead of $s_k$. Therefore, the only additional marginal posterior to calculate is $p(|\Lambda_k^{AB}|)$: the remaining three posteriors remain as given in the main text. Indeed, we have
\begin{equation}
\begin{split}
&p( \Lambda_k^A, \Lambda_k^B, s_k, \phi) \mathrm{d}\Lambda_k^A \mathrm{d}\Lambda_k^B \mathrm{d}s \,\mathrm{d}\phi \\
&= p( \Lambda_k^A, \Lambda_k^B, |\Lambda_k^{AB}|, \phi) \frac{\mathrm{d}s_k}{\mathrm{d}|\Lambda_k^{AB}|} \mathrm{d}\Lambda_k^A \mathrm{d}\Lambda_k^B \mathrm{d} |\Lambda_k^{AB}| \mathrm{d}\phi,
\end{split}
\end{equation}
and one can see that the marginals of $\Lambda_A$, $\Lambda_B$, and $\phi$ are the same using any of the coordinates sets, since integrating out $s_k$ is the same as integrating out $|\Lambda_k^{AB}|$. However, the marginals of $s_k$ versus $|\Lambda_k^{AB}|$ are different and can not be obtained from each other, as these two variables define different hyperplanes in the four-dimensional space.

\begin{figure}
    \centering
    \includegraphics{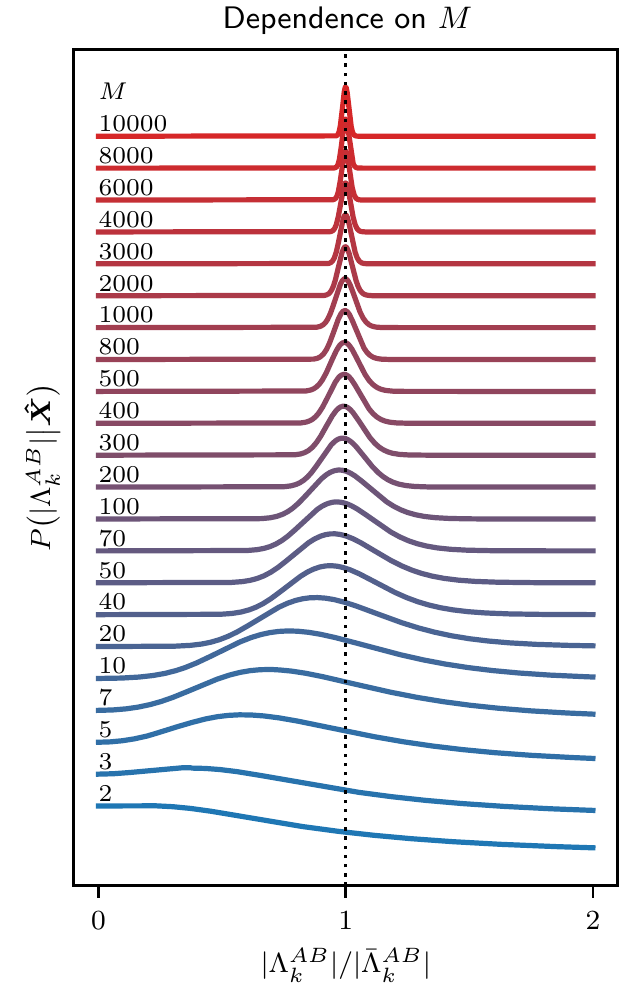}
    \caption{
    Distribution of the magnitude of unnormalized cross-correlations as a function of $M$ for $\bar{s}_k = 0.7$.
    The curves are shifted and scaled in the vertical axis for visibility.
}
    \label{fig:unnorm_ccor_theory_M}
\end{figure}

For completeness, we repeat the definition of the marginal,
\begin{equation}
\begin{split}
    &P\left(|\Lambda_k^{AB}| \Big| \bm{Z}_k\right)\\
    &= 
    \int_0^{\infty} d\Lambda_k^A
    \int_0^{\infty} d\Lambda_k^B
    \int_0^{2\pi} d\phi_k
 \overline{P(\Lambda_k^A \Lambda_k^B s_k\phi_k \bm{\nu}_k| \bm{Z}_k)}
    \\
    &
    \propto
    \int_0^{\infty} d\Lambda_k^A
    \int_0^{\infty} d\Lambda_k^B
        \int_0^{2\pi} d\phi_k
    \overline{P(\bm{Z}_k | \bm{\nu}_k\hLambda_k)} P(\hLambda_k).
    \label{eq:xxxxxx}
\end{split}
\end{equation}
We could cast it as the following one-dimensional integral

\begin{equation}
\begin{split}
    &P\left(|\Lambda_k^{AB}| \Big| \bm{Z}_k\right)
    \propto  \frac{1}{|\Lambda_k^{AB}|^{2d_k(M-\delta_{k,0})+1}  }
    \int_0^\infty dx \frac{x^{d_k(M-\delta_{k,0})}}{x^\frac{1}{2}(1+x)^\frac{3}{2}}
    \\
    & \times
    I_0\left( \frac{2 d_k M x}{|\Lambda_k^{AB}|/|\bar{\Lambda}_k^{AB}|} \right)
    K_0\left( \frac{2 d_k M \sqrt{x(x+1)}}{\bar{s}_k|\Lambda_k^{AB}|/|\bar{\Lambda}_k^{AB}|} \right)
    \label{eq:LA_cross_unnormalized},
\end{split}
\end{equation}
where $\bar{s}_k$ and $|\bar{\Lambda}_k^{AB}|$ are as defined in the main text, $K_n$ is the modified Bessel function of the second kind, and $I_n$ is the modified Bessel function of the first kind. Note how $|\bar{\Lambda}_k^{AB}|$ sets the scale for $\Lambda_k^{AB}$. In Fig.~\ref{fig:unnorm_ccor_theory_M} we illustrate this distribution for different values of $M$.

\section{Averaging overlapping points in log-log plots}

\label{app:pointAveraging}

The results of the estimation, the spectral plots, are often presented using the log-log scale, such as in Fig.~\ref{fig:acorr_clustered_AB}. It is useful in revealing a possible power-law frequency dependence of the spectral densities, which is often the case. On the other hand, it comes with a disadvantage of overcrowding the figure at higher frequencies: since the density of the points on the horizontal axis is constant on the linear scale, it grows exponentially on the log scale. Due to a finite resolution of the printer, computer screen, and the human eye, one can often see just an impenetrable mess in such plots above a certain frequency. Here we propose a solution of this problem.

The idea is to replace several neighboring points on the plot by a single representative point. While this idea seems too obvious, there are different ways how one can merge several points into one. We suggest the following way. We aim to represent a (dimensionfull) frequency interval $\mathcal{F} = \langle \kappa_\textrm{low}, \kappa_\mathrm{high} \rangle$, where for simplicity $\kappa_\mathrm{low}>0$.  It contains a set $\mathcal{K}$ of integer Fourier indexes $k$, that is, $k \in \mathcal{K} \Leftrightarrow \kappa_k \equiv k/n\Delta \in \mathcal{F}$. Let the number of integers in $\mathcal{K}$ be $K$. For each of these integers, one has the associated set $\{\alpha_k^{(m)}\}$. To get the single representative, one should merge these sets into one set containing $KM$ entries, $\{\alpha_\mathrm{eff}^{(m)}\} = \bigcup_{k\in\mathcal{K}}\{\alpha_k^{(m)}\}$. This merged set is interpreted as an input to Eq.~\eqref{eq:suff} corresponding to an effective batch size $KM$. Evaluating Eq.~\eqref{eq:periodogram} gives $\bLambda_\mathrm{eff} = (1/K)\sum_{k\in\mathcal{K}} \bLambda_{k}$, a simple average of the sufficient statistics. One then evaluates Eq.~\eqref{eq:autocorr} with effective parameters $\bLambda_k \to \bLambda_\mathrm{eff}$ and $M \to KM$, to obtain the representative estimating distribution. 

This procedure has the following advantages: 1) One can merge a variable number of Fourier components throughout the plot, for example, plotting non-merged values ($K=1$) at low frequencies, and increasing $K$ towards high frequencies. Therefore, one does not have to sacrifice the lowest frequencies, usually of primary interest, a deficiency which results from artificially splitting the data to shorter batches \cite{welch1967use}. 2) In the limit of a narrow frequency window (more specifically, in the limit where the spectral density is constant within the covered frequency interval), the resulting estimating distribution is \emph{exact}. 3) Since one gets an estimating distribution for the representative point, one still has the freedom of presenting it in various ways, for example, plotting its MLE or average, in a way which is consistent with the non-merged points on the plot. 4) Most crucially, due to 2), a rigorous error bar can be assigned to each representative point. 

Let us point out another paradox related to this approach. The number of points assigned to a single ``representative'' point grows for larger frequency. As one expects, and as we prove in App.~\ref{app:centralLimits} on the central limits, the estimating distributions shrink, in proportion to $1/\sqrt{K}$. It would seem that when such points are used in some subsequent estimation (say, parametric estimation of the continuous spectrum), these points with minuscule error bars will dominate the fit, making the low-frequency points irrelevant. The resolution of the paradox is in realizing that an ``effective point'' \emph{represents an interval} rather than a precise frequency, irrespective of where exactly it is placed. The interval width is proportional to $K$. Therefore, to use an ``effective point'' correctly, one needs to use error bars for both its coordinates, the vertical as well as the horizontal one. 

We have used this averaging in plotting Fig.~\ref{fig:acorr_clustered_AB}. We have split the horizontal axis to intervals of constant length on the log-scale and merged the frequencies within each such interval. This means the $K$ starts at 1 on that plot, becomes larger than 1 at $k=47$ and grows towards high frequencies up to 31 at the highest frequency.

\section{How to treat errors (finite precision) on input}
\label{app:errorsOnInput}

In this article, we have considered stochastic variables, denoted as $A$, $B$, or $C$. Their fluctuations (or noise) is what we aim at extracting. However, apart from their ``inherent'' fluctuations, the values that we use as input in the above formulas are often subject to additional errors. Namely, there is no realistic case where a continuous quantity, such as $A_j$, would be known precisely. In line with the spirit of this article, the values $A_j$ will be given as probability distributions rather than as exact numbers. These probability distributions are results of some measurement protocol, which is nothing else than another case of estimation. This estimation comes with its own precision, and thus non-zero width of those probability distributions on the input. We introduce the following notation,
\begin{equation}
\label{eq:errorA}
A^\prime_i = A_i +\delta A_i,
\end{equation}
with $A^\prime$ the available signal, $A$ the quantity of interest, and $\delta A$ the unwanted error.

To illustrate, consider that $A$ is the energy of a qubit. As there is no way to measure it directly, one needs to extract it from observing, for example, qubit precession, fitting it to damped harmonic oscillations. While there is no issue with the formulas in the main text, since they work whatever is the nature, character, or origin of the fluctuations in the signal, in this case one is interested in the inherent noise of the qubit energy, without the contribution from the imperfect energy-estimation. We now give a simple recipe how to subtract the latter contribution. The subtraction is possible if we can \emph{separately estimate the spectrum of the additional error $\delta A$} and \emph{this error is uncorrelated with the remaining part of the signal $A$}. In the following we give the necessary formulas for correcting the auto- and cross-spectra.

\subsection{Auto-correlation case}

In this scenario, one has individual estimates of the auto-correlation for the available input $A^\prime$ and for the error itself $\delta A$. The estimating distribution of the auto-correlation for the signal with the error removed is
\begin{align}
\tilde{P}(\Lambda_k) &\propto \int_0^\infty  \mathrm{d}\Lambda^\prime_k P^\prime\left(\Lambda^\prime_k\right)  \int_0^\infty  \mathrm{d}\lambda_k p\left(\lambda_k\right)
\nonumber\\&\qquad\qquad\qquad\qquad \quad \times
\delta\left(\Lambda_k+ \lambda_k -\Lambda^\prime_k\right) \label{eq:subtraction}\\
&=\int_0^\infty \mathrm{d}\lambda_k P^\prime(\lambda_k+\Lambda_k) p\left(\lambda_k \right) . \nonumber
\end{align}
Here, the probability distributions $P^\prime$ and $p$ are the estimating distributions given in Eq.~\eqref{eq:autocorr} or Eq.~\eqref{eq:LA_cross} for the error-polluted signal $A^\prime$ and the error $\delta A$, respectively. 
In the simple case of the error being white noise with variance $\sigma^2$, one can use $p(\lambda_k) = \delta(\lambda_k-\sigma^2)$ in the above, and the only calculation required is to get the overall normalization factor as a one-dimensional integral. 

In practice, one can not sample the error together with the signal (since that would mean one could simply subtract the error and use precise inputs). Therefore, the assumption on their zero correlation can not be checked, and will hardly be met in practice. For this reason, we do not delve on this topic more\footnote{First, one could consider more general relations, for example $\Lambda^\prime = \Lambda+ \lambda + s\sqrt{\Lambda\lambda}$ from where the correlation strength $s$ would be integrated out according to some prior or estimated distribution for $s$. But more importantly, Eq.~\eqref{eq:subtraction} is only an ad-hoc approximation; the rigorous removal requires a more complicated procedure, the explanation of which we leave for elsewhere.\label{fnt:adhoc}} and stop at Eq.~\eqref{eq:subtraction} as a rough approximation for the signal stripped of the additional error.
We put tilde on the resulting estimating distribution to emphasize that the result of Eq.~\eqref{eq:subtraction} relies on an explicit assumption (of zero correlation of the signal of interest and the error) which won't be met in practice, unlike all other estimating distributions derived in this article which are, in a well defined sense, exact.
Nevertheless, even if only approximate, we stress that Eq.~\eqref{eq:subtraction} produces a valid estimating distribution (a well-defined probability distribution), whatever is the relation of the signal and error: The error might be comparable or even much bigger than the typical value of signal auto-correlation as estimated from the available data.

\subsection{Cross-correlation case}
\label{app:errorsOnInput-b}

We now extend the input-error-removal procedure to cross-correlations. Equation \eqref{eq:errorA} then applies to each of two stochastic variables $A$ and $B$. The cross-correlator we can access from the error-polluted input data is
\begin{align}
\langle {A^\prime_i B^\prime_j}\rangle=\langle {A_i B_j}\rangle+\langle{\delta A_i \delta B_j}\rangle.
\end{align}
As before, this result assumes that errors are uncorrelated with signals. We now discuss two typical scenarios.

First, consider that $A$ and $B$ are estimated independently, either using separate experimental equipment or unrelated measurement and estimation procedures. In this case, one expects mutually independent errors $\langle{\delta A_i \delta B_j}\rangle=0$, and there is nothing to correct for.\footnote{First, this result still relies on the assumption of the errors being uncorrelated with signals. Otherwise the errors do become correlated through the correlation between the signals; such errors we simply can not correct for. Second, the result is counter intuitive, since acquiring additional relevant information must change the Bayes posterior. The lack of such correction uncovers a deficiency of the approximate procedure expressed by Eq.~\eqref{eq:subtraction} and its analogues. See Footnote \ref{fnt:adhoc}.}

Second, consider that $A$ and $B$ are composite quantities: Not directly accessible, they are derived from other measured or estimated variables. In this situation the errors will often be correlated by common elements in that derivation. Assuming that one can somehow estimate the cross-spectrum $\langle{\delta A_i \delta B_j}\rangle$, it can be removed analogously as before for auto-correlations. 
The main difference is that cross-correlations are complex quantities, having both phase and magnitude. The subtraction needs to be done for estimating distributions describing both, that is, before any these two variables (phase or magnitude) is marginalized out. Let us consider that the removal is done for the estimating distribution expressed using  the unnormalized cross-correlation strength $|\Lambda_k|$ and the phase $\phi_k$. It corresponds to Eq.~\eqref{eq:LA_cross_unnormalized} before integrating out the phase and reads
\begin{align}
&P\left(\Lambda_k\right)
    \propto  \frac{1}{|\Lambda_k|^{2d_k(M-\delta_{k,0})+1}  }
    \int_0^\infty dx \frac{x^{d_k(M-\delta_{k,0})}}{x^\frac{1}{2}(1+x)^\frac{3}{2}}
    \nonumber\\
    & \times
    \exp\left( \frac{2 d_k M \cos(\phi_k-\bar{\phi}_k)}{|\Lambda_k|/|\bar{\Lambda}_k|}x \right)
    K_0\left( \frac{2 d_k M \sqrt{x(x+1)}}{\bar{s}_k|\Lambda_k|/|\bar{\Lambda}_k|} \right)
    \label{eq:LA_phi_cross_unnormalized},
\end{align}
where the complex number $\Lambda_k = |\Lambda_k| \exp(i\phi_k)$. We now use this distribution in the following variants: $P^\prime(\Lambda^\prime)$ for the cross-correlation evaluated with the available---error-polluted---data, $p(\lambda)$ for the cross-correlation of the errors, and $P(\Lambda)$ for the cross-correlation of interest. Using tilde with the same meaning as explained below Eq.~\eqref{eq:subtraction}, the desired estimating distribution is 
\begin{align}
\label{eq:subtraction_cross}
\tilde{P}(\Lambda_k)\propto\int \mathrm{d}\lambda_k  P^\prime(\lambda_k+\Lambda_k) p(\lambda_k) ,
\end{align}
with the integration being over the complex plane. Otherwise, the equation is completely analogous to Eq.~\eqref{eq:subtraction}. The marginal distribution for the magnitude is obtained by integrating out the phase in $\tilde{P}(\Lambda_k)$, and vice-versa. 

Let us finish with comments on using Eq.~\eqref{eq:subtraction_cross} in practice. The formula contains two explicit one-dimensional integrals, while the distributions $P^\prime$ and $p$ require another integral each (they are given by two instances of Eq.~\eqref{eq:LA_phi_cross_unnormalized}, which do not have to be normalized individually as only the final distribution needs to be normalized). Counting also the integration required for the marginalized distributions, we get a procedure requiring five nested integrals, making it computationally demanding. Still, this is the most favorable scenario: The removal can be applied using alternative parametrizations, for example, the set $\{\Lambda_k^A, \Lambda_k^B, \mathfrak{Re}\ \Lambda_k^{AB},\mathfrak{Im}\ \Lambda_k^{AB}\}$. However, using normalized correlation strength $s_k$ among the parameters, the presence of auto-correlations in the denominator [see Eq.~\eqref{eq:sk}] makes the procedure too complex to be practical in our experience. Therefore, we do not give formulas for that variant: the errors on input can be removed only for the cross-correlation parametrized with the unnormalized correlation strength described in App.~\ref{app:unnormalizedCrossCorrelation}.

\section{Central-limit results}

\label{app:centralLimits}

Here we consider how estimating distributions scale in the limit $M\to \infty$. Compared to exact expressions, the limiting approximations have two benefits. First, this limit corresponds to a large number of observations for which the formulas should display the ``central-limit'' behavior: each estimating distribution should become a Gaussian with the variance falling of as $1/M$. Deriving such variance stands for a consistency check and gives insight on how does the distribution width (the uncertainty) depend on sufficient statistics. Second benefit is practical: the exact expressions are complicated functions, often given only as integrals, which become exceedingly unwieldy in this limit. Standard numerical libraries might fail to reliably evaluate hypergeometric functions with parameters of order thousand or more. While one rarely has so many data batches available, very high values of $M$ easily result when plotting representative values according to App.~\ref{app:pointAveraging}. Here, the limit formula might become numerically more precise than the exact one (unless the latter is evaluated in a special way). Perhaps even more importantly, the Gaussian-like posteriors drastically simplify the evaluation of the figure of merit in a parametric fit of the spectrum: Instead of possibly involving complicated integrals, the logarithm of Eq.~\eqref{eq:likelihood} becomes a simple sum of weighted squared differences.\footnote{For example, in the case of correlation spectrum $\Lambda_k$, Eq.~\eqref{eq:likelihood} becomes $\log L \propto \sum_k \sigma_{\Lambda_k}^{-2} [\bLambda_k- 
    \mathcal{F} [\mathcal{C}_{\vec{p}}]\left(\kappa_k\right)]^2$, see Eq.~\eqref{eq:sigmaLambdaA}, and analogously for other variables.} Therefore, the central-limit expressions can be used to quickly locate the approximate position of the minimum for a parametric fit of spectra, followed by a refinement using exact formulas.

In this part, we prefer simplicity to rigor: We are interested only in the leading order results. Among other, we neglect factors of order one compared to $M$, so that we drop any small integer, $\delta_{k,0}$ among them, accompanying $M$ in sum. We skip the derivations, give results only for the most important estimating distributions, and do not specify the error arising from the approximations. Concerning the last point, it is important to understand the following: the results given below are valid in the limit $M\to\infty$ unless the sufficient statistics are singular, for example $\bar{s}=0$ or $\bar{s}=1$. The singular cases might not display the central-limit behavior. This fact implies that when $M$ is fixed and the estimating distributions are considered as functions of their sufficient statistics, the limit expressions will fail once the sufficient statistics become close enough to the singular value, for example once $\bar{s}$ becomes small enough. In other words, at what value the parameter $M$ can be considered ``large'' so that the formulas below apply, depends not only on $M$ itself but also on the values of sufficient statistics.

\subsection{The auto-correlation}

\begin{figure}
    \centering
    \includegraphics{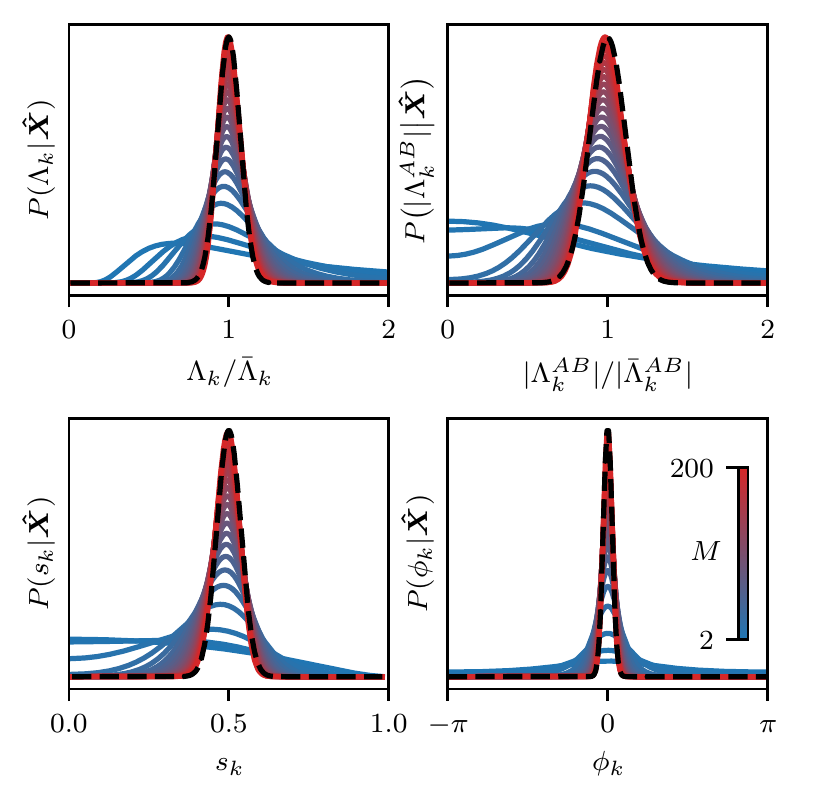}
    \caption{
    Central-limit (large-$M$) approximations to estimating distributions. In each panel, exact estimating distributions for the variable given on the x axis are plotted for several values of $m$ as given by the color bar. The black dashed curves give the approximate (central-limit) estimating distributions for $m=200$.
}
    \label{fig:Central_limit}
\end{figure}

The estimating distribution in Eq.~\eqref{eq:autocorr} is the Poisson distribution, simple enough to work with so that no approximation is needed. Nevertheless, to be consistent with the remainder of this section,
we approximate it by the following form valid in the limit $M\to \infty$:\footnote{The exact expression for the density $p(\lambda;m)$ contains an additional factor $1/\lambda^{\delta_{k_0}/2+1}$.}
\begin{align}
    \label{eq:autocorrCL}
     P(\Lambda_k | \bm{\alpha}_k)\, \mathrm{d}\Lambda_k &\equiv p(\lambda; m)\, \mathrm{d}\lambda,
\end{align}
where we introduced the following variables
\begin{subequations}
\label{eq:autocorrCLForm}
\begin{align}
     \label{eq:plm}
        p(\lambda; m) &= \frac{(m\lambda)^m}{m!} \exp(-m\lambda),\\
    \label{eq:newVariableLambda}
        \lambda &= \frac{\bLambda_k}{\Lambda_k},\\
            \label{eq:newVariableM}
        m &=  M d_k.
\end{align}
\end{subequations}
The distribution $p(\lambda; m)$ in Eq.~\eqref{eq:plm} is the probability to observe $m$ events over time $m$ for a Poisson process with rate $\lambda$. It peaks at rate $\lambda = 1$, corresponding to $\Lambda_k=\bLambda_k$. At large $m$, the distribution is well approximated by
\begin{align}
\label{eq:sigmaLambdaA}
\begin{split}
P(\Lambda_k | \bm{\alpha}_k) &\simeq \frac{1}{\sqrt{2\pi}\sigma_{\Lambda_k}} \exp\left( - \frac{(\Lambda_k - \bLambda_k)^2}{2\sigma_{\Lambda_k}^2} \right),\\
\sigma_{\Lambda_k}^2 &= \frac{ \bLambda_k^2 }{M d_k}.
\end{split}
\end{align}
In the first panel of Fig.~\ref{fig:Central_limit}, we compare Eq.~\eqref{eq:autocorr} to its central-limit approximation in Eq.~\eqref{eq:sigmaLambdaA}.

While it was not exact for  Eq.~\eqref{eq:autocorr}, the appeal of Eq.~\eqref{eq:autocorrCLForm} is that it covers also the case of correlation measurement considered in Sec.~\ref{sec:crosscorr_estimation}. Namely, Eq.~\eqref{eq:LA_cross} reduces to it in the limit $M\to\infty$,
\begin{align}
    \label{eq:autocorrCL2}
     P(\Lambda_k^A | \bm{Z}_k)\, \mathrm{d}\Lambda_k^A &\equiv p(\lambda; m)\, \mathrm{d}\lambda;
\end{align}
the only change needed is to put the upper subscript $A$ on the two quantities on the right hand side of Eq.~\eqref{eq:newVariableLambda}.

\subsection{The normalized cross-correlation}

For large $M$, the hypergeometric function in Eq.~\eqref{eq:skphik} can be set to 1, which greatly simplifies the expression. Integrating out the variable $s_k$, we got
\begin{align}
\label{eq:phiCL1}
P\left(\phi_k \Big| \bm{Z}_k\right)  \propto \exp\left( - M d_k \frac{\bar{s}_k^2}{1-\bar{s}_k^2} \sin^2(\phi_k-\bar{\phi}_k)\right),
\end{align}
a form which is suitable when the explicit periodicity in the phase variable is beneficial. Alternatively, the expression explicitly Gaussian in the phase variable is 
\begin{align}
\label{eq:phiCL2}
\begin{split}
P\left(\phi_k \Big| \bm{Z}_k\right) & \simeq \frac{1}{\sqrt{2\pi}\sigma_{\phi_k}} \exp\left( - \frac{(\phi_k^{AB} -\bar{\phi}_k)^2}{2\sigma_{\phi_k}^2} \right),\\
\sigma_{\phi_k}^2 &= \frac{\bar{s}_k^{-2}-1}{2M d_k}=\frac{1}{2M d_k}\frac{\bLambda_k^A \bLambda_k^B - |\bLambda_k^{AB}|^2}{|\bLambda_k^{AB}|^2}.
\end{split}
\end{align}
Similarly, integrating out the phase gives
\begin{align}
\label{eq:sCL}
\begin{split}
&P\left(s_k \Big| \bm{Z}_k\right)  \simeq \frac{1}{\sqrt{2\pi}\sigma_{s_k}} \exp\left( - \frac{(s_k -\bar{s}_k)^2}{2\sigma_{s_k}^2} \right),\\
&\,\,\,\sigma_{s_k}^2 =\frac{(1-\bar{s}_k^{2})^2}{2M d_k}=\frac{1}{2M d_k}\left( \frac{\bLambda_k^A \bLambda_k^B - |\bLambda_k^{AB}|^2}{\bLambda_k^A \bLambda_k^B} \right)^2.
\end{split}
\end{align}
The two approximations are compared to the exact distributions in the lower two panels of Fig.~\ref{fig:Central_limit}.

\subsection{The unnormalized cross-correlation}

We could cast Eq.~\eqref{eq:LA_cross_unnormalized} into the following form
\begin{subequations}
\label{eq:crosscorrCL}
\begin{align}
\label{eq:crosscorrCL1}
P\left(|\Lambda_k^{AB}| \Big| \bm{Z}_k\right) &\simeq \frac{1}{\sqrt{2\pi}\sigma_{\Lambda_k^{AB}}} \exp\left( - \frac{(|\Lambda_k^{AB}| -|\bLambda_k^{AB}|)^2}{2\sigma_{\Lambda_k^{AB}}^2} \right).
\end{align}
We were not able to obtain an analytic estimate for the variance $\sigma_{\Lambda_k^{AB}}$ directly from Eq.~\eqref{eq:LA_cross_unnormalized}. However, our numerical investigations suggested
\begin{equation}
\label{eq:crosscorrCL2}
\sigma_{\Lambda_k^{AB}}^2 = \frac{\bLambda_k^A \bLambda_k^B}{2M d_k} \left( 1 + \bar{s}_k^2 \right).
\end{equation}
\end{subequations}
As we could derive the same formula from Eq.~\eqref{eq:LA_cross_unnormalized} in the limit $\bar{s}_k \to 0$, we believe that it is valid generally.
The correspondence between the exact distribution and its approximate form is illustrated in the upper right panel of Fig.~\ref{fig:Central_limit}.
Among other, Eq.~\eqref{eq:crosscorrCL} implies that one needs $M \sim \Lambda_k^A \Lambda_k^B / (\Lambda_k^{AB})^2$ batches to discriminate a finite value of $\Lambda_k^{AB}$ from zero, that is, to detect a cross-correlation.

\bibliographystyle{apsrev4-1}
\bibliography{2021-Angel-Spectral-Estimation.out,spectral_analysis.out}

\end{document}